\documentclass[10pt, doublecolumn, nofonttune]{IEEEtran}
\usepackage{amsmath,amsfonts}
\usepackage{algorithmic}
\usepackage{array}
\usepackage{enumitem}
\usepackage{textcomp}
\usepackage{stfloats}
\usepackage{url}
\usepackage{verbatim}
\usepackage{graphicx}
\usepackage{tikz}
\usepackage{pgfplots}
\usepgfplotslibrary{groupplots}
\pgfplotsset{compat=1.18}
\pgfplotsset{every tick label/.append style={font=\scriptsize}}
\usepackage{amsmath}
\usepackage[acronyms,nonumberlist,nopostdot,nomain,nogroupskip,acronymlists={hidden}]{glossaries}
\usepackage{xspace}
\usepackage{float}
\newglossary[algh]{hidden}{acrh}{acnh}{Hidden Acronyms}
\usepackage{soul}
\usepackage{color}
\usepackage{cite}
\usepackage{booktabs}
\usepackage{pgf}
\usepackage{float}
\usepackage[font=small]{caption}
\usepackage[font=footnotesize]{subcaption}
\usepackage{xcolor}
\usepackage{listings}
\usepackage{tabularx}
\usepackage{multirow}
\usepackage{mdframed}
\usepackage[subtle]{savetrees}
\glsdisablehyper

\usepackage{circledsteps}

\usepackage{tcolorbox}
\tcbuselibrary{skins}

\newtcolorbox{promptbox}{
  enhanced,
  colback=gray!10,
  colframe=gray!50,
  arc=0mm,
  title=Prompt,
  fonttitle=\bfseries
}

\newtcolorbox{responsebox}{
  enhanced,
  colback=blue!5,
  colframe=blue!30,
  arc=0mm,
  title=Response,
  fonttitle=\bfseries
}

\definecolor{BlueKey}{rgb}{0.1, 0.3, 0.8}      

\lstdefinestyle{mystyle-json}{
      commentstyle=\color{gray},           
  keywordstyle=\color{BlueKey},        
  numberstyle=\tiny\color{gray},       
  stringstyle=\color{BlueKey},      
  basicstyle=\color{BlueKey}\ttfamily\scriptsize, 
  rulecolor=\color{black},
  breakatwhitespace=true,         
  breaklines=true,                 
  captionpos=b,
  frame=tb,
  keepspaces=true,                 
  numbers=left,                    
  numbersep=5pt,                  
  showspaces=false,                
  showstringspaces=false,
  showtabs=false,                  
  tabsize=2,
  xleftmargin=10pt,
  belowskip=-10pt,
}

\definecolor{luacomment}{RGB}{0,128,0}
\definecolor{luastring}{RGB}{163,21,21}
\definecolor{luakeyword}{RGB}{0,0,255}

\lstdefinestyle{LuaStyle}{
  commentstyle=\color{gray},           
  keywordstyle=\color{BlueKey},        
  numberstyle=\tiny\color{gray},       
  stringstyle=\color{BlueKey},         
  basicstyle=\color{black}\ttfamily\scriptsize, 
  rulecolor=\color{black},
  breakatwhitespace=true,         
  breaklines=true,                 
  captionpos=b,
  frame=tb,                            
  keepspaces=true,                 
  numbers=left,                        
  numbersep=5pt,                  
  showspaces=false,                
  showstringspaces=false,
  showtabs=false,                  
  tabsize=2,
  xleftmargin=10pt,
  belowskip=-10pt,
  morekeywords={and,break,do,else,elseif,end,false,for,function,if,in,local,nil,not,or,repeat,return,then,true,until,while},
  morecomment=[l]{--},                 
}

\lstdefinestyle{myLuastyle}
{
  language         = {[5.0]Lua},
  basicstyle       = \ttfamily,
  showstringspaces = false,
  upquote          = true,
}

\lstdefinestyle{mystyle-lua}{
    commentstyle=\color{green!50!black},
    keywordstyle=\color{blue},
    stringstyle=\color{red},
    basicstyle=\ttfamily\small,
    breakatwhitespace=false,
    breaklines=true,
    captionpos=b,
    keepspaces=true,
    numbers=left,
    numbersep=5pt,
    showspaces=false,
    showstringspaces=false,
    showtabs=false,
    tabsize=2
}

\lstdefinestyle{mystyle-json-no-numbers}{
  commentstyle=\color{gray},           
  keywordstyle=\color{BlueKey},        
  numberstyle=\tiny\color{gray},       
  stringstyle=\color{BlueKey},      
  basicstyle=\color{BlueKey}\ttfamily\scriptsize, 
  rulecolor=\color{black},
  breakatwhitespace=true,         
  breaklines=true,                 
  captionpos=b,
  frame=tb,
  keepspaces=true,                 
  numbersep=5pt,                  
  showspaces=false,                
  showstringspaces=false,
  showtabs=false,                  
  tabsize=2,
  xleftmargin=10pt,
  belowskip=-10pt,
}

\lstdefinelanguage{json}{
  alsoletter={-},
  keywords={listen,on,off,dst,periodic},
  sensitive=false,
  comment=[l]{\#},
  morecomment=[s]{/*}{*/},
  moredelim=[l][\color{orange}]{\&},
  moredelim=[l][\color{magenta}]{*},
  moredelim=**[il][\color{purple}{:}\color{blue}]{:},   
  morestring=[b]',
  morestring=[b]",
}

\newcommand{\framework}{ALLSTaR\xspace}

\newcommand{\slot}{t} 

\newcommand{\CQI}{\text{CQI}} 
\newcommand{\CQIuser}[1]{\text{CQI}_{#1}} 

\newcommand{\Rinst}{R_{\text{inst}}} 
\newcommand{\Rinstuser}[1]{R_{\text{inst},{#1}}} 
\newcommand{\Rave}{R_{\text{avg}}} 
\newcommand{\Raveuser}[1]{R_{\text{avg},{#1}}} 

\newcommand{\MCS}{\text{MCS}} 
\newcommand{\MCSmax}{\text{MCS}_{\text{max}}} 
\newcommand{\BLER}{\text{BLER}\xspace} 

\newcommand{\Buff}{B} 
\newcommand{\MaxBuff}{B_{\mathrm{max}}} 
\newcommand{\Buffuser}[1]{B_{#1}} 
\newcommand{\HOLdelay}{D_{\text{HOL}}} 
\newcommand{\HOLdelayAvg}{\bar{D}_{{\text{HOL}}}} 
\newcommand{\HOLdelayuser}[1]{D_{\text{HOL},{#1}}} 

\newcommand{\PRBmax}{N^{\text{PRB}}_{\text{max}}} 

\newcommand{\FQI}{\text{5QI}} 
\newcommand{\PDB}{\tau} 
\newcommand{\PDBuser}[1]{\tau_{#1}} 
\newcommand{\DelayThresholdUser}[1]{\delta_{#1}}
\newcommand{\DelayThreshold}{\delta}

\newcommand{\TraffType}{T}

\def\BibTeX{{\rm B\kern-.05em{\sc i\kern-.025em b}\kern-.08em
    T\kern-.1667em\lower.7ex\hbox{E}\kern-.125emX}}
\usepackage{balance}

\makeatletter
\patchcmd{\@maketitle}
  {\addvspace{0.5\baselineskip}\egroup}
  {\addvspace{-1\baselineskip}\egroup}
  {}
  {}
\makeatother

\begin{document}
\bstctlcite{IEEEexample:BSTcontrol}
\newacronym{isac}{ISAC}{Integrated Sensing and Communication}
\newacronym{re}{RE}{Resource Element}
\newacronym{ewma}{EWMA}{Exponentially Weighted Moving Average}
\newacronym{gpfb}{GPFB}{Generalized Proportional Fair Buffer}
\newacronym{drb}{DRB}{Data Radio Bearer}
\newacronym{hol}{D-HoL}{Head-of-Line Delay}
\newacronym{bcqi}{BCQI}{Best CQI}
\newacronym{ft}{FT}{Fair Throughput}
\newacronym{rr}{RR}{Round-Robin}
\newacronym{pf}{PF}{Proportional Fair}
\newacronym{x5g}{X5G}{}
\newacronym{ibs}{IBS}{Intent-Based Scheduling}
\newacronym{rapp}{rApp}{RAN Application}
\newacronym{xapp}{xApp}{eXtended Application}
\newacronym{dapp}{dApp}{Distributed Application}
\newacronym{ibn}{IBN}{Intent-Based Networking}
\newacronym{dlsch}{DLSCH}{Downlink Shared Channel}
\newacronym{ocr}{OCR}{Optical Character Recognition}
\newacronym{3gpp}{3GPP}{3rd Generation Partnership Project}
\newacronym{sriov}{SR-IOV}{Single Root I/O Virtualization}
\newacronym{vf}{VF}{Vitual Functions}
\newacronym{4g}{4G}{4th generation}
\newacronym{5g}{5G}{5th generation}
\newacronym{6g}{6G}{6th generation}
\newacronym{5gc}{5GC}{5G Core}
\newacronym{adc}{ADC}{Analog to Digital Converter}
\newacronym{aerpaw}{AERPAW}{Aerial Experimentation and Research Platform for Advanced Wireless}
\newacronym{ai}{AI}{Artificial Intelligence}
\newacronym{aimd}{AIMD}{Additive Increase Multiplicative Decrease}
\newacronym{am}{AM}{Acknowledged Mode}
\newacronym{amc}{AMC}{Adaptive Modulation and Coding}
\newacronym{amf}{AMF}{Access and Mobility Management Function}
\newacronym{aops}{AOPS}{Adaptive Order Prediction Scheduling}
\newacronym{api}{API}{Application Programming Interface}
\newacronym{apn}{APN}{Access Point Name}
\newacronym{ap}{AP}{Application Protocol}
\newacronym{aqm}{AQM}{Active Queue Management}
\newacronym{ausf}{AUSF}{Authentication Server Function}
\newacronym{avc}{AVC}{Advanced Video Coding}
\newacronym{awgn}{AGWN}{Additive White Gaussian Noise}
\newacronym{balia}{BALIA}{Balanced Link Adaptation Algorithm}
\newacronym{bbu}{BBU}{Base Band Unit}
\newacronym{bdp}{BDP}{Bandwidth-Delay Product}
\newacronym{ber}{BER}{Bit Error Rate}
\newacronym{bf}{BF}{Beamforming}
\newacronym{bler}{BLER}{Block Error Rate}
\newacronym{brr}{BRR}{Bayesian Ridge Regressor}
\newacronym{bs}{BS}{Base Station}
\newacronym{bsr}{BSR}{Buffer Status Report}
\newacronym{bss}{BSS}{Business Support System}
\newacronym{ca}{CA}{Carrier Aggregation}
\newacronym{caas}{CaaS}{Connectivity-as-a-Service}
\newacronym{cb}{CB}{Code Block}
\newacronym{cc}{CC}{Congestion Control}
\newacronym{ccid}{CCID}{Congestion Control ID}
\newacronym{cco}{CC}{Carrier Component}
\newacronym{cd}{CD}{Continuous Delivery}
\newacronym{cdd}{CDD}{Cyclic Delay Diversity}
\newacronym{cdf}{CDF}{Cumulative Distribution Function}
\newacronym{cdn}{CDN}{Content Distribution Network}
\newacronym{cli}{CLI}{Command-line Interface}
\newacronym{cn}{CN}{Core Network}
\newacronym{codel}{CoDel}{Controlled Delay Management}
\newacronym{comac}{COMAC}{Converged Multi-Access and Core}
\newacronym{cord}{CORD}{Central Office Re-architected as a Datacenter}
\newacronym{cornet}{CORNET}{COgnitive Radio NETwork}
\newacronym{cosmos}{COSMOS}{Cloud Enhanced Open Software Defined Mobile Wireless Testbed for City-Scale Deployment}
\newacronym{cots}{COTS}{Commercial Off-the-Shelf}
\newacronym{cp}{CP}{Control Plane}
\newacronym{cyp}{CP}{Cyclic Prefix}
\newacronym{up}{UP}{User Plane}
\newacronym{cpu}{CPU}{Central Processing Unit}
\newacronym{cqi}{CQI}{Channel Quality Information}
\newacronym{cr}{CR}{Cognitive Radio}
\newacronym{cran}{CRAN}{Cloud RAN}
\newacronym{crs}{CRS}{Cell Reference Signal}
\newacronym{csi}{CSI}{Channel State Information}
\newacronym{csirs}{CSI-RS}{Channel State Information - Reference Signal}
\newacronym{cu}{CU}{Central Unit}
\newacronym{cubb}{cuBB}{CUDA Baseband}
\newacronym{d2tcp}{D$^2$TCP}{Deadline-aware Data center TCP}
\newacronym{d3}{D$^3$}{Deadline-Driven Delivery}
\newacronym{dac}{DAC}{Digital to Analog Converter}
\newacronym{dag}{DAG}{Directed Acyclic Graph}
\newacronym{das}{DAS}{Distributed Antenna System}
\newacronym{dash}{DASH}{Dynamic Adaptive Streaming over HTTP}
\newacronym{dc}{DC}{Dual Connectivity}
\newacronym{dccp}{DCCP}{Datagram Congestion Control Protocol}
\newacronym{dce}{DCE}{Direct Code Execution}
\newacronym{dci}{DCI}{Downlink Control Information}
\newacronym{dctcp}{DCTCP}{Data Center TCP}
\newacronym{dl}{DL}{Downlink}
\newacronym{dmr}{DMR}{Deadline Miss Ratio}
\newacronym{dmrs}{DMRS}{DeModulation Reference Signal}
\newacronym{drlcc}{DRL-CC}{Deep Reinforcement Learning Congestion Control}
\newacronym{drs}{DRS}{Discovery Reference Signal}
\newacronym{du}{DU}{Distributed Unit}
\newacronym{e2e}{E2E}{end-to-end}
\newacronym{earfcn}{EARFCN}{E-UTRA Absolute Radio Frequency Channel Number}
\newacronym{ecaas}{ECaaS}{Edge-Cloud-as-a-Service}
\newacronym{ecn}{ECN}{Explicit Congestion Notification}
\newacronym{edf}{EDF}{Earliest Deadline First}
\newacronym{embb}{eMBB}{Enhanced Mobile Broadband}
\newacronym{empower}{EMPOWER}{EMpowering transatlantic PlatfOrms for advanced WirEless Research}
\newacronym{enb}{eNB}{evolved Node Base}
\newacronym{endc}{EN-DC}{E-UTRAN-\gls{nr} \gls{dc}}
\newacronym{epc}{EPC}{Evolved Packet Core}
\newacronym{eps}{EPS}{Evolved Packet System}
\newacronym{es}{ES}{Edge Server}
\newacronym{etsi}{ETSI}{European Telecommunications Standards Institute}
\newacronym[firstplural=Estimated Times of Arrival (ETAs)]{eta}{ETA}{Estimated Time of Arrival}
\newacronym{eutran}{E-UTRAN}{Evolved Universal Terrestrial Access Network}
\newacronym{faas}{FaaS}{Function-as-a-Service}
\newacronym{fapi}{FAPI}{Functional Application Platform Interface}
\newacronym{fdd}{FDD}{Frequency Division Duplexing}
\newacronym{fdm}{FDM}{Frequency Division Multiplexing}
\newacronym{fdma}{FDMA}{Frequency Division Multiple Access}
\newacronym{fed4fire}{FED4FIRE+}{Federation 4 Future Internet Research and Experimentation Plus}
\newacronym{fir}{FIR}{Finite Impulse Response}
\newacronym{fit}{FIT}{Future \acrlong{iot}}
\newacronym{fpga}{FPGA}{Field Programmable Gate Array}
\newacronym{fr2}{FR2}{Frequency Range 2}
\newacronym{fs}{FS}{Fast Switching}
\newacronym{fscc}{FSCC}{Flow Sharing Congestion Control}
\newacronym{ftp}{FTP}{File Transfer Protocol}
\newacronym{fw}{FW}{Flow Window}
\newacronym{ge}{GE}{Gaussian Elimination}
\newacronym{gnb}{gNB}{Next Generation Node Base}
\newacronym{gop}{GOP}{Group of Pictures}
\newacronym{gpr}{GPR}{Gaussian Process Regressor}
\newacronym{gpu}{GPU}{Graphics Processing Unit}
\newacronym{gtp}{GTP}{GPRS Tunneling Protocol}
\newacronym{gtpc}{GTP-C}{GPRS Tunnelling Protocol Control Plane}
\newacronym{gtpu}{GTP-U}{GPRS Tunnelling Protocol User Plane}
\newacronym{gtpv2c}{GTPv2-C}{\gls{gtp} v2 - Control}
\newacronym{gw}{GW}{Gateway}
\newacronym{harq}{HARQ}{Hybrid Automatic Repeat reQuest}
\newacronym{hetnet}{HetNet}{Heterogeneous Network}
\newacronym{hh}{HH}{Hard Handover}
\newacronym{hqf}{HQF}{Highest-quality-first}
\newacronym{hss}{HSS}{Home Subscription Server}
\newacronym{http}{HTTP}{HyperText Transfer Protocol}
\newacronym{ia}{IA}{Initial Access}
\newacronym{iab}{IAB}{Integrated Access and Backhaul}
\newacronym{ic}{IC}{Incident Command}
\newacronym{ietf}{IETF}{Internet Engineering Task Force}
\newacronym{imsi}{IMSI}{International Mobile Subscriber Identity}
\newacronym{imt}{IMT}{International Mobile Telecommunication}
\newacronym{iot}{IoT}{Internet of Things}
\newacronym{ip}{IP}{Internet Protocol}
\newacronym{itu}{ITU}{International Telecommunication Union}
\newacronym{kpi}{KPI}{Key Performance Indicator}
\newacronym{kpm}{KPM}{Key Performance Measurement}
\newacronym{kvm}{KVM}{Kernel-based Virtual Machine}
\newacronym{los}{LOS}{Line-of-Sight}
\newacronym{lsm}{LSM}{Link-to-System Mapping}
\newacronym{lstm}{LSTM}{Long Short Term Memory}
\newacronym{lte}{LTE}{Long Term Evolution}
\newacronym{lxc}{LXC}{Linux Container}
\newacronym{m2m}{M2M}{Machine to Machine}
\newacronym{mac}{MAC}{Medium Access Control}
\newacronym{manet}{MANET}{Mobile Ad Hoc Network}
\newacronym{mano}{MANO}{Management and Orchestration}
\newacronym{mc}{MC}{Multi-Connectivity}
\newacronym{mcc}{MCC}{Mobile Cloud Computing}
\newacronym{mchem}{MCHEM}{Massive Channel Emulator}
\newacronym{mcs}{MCS}{Modulation and Coding Scheme}
\newacronym{mec}{MEC}{Multi-access Edge Computing}
\newacronym{mec2}{MEC}{Mobile Edge Cloud}
\newacronym{mfc}{MFC}{Mobile Fog Computing}
\newacronym{mgen}{MGEN}{Multi-Generator}
\newacronym{mi}{MI}{Mutual Information}
\newacronym{mib}{MIB}{Master Information Block}
\newacronym{miesm}{MIESM}{Mutual Information Based Effective SINR}
\newacronym{mimo}{MIMO}{Multiple Input, Multiple Output}
\newacronym{ml}{ML}{Machine Learning}
\newacronym{mlr}{MLR}{Maximum-local-rate}
\newacronym[plural=\gls{mme}s,firstplural=Mobility Management Entities (MMEs)]{mme}{MME}{Mobility Management Entity}
\newacronym{mmtc}{mMTC}{Massive Machine-Type Communications}
\newacronym{mmwave}{mmWave}{millimeter wave}
\newacronym{mpdccp}{MP-DCCP}{Multipath Datagram Congestion Control Protocol}
\newacronym{mptcp}{MPTCP}{Multipath TCP}
\newacronym{mr}{MR}{Maximum Rate}
\newacronym{mrdc}{MR-DC}{Multi \gls{rat} \gls{dc}}
\newacronym{mse}{MSE}{Mean Square Error}
\newacronym{mss}{MSS}{Maximum Segment Size}
\newacronym{mt}{MT}{Mobile Termination}
\newacronym{mtd}{MTD}{Machine-Type Device}
\newacronym{mtu}{MTU}{Maximum Transmission Unit}
\newacronym{mumimo}{MU-MIMO}{Multi-user \gls{mimo}}
\newacronym{mvno}{MVNO}{Mobile Virtual Network Operator}
\newacronym{nalu}{NALU}{Network Abstraction Layer Unit}
\newacronym{nas}{NAS}{Network Attached Storage}
\newacronym{nat}{NAT}{Network Address Translation}
\newacronym{nbiot}{NB-IoT}{Narrow Band IoT}
\newacronym{nfv}{NFV}{Network Function Virtualization}
\newacronym{nfvi}{NFVI}{Network Function Virtualization Infrastructure}
\newacronym{ni}{NI}{Network Interfaces}
\newacronym{llm}{LLM}{Large Language Model}
\newacronym{dpdk}{DPDK}{Data Plane Development Kit}
\newacronym{cicd}{CI/CD}{Continuous Integration and Continuous Delivery/Deployment}
\newacronym{nic}{NIC}{Network Interface Card}
\newacronym{nlos}{NLOS}{Non-Line-of-Sight}
\newacronym{now}{NOW}{Non Overlapping Window}
\newacronym{nsm}{NSM}{Network Service Mesh}
\newacronym[type=hidden]{nr}{NR}{New Radio}
\newacronym[type=hidden]{ota}{OTA}{Over-The-Air}
\newacronym{nrf}{NRF}{Network Repository Function}
\newacronym{nsa}{NSA}{Non Stand Alone}
\newacronym{nse}{NSE}{Network Slicing Engine}
\newacronym{nssf}{NSSF}{Network Slice Selection Function}
\newacronym{o2i}{O2I}{Outdoor to Indoor}
\newacronym{oai}{OAI}{OpenAirInterface}
\newacronym{oaicn}{OAI-CN}{\gls{oai} \acrlong{cn}}
\newacronym{oairan}{OAI-RAN}{\acrlong{oai} \acrlong{ran}}
\newacronym{oam}{OAM}{Operations, Administration and Maintenance}
\newacronym{ofdm}{OFDM}{Orthogonal Frequency Division Multiplexing}
\newacronym{olia}{OLIA}{Opportunistic Linked Increase Algorithm}
\newacronym{omec}{OMEC}{Open Mobile Evolved Core}
\newacronym{onap}{ONAP}{Open Network Automation Platform}
\newacronym{onf}{ONF}{Open Networking Foundation}
\newacronym{onos}{ONOS}{Open Networking Operating System}
\newacronym{oom}{OOM}{\gls{onap} Operations Manager}
\newacronym{opnfv}{OPNFV}{Open Platform for \gls{nfv}}
\newacronym{oran}{O-RAN}{Open RAN}
\newacronym{orbit}{ORBIT}{Open-Access Research Testbed for Next-Generation Wireless Networks}
\newacronym{os}{OS}{Operating System}
\newacronym{oss}{OSS}{Operations Support System}
\newacronym{pa}{PA}{Position-aware}
\newacronym{pase}{PASE}{Prioritization, Arbitration, and Self-adjusting Endpoints}
\newacronym{pawr}{PAWR}{Platforms for Advanced Wireless Research}
\newacronym{pbch}{PBCH}{Physical Broadcast Channel}
\newacronym{pcef}{PCEF}{Policy and Charging Enforcement Function}
\newacronym{pcfich}{PCFICH}{Physical Control Format Indicator Channel}
\newacronym{pcrf}{PCRF}{Policy and Charging Rules Function}
\newacronym{pdcch}{PDCCH}{Physical Downlink Control Channel}
\newacronym{pdcp}{PDCP}{Packet Data Convergence Protocol}
\newacronym{pdf}{PDF}{Probability Density Function}
\newacronym{pdsch}{PDSCH}{Physical Downlink Shared Channel}
\newacronym{pdu}{PDU}{Packet Data Unit}
\newacronym{pgw}{PGW}{Packet Gateway}
\newacronym{phich}{PHICH}{Physical Hybrid ARQ Indicator Channel}
\newacronym{phy}{PHY}{Physical}
\newacronym{pmch}{PMCH}{Physical Multicast Channel}
\newacronym{pmi}{PMI}{Precoding Matrix Indicators}
\newacronym{powder}{POWDER}{Platform for Open Wireless Data-driven Experimental Research}
\newacronym{ppo}{PPO}{Proximal Policy Optimization}
\newacronym{ppp}{PPP}{Poisson Point Process}
\newacronym{prach}{PRACH}{Physical Random Access Channel}
\newacronym{prb}{PRB}{Physical Resource Block}
\newacronym{psnr}{PSNR}{Peak Signal to Noise Ratio}
\newacronym{pss}{PSS}{Primary Synchronization Signal}
\newacronym{pucch}{PUCCH}{Physical Uplink Control Channel}
\newacronym{pusch}{PUSCH}{Physical Uplink Shared Channel}
\newacronym{qam}{QAM}{Quadrature Amplitude Modulation}
\newacronym{qci}{QCI}{\gls{qos} Class Identifier}
\newacronym{qoe}{QoE}{Quality of Experience}
\newacronym{qos}{QoS}{Quality of Service}
\newacronym{quic}{QUIC}{Quick UDP Internet Connections}
\newacronym{rach}{RACH}{Random Access Channel}
\newacronym{ran}{RAN}{Radio Access Network}
\newacronym[firstplural=Radio Access Technologies (RATs)]{rat}{RAT}{Radio Access Technology}
\newacronym{rbg}{RBG}{Resource Block Group}
\newacronym{rcn}{RCN}{Research Coordination Network}
\newacronym{rc}{RC}{RAN Control}
\newacronym{rec}{REC}{Radio Edge Cloud}
\newacronym{red}{RED}{Random Early Detection}
\newacronym{renew}{RENEW}{Reconfigurable Eco-system for Next-generation End-to-end Wireless}
\newacronym{rf}{RF}{Radio Frequency}
\newacronym{rfc}{RFC}{Request for Comments}
\newacronym{rfr}{RFR}{Random Forest Regressor}
\newacronym{ric}{RIC}{RAN Intelligent Controller}
\newacronym{nrric}{Near-RT RIC}{Near-Real-Time \gls{ran} Intelligent Controller}
\newacronym{rlc}{RLC}{Radio Link Control}
\newacronym{rlf}{RLF}{Radio Link Failure}
\newacronym{rlnc}{RLNC}{Random Linear Network Coding}
\newacronym{rmr}{RMR}{RIC Message Router}
\newacronym{rmse}{RMSE}{Root Mean Squared Error}
\newacronym{rnis}{RNIS}{Radio Network Information Service}
\newacronym{rrc}{RRC}{Radio Resource Control}
\newacronym{rrm}{RRM}{Radio Resource Management}
\newacronym{rru}{RRU}{Remote Radio Unit}
\newacronym{rs}{RS}{Remote Server}
\newacronym{rsrp}{RSRP}{Reference Signal Received Power}
\newacronym{rsrq}{RSRQ}{Reference Signal Received Quality}
\newacronym{rss}{RSS}{Received Signal Strength}
\newacronym{rssi}{RSSI}{Received Signal Strength Indicator}
\newacronym{rtt}{RTT}{Round Trip Time}
\newacronym{ru}{RU}{Radio Unit}
\newacronym{rw}{RW}{Receive Window}
\newacronym{rx}{RX}{Receiver}
\newacronym{s1ap}{S1AP}{S1 Application Protocol}
\newacronym{sa}{SA}{standalone}
\newacronym{sack}{SACK}{Selective Acknowledgment}
\newacronym{sap}{SAP}{Service Access Point}
\newacronym{sc2}{SC2}{Spectrum Collaboration Challenge}
\newacronym{scef}{SCEF}{Service Capability Exposure Function}
\newacronym{sch}{SCH}{Secondary Cell Handover}
\newacronym{scoot}{SCOOT}{Split Cycle Offset Optimization Technique}
\newacronym{sctp}{SCTP}{Stream Control Transmission Protocol}
\newacronym{sdap}{SDAP}{Service Data Adaptation Protocol}
\newacronym{sdk}{SDK}{Software Development Kit}
\newacronym{sdm}{SDM}{Space Division Multiplexing}
\newacronym{sdma}{SDMA}{Spatial Division Multiple Access}
\newacronym{sdl}{SDL}{Shared Data Layer}
\newacronym{sdn}{SDN}{Software-defined Networking}
\newacronym{sdr}{SDR}{Software-defined Radio}
\newacronym{seba}{SEBA}{SDN-Enabled Broadband Access}
\newacronym{sgsn}{SGSN}{Serving GPRS Support Node}
\newacronym{sgw}{SGW}{Service Gateway}
\newacronym{si}{SI}{Study Item}
\newacronym{sib}{SIB}{Secondary Information Block}
\newacronym{sinr}{SINR}{Signal to Interference plus Noise Ratio}
\newacronym{sip}{SIP}{Session Initiation Protocol}
\newacronym{siso}{SISO}{Single Input, Single Output}
\newacronym{sla}{SLA}{Service Level Agreement}
\newacronym{sm}{SM}{Service Model}
\newacronym{e2sm}{E2SM}{E2 Service Model}
\newacronym{e2ap}{E2AP}{E2 Application Protocol}
\newacronym{smf}{SMF}{Session Management Function}
\newacronym{smo}{SMO}{Service Management and Orchestration}
\newacronym{sms}{SMS}{Short Message Service}
\newacronym{smsgmsc}{SMS-GMSC}{\gls{sms}-Gateway}
\newacronym{snr}{SNR}{Signal-to-Noise-Ratio}
\newacronym{son}{SON}{Self-Organizing Network}
\newacronym{sptcp}{SPTCP}{Single Path TCP}
\newacronym{srb}{SRB}{Service Radio Bearer}
\newacronym{srn}{SRN}{Standard Radio Node}
\newacronym{srs}{SRS}{Sounding Reference Signal}
\newacronym{ss}{SS}{Synchronization Signal}
\newacronym{sss}{SSS}{Secondary Synchronization Signal}
\newacronym{st}{ST}{Spanning Tree}
\newacronym{svc}{SVC}{Scalable Video Coding}
\newacronym{tb}{TB}{Transport Block}
\newacronym{tcp}{TCP}{Transmission Control Protocol}
\newacronym{tdd}{TDD}{Time Division Duplexing}
\newacronym{tdm}{TDM}{Time Division Multiplexing}
\newacronym{tdma}{TDMA}{Time Division Multiple Access}
\newacronym{tfl}{TfL}{Transport for London}
\newacronym{tfrc}{TFRC}{TCP-Friendly Rate Control}
\newacronym{tft}{TFT}{Traffic Flow Template}
\newacronym{tgen}{TGEN}{Traffic Generator}
\newacronym{tip}{TIP}{Telecom Infra Project}
\newacronym{tm}{TM}{Transparent Mode}
\newacronym{to}{TO}{Telco Operator}
\newacronym{tr}{TR}{Technical Report}
\newacronym{trp}{TRP}{Transmitter Receiver Pair}
\newacronym{ts}{TS}{Technical Specification}
\newacronym{tti}{TTI}{Transmission Time Interval}
\newacronym{ttt}{TTT}{Time-to-Trigger}
\newacronym{tx}{TX}{Transmitter}
\newacronym{uas}{UAS}{Unmanned Aerial System}
\newacronym{uav}{UAV}{Unmanned Aerial Vehicle}
\newacronym{udm}{UDM}{Unified Data Management}
\newacronym{udp}{UDP}{User Datagram Protocol}
\newacronym{udr}{UDR}{Unified Data Repository}
\newacronym{ue}{UE}{User Equipment}
\newacronym{uhd}{UHD}{\gls{usrp} Hardware Driver}
\newacronym{ul}{UL}{Uplink}
\newacronym{um}{UM}{Unacknowledged Mode}
\newacronym{uml}{UML}{Unified Modeling Language}
\newacronym{upa}{UPA}{Uniform Planar Array}
\newacronym{upf}{UPF}{User Plane Function}
\newacronym{urllc}{URLLC}{Ultra Reliable and Low Latency Communications}
\newacronym{usa}{U.S.}{United States}
\newacronym{usim}{USIM}{Universal Subscriber Identity Module}
\newacronym{usrp}{USRP}{Universal Software Radio Peripheral}
\newacronym{utc}{UTC}{Urban Traffic Control}
\newacronym{vim}{VIM}{Virtualization Infrastructure Manager}
\newacronym{vm}{VM}{Virtual Machine}
\newacronym{vnf}{VNF}{Virtual Network Function}
\newacronym{volte}{VoLTE}{Voice over \gls{lte}}
\newacronym{voltha}{VOLTHA}{Virtual OLT HArdware Abstraction}
\newacronym{vr}{VR}{Virtual Reality}
\newacronym{vran}{vRAN}{Virtualized RAN}
\newacronym{vss}{VSS}{Video Streaming Server}
\newacronym{wbf}{WBF}{Wired Bias Function}
\newacronym{wf}{WF}{Waterfilling}
\newacronym{wg}{WG}{Working Group}
\newacronym{wlan}{WLAN}{Wireless Local Area Network}
\newacronym{osm}{OSM}{Open Source \gls{nfv} Management and Orchestration}
\newacronym{pnf}{PNF}{Physical Network Function}
\newacronym{drl}{DRL}{Deep Reinforcement Learning}
\newacronym{mtc}{MTC}{Machine-type Communications}
\newacronym{osc}{OSC}{O-RAN Software Community}
\newacronym{mns}{MnS}{Management Services}
\newacronym{ves}{VES}{\gls{vnf} Event Stream}
\newacronym{ei}{EI}{Enrichment Information}
\newacronym{fh}{FH}{Fronthaul}
\newacronym{fft}{FFT}{Fast Fourier Transform}
\newacronym{laa}{LAA}{Licensed-Assisted Access}
\newacronym{plfs}{PLFS}{Physical Layer Frequency Signals}
\newacronym{ptp}{PTP}{Precision Time Protocol}
\newacronym{ntp}{NTP}{Network Time Protocol}
\newacronym{cbrs}{CBRS}{Citizen Broadband Radio Service}
\newacronym{rnti}{RNTI}{Radio Network Temporary Identifier}
\newacronym{tbs}{TBS}{Transport Block Size}
\newacronym{nfd}{NFD}{Node Feature Discovery}
\newacronym{mcp}{MCP}{Machine Configuration Pool}
\newacronym{vpn}{VPN}{Virtual Private Network}
\newacronym{onr}{ONR}{Office of Naval Research}
\newacronym{afosr}{AFOSR}{Air Force Office of Scientific Research}
\newacronym{afrl}{AFRL}{Air Force Research Laboratory}
\newacronym{arl}{ARL}{Army Research Laboratory}

\newacronym{arc}{ARC}{Aerial Research Cloud}

\newacronym{ct}{CT}{Continuous Testing}
\newacronym{mno}{MNO}{Mobile Network Operator}
\newacronym{oci}{OCI}{Open Container Initiative}
\newacronym{macsec}{MACsec}{Media Access Control Security}
\newacronym{pt}{PT}{Plain Text}
\newacronym{cuda}{CUDA}{Compute Unified Device Architecture}
\newacronym{dsp}{DSP}{Digital Signal Processing}

\newacronym{cus}{CUS}{Control, User, Synchronization}
\newacronym{dpd}{DPD}{Digital Pre-Distorsion}
\newacronym{cfr}{CFR}{Crest Factor Reduction}
\newacronym{pci}{PCIe}{Peripheral Component Interconnect Express}
\newacronym{dpu}{DPU}{Data Processing Unit}
\newacronym{rfsoc}{RFSoC}{Radio Frequency System-on-Chip}
\newacronym{if}{IF}{Intermediate Frequency}
\newacronym{nyu}{NYU}{New York University}
\newacronym{gh}{GH}{Grace Hopper}
\newacronym{trl}{TRL}{Technology Readiness Level}
\newacronym{srfa}{SRFA}{Special Research Focus Area}
\newacronym{qsfp}{QSFP}{quad small form factor pluggable}
\newacronym{pse}{PSE}{Performance Specialized Engine}
\newacronym{cae}{CAE}{Cognitive Analysis Engine}
\newacronym{simd}{SIMD}{Single Instruction/Multiple Data}
\newacronym{rt}{RT}{Real-Time}
\newacronym{nrt}{NRT}{Non-Real-Time}
\newacronym{asm}{ASM}{Advanced Sleep Mode}
\newacronym{aoa}{AoA}{Angle of Arrival}
\newacronym{eaxcid}{eAxC\_ID}{extended Antenna-Carrier Identifier}
\newacronym{bwp}{BWP}{Bandwidth Part}
\newacronym{dfe}{DFE}{Digital Front-End}
\newacronym{spi}{SPI}{Serial Peripheral Interface}
\newacronym{gpio}{GPIO}{General Purpose Input/Output}
\newacronym{nco}{NCO}{Numerically Controlled Oscillator}
\newacronym{lo}{LO}{Local Oscillator}
\newacronym{lna}{LNA}{Low-Noise Amplifier}
\newacronym{pll}{PLL}{Phased-Locked Loop}
\newacronym{som}{SOM}{System-on-Module}
\newacronym{papr}{PAPR}{Peak-to-Average Power Ratio}
\newacronym{pcb}{PCB}{Printed Circuit Board}
\newacronym{gcpw}{GCPW}{Grounded Co-Planar Waveguide}
\newacronym{cnn}{CNN}{Convolutional Neural Network}
\newacronym{gmp}{GMP}{Generalized Memory Polynomial}
\newacronym{ngrg}{nGRG}{next Generation Research Group}
\newacronym{mrl}{MRL}{Manufacturing Readiness Level}
\newacronym{fr}{FR}{Frequency Range}

\newacronym{sbom}{SBOM}{Software Bill of Materials}
\newacronym{hbom}{HBOM}{Hardware Bill of Materials}
\newacronym{vex}{VEX}{Vulnerability Exploitability eXchange}
\newacronym{dos}{DoS}{Denial of Service}
\newacronym{sme}{SME}{Small-Medium Enterprise}

\newacronym{ulpi}{ULPI}{Uplink Performance Improvement}
\newacronym{oem}{OEM}{Original Equipment Manufacturer}
\newacronym{nsin}{NSIN}{National Security Innovation Network}
\newacronym{dod}{DoD}{Department of Defense}
\newacronym{arpu}{ARPU}{Average Revenue per User}
\newacronym{opex}{OPEX}{operational expenses}
\newacronym{txb}{TXB}{Transmit Beam}
\newacronym{cve}{CVE}{Common Vulnerabilities and Exposure}
\newacronym{json}{JSON}{JavaScript Object Notation}
\newacronym{it}{IT}{Information Technology}
\newacronym{ci}{CI}{Continuous Integration}
\newacronym{nlp}{NLP}{Natural Language Processing}
\newacronym{nl}{NL}{Natural Language}
\newacronym{rl}{RL}{Reinforcement Learning}

\title{\framework\xspace --- Automated LLM-Driven Scheduler Generation and Testing for Intent-Based RAN}
\author{Maxime Elkael, Michele Polese, Reshma Prasad, Stefano Maxenti, Tommaso Melodia
\thanks{The authors are with the Institute for the Wireless Internet of Things, Northeastern University, Boston, MA 02115. Email: \{m.elkael, m.polese, re.prasad, maxenti.s, melodia\}@northeastern.edu.}
\thanks{This article is based upon work partially supported by the National Telecommunications and Information Administration (NTIA)'s Public Wireless Supply Chain Innovation Fund (PWSCIF) under Award No. 25-60-IF054 and by OUSD(R\&E) through Army Research Laboratory Cooperative Agreement Number W911NF-24-2-0065. The views and conclusions contained in this document are those of the authors and should not be interpreted as representing the official policies, either expressed or implied, of the Army Research Laboratory or the U.S. Government. The U.S. Government is authorized to reproduce and distribute reprints for Government purposes notwithstanding any copyright notation herein. This work is also partially supported by the U.S. NSF under award CNS-2112471 and by Qualcomm, Inc. The authors disclose that AI tools (ChatGPT, Claude) have been used to perform minor edits and reformulations of the text.}}


\maketitle

\begin{abstract}
The evolution toward open, programmable O-RAN and AI-RAN 6G networks creates unprecedented opportunities for \gls{ibn} to dynamically optimize \gls{ran} operations based on dynamic operators requirements. However, applying \gls{ibn} effectively to the \gls{ran} scheduler - a critical component determining resource allocation and system performance - remains a significant challenge. Current approaches predominantly rely on coarse-grained network slicing, lacking the granularity for dynamic adaptation to individual user conditions and traffic patterns. Despite the existence of a vast body of  scheduling algorithms that could potentially translate high-level intents into executable policies, their practical utilization is hindered by implementation heterogeneity, insufficient systematic evaluation in production environments, and the complexity of developing high-performance scheduler implementations. This necessitates a more granular, flexible, and verifiable approach to align scheduler behavior with operator-defined intents.

To address these limitations, we propose \framework (Automated LLm-driven Scheduler generation and Testing for intent-based \gls{ran}), a novel framework leveraging \glspl{llm} for automated, intent-driven scheduler design, implementation, and evaluation. \framework interprets natural language intents, automatically generates functional scheduler code from the research literature using \gls{ocr} and \glspl{llm}, and intelligently matches operator intents to the most suitable scheduler(s). Our implementation deploys these schedulers as O-RAN dApps, enabling on-the-fly deployment and comprehensive testing on a production-grade, multi-vendor 5G-compliant testbed. This approach has enabled the largest-scale \gls{ota} experimental comparison of 18 scheduling algorithms automatically synthesized from the academic literature. The resulting performance profiles serve as the input for our \gls{ibs} framework, which  dynamically selects and deploys appropriate schedulers that optimally satisfy operator intents. We validate our approach through multiple use cases unattainable with current slicing-based optimization techniques, demonstrating fine-grained control based on buffer status, physical layer conditions, and heterogeneous traffic types. 
\end{abstract}


\glsresetall
\glsunset{xapp}
\glsunset{rapp}
\glsunset{dapp}

\vspace{-10pt} 
\section{Introduction}
\label{sec:introduction}

To support an ever-growing variety of devices and use cases envisioned for \gls{6g}~\cite{tataria20216g}, cellular networks are shifting toward programmable, software-based, open architectures~\cite{polese2022understanding}. In this context, the \gls{oran} paradigm, and the O-RAN ALLIANCE specifications, define a set of disaggregated network elements connected by open interfaces. 
This approach, combining software and disaggregation, addresses the lack of observability and programmability of traditional monolithic, black-box \gls{ran}.
Notably, the O-RAN architecture introduces two types of applications, namely, \glspl{rapp}, which optimize the network with a control loop of one second or more (non-real time), and \glspl{xapp}, for which the control loop delay is between $10\:\mathrm{ms}$ and $1\:\mathrm{s}$ (near-real time). 
\glspl{dapp}~\cite{lacava2025dapps} have also been proposed as a third class of applications for the next generation of the O-RAN architecture. Here, the control loop is below $10$ ms, enabling real-time, programmable control embedded on \gls{ran}. In parallel, the AI-RAN Alliance is studying solutions for cellular networks based on \gls{ai}, to increase spectral efficiency, flexibility, and automation of the systems~\cite{kundu2025ai}. 

Programmability, open interfaces, and \gls{ai} present opportunities to develop \gls{ibn}, i.e., to let operators dynamically customize, control, and optimize their network based on different use cases or evolving requirements, expressed through high-level intents~\cite{leivadeas2023survey}. The concept of \gls{ibn} in the \gls{ran}, however, has still not reached a full realization, especially when considering the key component of cellular base stations, i.e., the scheduler. 
Prior work that combines \gls{oran} and \gls{ibn} typically focuses on resource allocation for slices, i.e., groups of \glspl{ue} which are set a-priori,
and where an \gls{xapp} can control the number of \gls{prb} resources each slice receives, based on network \glspl{kpi}~\cite{nahum2024intent}. 
While this approach allows for a near- or non-real time adaptation of the network, it typically suffers from (i) the need to divide the \glspl{ue} a-priori, preventing intents based on each \gls{ue}'s dynamic conditions (e.g., mobility patterns or traffic flows and buffering within a slice~\cite{zhong2017heterogeneous}); and (ii) relying on aggregated metrics which prevents more granular allocation based on channel condition, and typically leads to designing model-free resource allocation techniques with limited interpretability. 
This lack of understandability makes such methods hard to trust for scenarios such as \gls{urllc} and more generally high stakes use cases where extremely high levels of reliability is required. Furthermore, those techniques also require extensive retraining for each new intent.

On the other hand, historically, a large body of research on \gls{ran} resource allocation has been dedicated to scheduling, making a trove of scheduling algorithms available~\cite{capozzi2013downlink,mamane2022scheduling}.
While the variety of scheduler existing in the literature allows, in principle, the mapping to a diverse set of intents that an operator may express, this potential has remained untapped. First, the heterogeneity of techniques, evaluation methods, and input/output parameters makes it difficult to systematically understand how different schedulers would perform in the same network conditions and how they would compare in expressing an intent. Second, most studies focus on theoretical analysis and/or simulation for the scheduler evaluation, complicating the understanding of their performance in a production setup.
Third, as scheduling is a real-time process (in \gls{5g}, it happens every slot, e.g., every $0.5$ ms),  {operators may be risk-averse toward} dynamic adaptations based on high-level intents, considering the strict timing and the need for performance guarantees. Finally, the scheduler does not currently have well-defined interfaces for parameter or logic updates, and, in most implementations, it is a proprietary, vendor-specific piece of software.

To summarize, dynamically adapting schedulers to reflect operators' requirements and intents calls for a more granular and dynamic approach than slicing, and for more flexible and production-tested intent-based schedulers.

\begin{figure}[t]
    \centering
    \includegraphics[width=1.\linewidth]{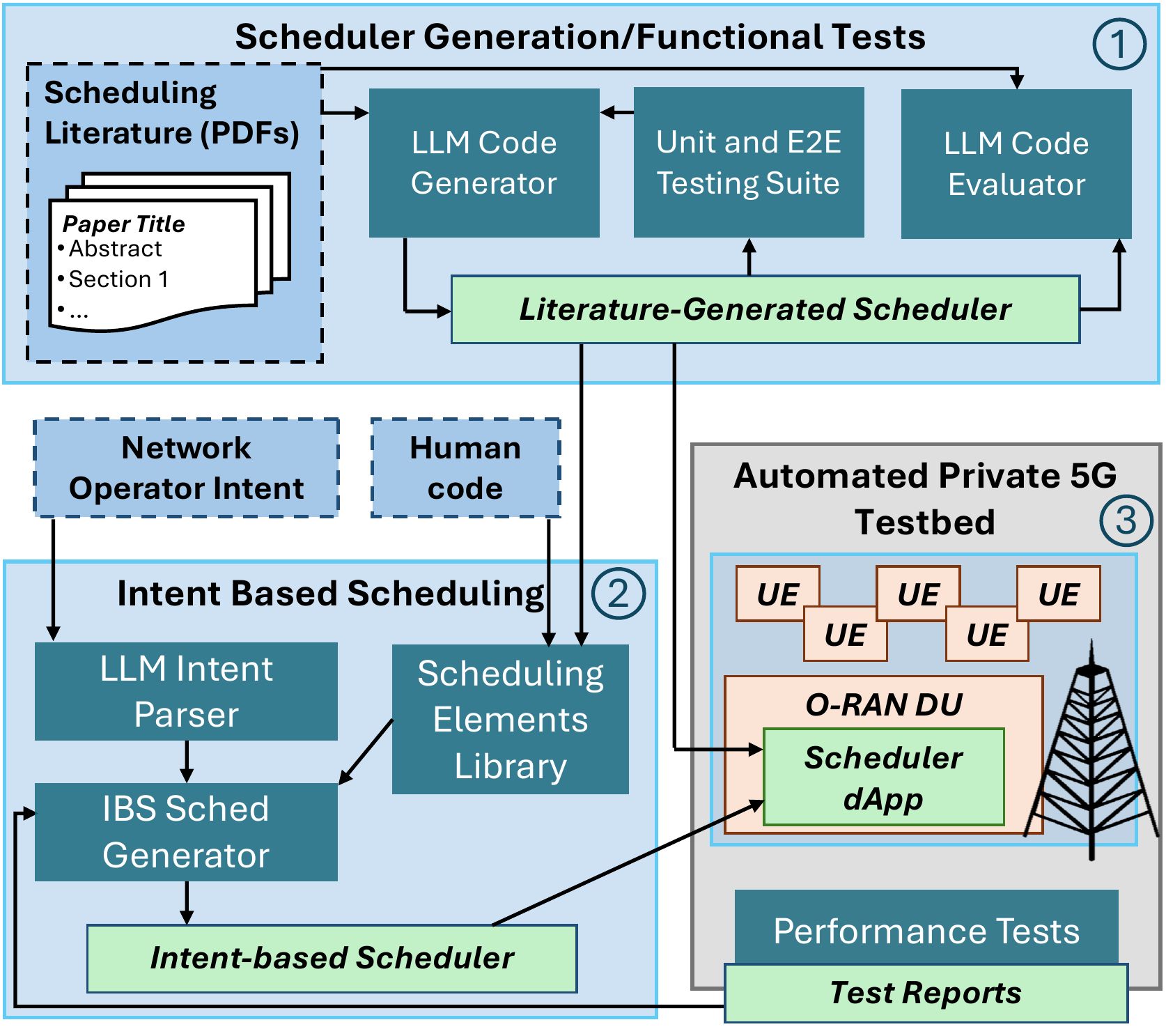}
    \setlength{\abovecaptionskip}{-.2cm}
    \setlength{\belowcaptionskip}{-.6cm}
    \caption{High-level architecture of \framework. LLM pipelines are in blue. Outputs are in green. The 5G infrastructure is gray.}
    \label{fig:framework_hl}
\end{figure}

\textbf{\framework.}
To address this, we propose a radically different way to design, implement, and evaluate schedulers for cellular networks, based on a series of \glspl{llm} agents that can (i) interpret the operator's intent; (ii) automatically generate functional schedulers from papers and reports in the literature; and (iii) match the intent to one or multiple schedulers that have the profile to satisfy the operator request. This, as shown in Fig.~\ref{fig:framework_hl}, is combined with a robust automation framework that provides guardrails in the \gls{llm} pipelines and allows for automated testing of the \gls{llm}-generated schedulers on an open, programmable, and production-ready 5G network.

Specifically, our contributions are as follows:
\begin{itemize}[leftmargin=*]
    \item We  {address the lack of dynamic adaptability of schedulers by} designing the scheduler as an O-RAN dApp, defining input and output of the scheduler logic and a prototype implementation that allows swapping different schedulers in a running \gls{oran} network on-the-fly;
    \item We design and prototype an automated platform for large-scale experimental evaluation of dApp-based-schedulers (part \Circled{3} in Fig.~\ref{fig:framework_hl}), integrated in a private 5G network based on multi-vendor components from NVIDIA, Foxconn, Sierra Wireless, and open-source software from \gls{oai} and Open5Gs;
    \item Based on this scheduler-as-\gls{dapp} approach and the evaluation framework, we design and prototype an automated scheduler evaluation pipeline, which uses \gls{ocr} and \glspl{llm} to generate and test scheduling algorithms solely from research articles or reports describing the algorithm (part \Circled{1} of Fig.~\ref{fig:framework_hl}). 
    This enables us to carry out the largest scale (to our knowledge) experimental \gls{ota} scheduler comparison study to date, evaluating 18 algorithms~\cite{kaltenberger2019openairinterface,kelly1997charging,saglam20195g,sadiq2010delay,mamane2021proportional,ramjee2006generalized,afifi2021novel,sadiq2009downlink,monghal2010dynamic,brehm2013overload,iturralde2012resource,basukala2009performance,rhee2003scheduling,piro2011two,mahfoudi2015new,husain2020efficient};
    \item We then leverage the thorough profiling and evaluation of the schedulers to design a general \gls{ibs} framework, in which a custom-tailored scheduler is generated based on our library of schedulers from the literature and   {an operator}-provided natural language intent, without requiring retraining ({part \Circled{2} of Fig.}~\ref{fig:framework_hl}); 
    \item Finally, we validate the correctness of our \gls{ibs} framework on multiple use cases which are fundamentally unadressable by current \gls{ibn} approaches, such as intent-based slicing. This includes use cases that deliver different \gls{qos} based on the buffer status (for example, prioritizing users with currently bursty traffic), on the physical layer indicators, which enables selective scheduling based on the mobility pattern, and on traffic classification, to selectively adapt resource allocation depending on the current traffic type.
\end{itemize}

Our results indicate that our framework can generate code that is highly faithful to the original paper, while accommodating up to hundreds of \glspl{ue}. This enables us to unveil large differences between scheduler behaviors claimed in simulation, compared to our \gls{ota} deployment, enabling us to select the best \gls{qos}-agnostic and \gls{qos}-aware schedulers, achieving controllable fairness, and tightly respecting delay constraints. Finally, we show how these building blocks can be used in scenarios such as \gls{isac}-guided scheduling and joint traffic/delay aware scheduling. In delay-aware scenarios, our framework enables us to select the only algorithm which reliably reduces the $99^{th}$ percentile of delay below a target of $50$ ms, an order of magnitude lower than the $1$ second range experienced with \gls{pf} in the same scenario.

The rest of this paper is organized as follows. In Sec.~\ref{sec:soa}, we discuss the state of the art on AI- and \gls{llm}-driven intent-based scheduling. In Sec.~\ref{sec:framework}, we introduce the components of \framework. This is followed a review of the experimental platform we use for evaluating the schedulers, which are then
reviewed in Sec.~\ref{sec:review}. Sec.~\ref{sec:qualitativeevaluation} is then devoted to the qualitative evaluation of \framework as a tool for scheduler testing, and Sec.~\ref{sec:num-results} compares the different schedulers numerically. We finally present the use cases of \gls{ibs} in Sec.~\ref{sec:ibs-results}, and we conclude the paper in Sec.~\ref{sec:conclusions}.





\vspace{-5pt} 
\section{Related Works}
\label{sec:soa}

In the recent years, intent-based networking ({i.e.}, controlling network parameters based on high level, human formulated intents) has attracted increasing research attention \cite{leivadeas2022survey,wei2020intent}. A significant portion of the literature focuses on wired networks, due to the maturity of flexible approaches such as \gls{nfv} \cite{mijumbi2015network} and \gls{sdn} \cite{xia2014survey}. Notable works in that domain include \cite{han2016intent}, where a hierarchical architecture is proposed that combines a hypervisor and an SDN controller, providing a high-level abstraction to control the network. Other similar works on the architecture of \gls{sdn}-based \gls{ibn} implementations include \cite{pham2016sdn} and \cite{szyrkowiec2018automatic}.

On the other hand, the domain of cellular networks has historically been lagging behind, due to the historically vendor-locked networks and high cost to entry. However, this has changed in recent years, with the \gls{oran} ALLIANCE \cite{polese2022understanding} developing open interfaces for cellular networks. Among them, the E2 interface exposes \glspl{kpi} and control knobs at a $10$ ms granularity. 
The most studied control parameter in the \gls{ibn} community relates to the 5G-introduced slicing feature \cite{foukas2017network, cheng2024oranslice}, where upon creating \gls{pdu} sessions, \glspl{ue} can be separated into groups to be treated differently.
Notable works leveraging the E2 interface and/or slicing to optimize the slice resource allocation to meet different \gls{qos} requirements include FlexSlice \cite{chen2023flexslice}, PandORA \cite{tsampazi2024pandora}, and ColO-RAN \cite{polese2022colo}. Other notable papers performing such optimizations include \cite{yeh2023deep, kouchaki2022actor, filali2023communication}. Note that the output of these works is an aggregated resource allocation ({i.e.}, per-slice allocation) that is then sent to the scheduler. Indeed, the E2 interface does not provide the real-time granularity which would allow the algorithm to act on each scheduling decision. Note that besides \gls{oran}, there is a large history of studying resource allocation for wireless communication for various objectives through optimizing the scheduler directly, both with classical optimization (we defer the details of schedulers in this category, which are of interest to the present work, to Section \ref{sec:review}) and \gls{ml} techniques \cite{fattah2002overview, hu2021distributed}. Among these, one recent work \cite{apostolakis2023athena} makes steps towards bridging real-life scheduler programmability and \gls{ai}, by training a \gls{dl}/\gls{rl} algorithms for optimizing the \gls{ul} throughput with respect to available cloud resources. \cite{chinchali2018cellular} is also a notable work which trains a \gls{dl} scheduling policy in a simulation environment.

The aforementioned \gls{ml} approaches typically leverage specialized models trained for a subset of intents  {(i.e., one specific reward function)} and associated optimization objectives.  {Furthermore, each reward function needs to be carefully hand-crafted by a human, as reward shaping for \gls{rl} is itself an entire research field \cite{eschmann2021reward}.} On the other hand, the recent advent of \glspl{llm} has motivated researchers to explore their application to telecommunications, their main appeal being zero-shot and in context learning, in which a \gls{llm} is capable of solving a problem from a \gls{nl} prompt, not requiring a large dataset {, reward shaping} and retraining. Notorious works include specification assistants \cite{nikbakht2024tspec, saraiva2024telco, zou2024telecomgpt}, the LLM-xApp framework~\cite{wu2025llm}, which leverages an \gls{llm} for slice resource allocation, and an \gls{llm} for energy-optimization of \gls{oran} networks~\cite{bao2025llmguidedopenranempowering}. \glspl{llm} have also been leveraged for service management and deployment in works such as the AutoRAN framework \cite{maxenti2025autoranautomatedzerotouchopen} and the intent-driven service deployer of \cite{mekrache2024llm, mekrache2024intent}.
Other relevant approaches for network configuration, but not specifically tailored to cellular networks, include NetConfEval~\cite{changjie2024netconfeval} and GeNet~\cite{ifland2024genet}.

Finally, one of the biggest breakthrough of current \glspl{llm} is general-purpose code generation, with models such as GPT O3, Claude 3.7 Sonnet, and DeepSeek R1 being able to generate high-quality code based on human-provided prompts. This has created opportunities for research such as the Paper2Code agent \cite{seo2025paper2code} which generates Python implementations of \gls{ml} research articles.

All in all, the current state of \gls{ibn} resource allocation is fragmented between recent \gls{llm}-based zero-shot approaches for wired \gls{sdn} networks, non real-time practical wireless resource allocation based on classical \gls{dl}, and real-time simulation-based scheduling.
We leverage \glspl{llm} and \glspl{dapp} to close that gap, bringing both real-time resource allocation and zero-shot \gls{ibn} programmability to a production \gls{ran}. 
 {Zero-shot programmability through \gls{nl} is particularly desirable, as it cuts the need for the operator to (i) formalize the problem mathematically, instead using natural language and (ii) retrain their algorithm each time the intent changes or needs to be tweaked (which, for most approaches such as \gls{rl} is done via updating the reward function and retraining).}



%

\vspace{-5pt} 
\section{\framework Architecture and Design}
\label{sec:framework}

In this section, we detail the architecture and design of \framework. As shown in Fig. \ref{fig:framework_hl}, at a high level, \framework is comprised of two main components.

First, the \emph{Scheduler Generation and Testing} pipeline \Circled{1} is tasked with generating fully functional scheduler \glspl{dapp} from research articles. This first pipeline uses two \gls{llm} Agents along with a unit testing suite to generate the code of the schedulers.
As will be described in further details in section \ref{sec:testing}, the pipeline leverages the testing suite to validate the correctness of the generated code. Then, once validated, the code is pushed to our automated \gls{ota} private 5G testbed (\Circled{3} in Fig.~\ref{fig:framework_hl}), which orchestrates experiments to assess the performance of the algorithm. The code is also fed to another \gls{llm} agent, which ensures the code correctly implements the algorithm from the paper.

Then, these reports along with the scheduler code and optional additional human-written code are passed to our second pipeline for \emph{Intent-Based Scheduling} (described in details in Section \ref{sec:ibs}, and illustrated as \Circled{2} in Fig.~\ref{fig:framework_hl}). The \gls{ibs} allows network operators to specify required configurations and performance levels expected from the scheduler, and to translate them automatically to a scheduling policy \emph{and} implementation. This is based on the insights collected by analyzing the wide set of existing scheduler proposals with the \emph{Testing and Scheduler Generation} pipeline. The characteristics of the scheduler are pulled from a database when the operator prompts the \gls{ibs} pipeline with a scheduling intent, and fed to an \gls{llm} agent which generates a new scheduler fitting the intent requested. Because of using pre-tested elements, this enables us to fulfill the request without needing to redo the long and extensive testing process.

At its core, \framework leverages the concept of dApp to dynamically embed and update the scheduler within the \gls{mac} layer of the \gls{du}. dApps are programmable logic components that can connect and interact with \glspl{du} and \glspl{cu} in real-time~\cite{lacava2025dapps,dapp-ngrg-report}, through interfaces that enable exposure of \gls{ran} \glspl{kpi} and feedback of control. This concept is now being discussed for integration within the next-generation of the O-RAN architecture by the O-RAN ALLIANCE~\cite{dapp-ngrg-report}. Next, we discuss how the \framework dApp is designed.

\subsection{Scheduler Generation and Testing Problem Formulation}
Let us start by formalizing the scheduler generation and testing problem mathematically. We consider the scheduler as a function which solves an optimization problem at each \gls{dl} slot $t$:
\begin{equation}
\begin{aligned}
\max_{\theta(t)} \quad & \mathcal{V}(\theta(t), \mathbf{K}(t)) \\
\text{subject to} \quad & \sum_{i=1}^{N} \theta_i(t) \leq B_{\text{max}} \\
& l_i(K_i(t)) \leq \theta_i(t) \leq r_i(K_i(t)), & \forall i \in \{1, \ldots, N\} \\
& 0 \leq \theta_i(t) \leq B_{\text{max}}, &\text{for all $N$ \glspl{ue} $i$}
\end{aligned}
\label{initialPB}
\end{equation}
where $\theta(t)$ is the vector of $N$ $\theta_i(t)$ decision variables (one per \gls{ue}) which decide how many \gls{prb} each \gls{ue} $i$ receives at slot $t$. We call $\theta^*(t)$ the optimal value of $\theta(t)$.
$\mathbf{K}(t)$ is the matrix of per \gls{ue} metrics provided by the \gls{ran} at each slot, as detailed in Listing \ref{listing:datastruct}. From this matrix, for each user $i$, we can derive the vector $K_i$ associated with user $i$. 
We consider that, for all \glspl{ue}, the \gls{cqi} provided in $\mathbf{K}$ is the wideband \gls{cqi}. The objective function $\mathcal{V}(\theta(t), \mathbf{K}(t))$ represents the specific scheduling policy being implemented (e.g., maximizing sum-rate, proportional fairness, or other utility metrics). Finally, $N$ is the number of \glspl{ue}, $B_{\text{max}}$ is the maximum number of \glspl{prb} that can be allocated and $l_i(K_i(t))$ and $r_i(K_i(t))$ are functions providing dynamic lower and upper bounds on PRB allocation for UE $i$.

Due to the real-time execution requirement, we typically use approximate solutions to this optimization problem. The mathematical version of this approximate scheduler function $\zeta(\mathbf{K}(t))$ outputs a vector $\theta(t) \approx \theta^*(t)$, with one element $\theta_i(t)$ per \gls{ue}. This vector indicates how many \glspl{prb} should be allocated to each \gls{ue} (corresponding to Type 1 \gls{prb} allocation~\cite{3GPP-38214} in \gls{5g}). 
This mathematical function also exists in two other forms. First, we have $\zeta_{NL}$, which is a \gls{nl} description of the algorithm of $\zeta$ (e.g., a research paper or some part of it). Second, we have $\zeta_{code}$ which is the actual implementation of $\zeta$ as a piece of code.
From this, our objective is to design a two-stage pipeline:

\textbf{Stage 1: Code Generation Function} $\mathcal{G}$
\begin{equation}
\mathcal{G}: \zeta_{NL} \mapsto (\zeta_{code}, r, R)
\end{equation}
where:
\begin{itemize}
    \item $\zeta_{code}$ is the executable \gls{dapp} implementing the scheduling algorithm
    \item $r \in [0, 10]$ is a confidence score evaluating the semantic alignment between $\zeta_{code}$ and $\zeta_{NL}$
    \item $R$ is a natural language report justifying the score $r$
\end{itemize}

\textbf{Stage 2: Testing Function} $\mathcal{T}$
\begin{equation}
\mathcal{T}: (\zeta_{code}, \mathcal{D}_{test}) \mapsto (v, \mathcal{M})
\end{equation}
where:
\begin{itemize}
    \item $\mathcal{D}_{test}$ is a test dataset of input scenarios $\{\mathbf{K}(t)\}$
    \item $v \in \{\text{pass}, \text{fail}\}$ indicates whether $\zeta_{code}$ satisfies all constraints in our optimization problem
    \item $\mathcal{M}$ contains performance metrics including execution time, constraint violations, and functionality errors
\end{itemize}

The testing function $\mathcal{T}$ verifies that $\zeta_{code}$ produces valid solutions $\theta(t)$ that satisfy all constraints from our optimization problem, ensuring the scheduler is deployable in a real \gls{mac} stack. While $\mathcal{G}$ focuses on the semantic correctness of the NL-to-code translation, $\mathcal{T}$ guarantees functional correctness and \gls{rt} performance.

\color{black}

\vspace{-.2cm}
\subsection{Design of \gls{du} Schedulers as \glspl{dapp}}
\vspace{-.1cm}
\label{sec:dapp}


\begin{figure}
    \centering
    \includegraphics[width=1.\linewidth]{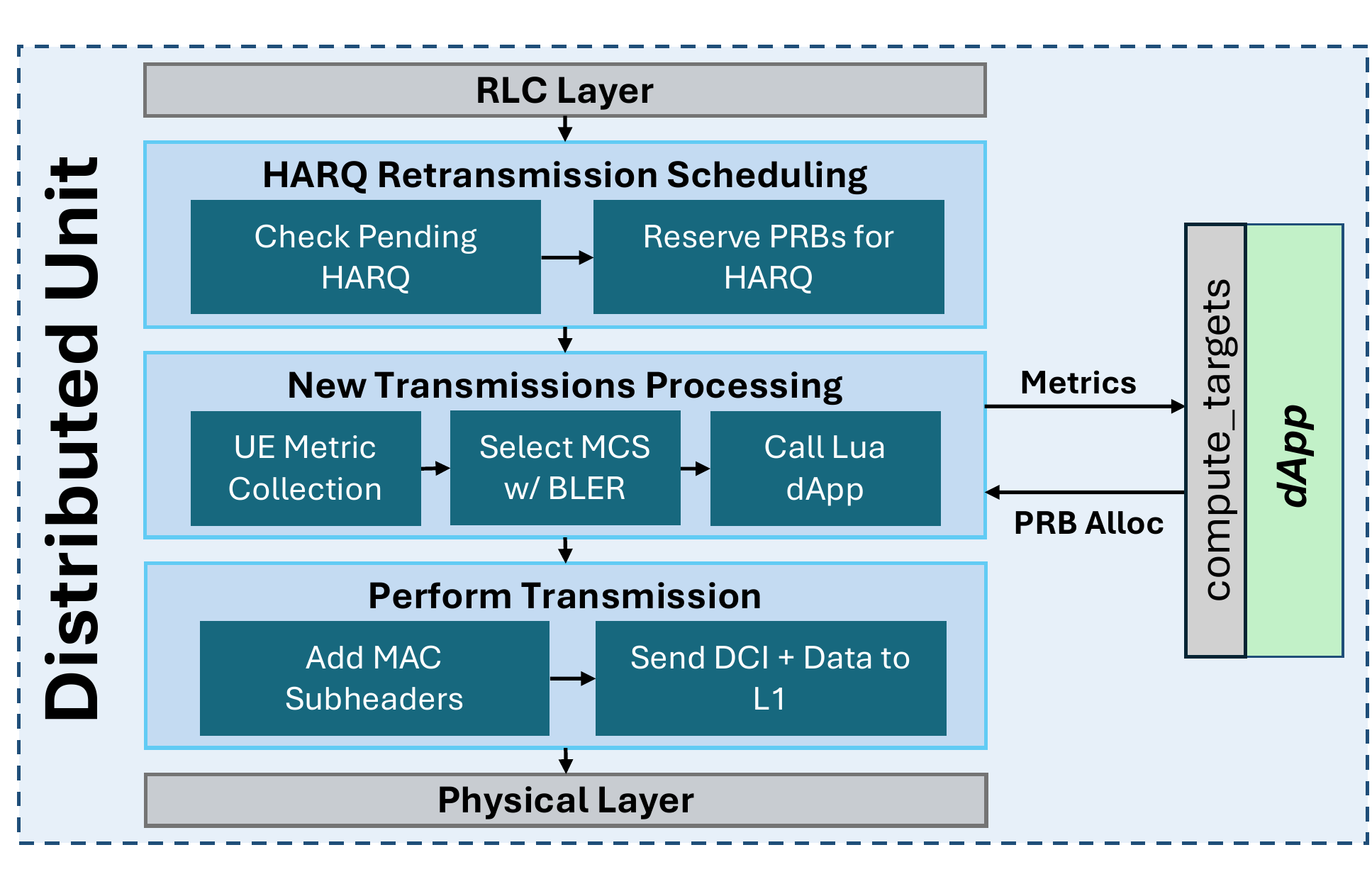}        \setlength\abovecaptionskip{-.3cm}
    \setlength\belowcaptionskip{-.3cm}
\caption{\framework MAC scheduler and interaction with the dApp.}
    
    \label{fig:OAI_stack}
\end{figure}

This section discusses the \framework flexible \gls{dapp} implementation, which interacts with procedures on the \gls{dlsch} to shift the scheduler from an integrated \gls{du} component to an easily swappable and optimizable function. 
In most \gls{ran} products, the scheduler is implemented in a compiled programming language (e.g., C/C++ or telecom-specific languages) and optimized for high execution speed and reliability. This is because of its time-sensitive nature (5G slots are scheduled at a periodicity ranging from $0.0625$ ms to $1$ ms) and the need for predictable performance. 
This limits the \gls{ran} to a few battle-tested algorithms (such as the well known \gls{pf} in, e.g., \gls{oai}) or manually tuned solutions from major vendors, hindering flexibility and adaptability. 

A scheduler typically follows a sequential process: first, it prioritizes \gls{harq} retransmissions to ensure reliability; second, it computes the \gls{mcs} of each \gls{ue} leveraging the \gls{bler}. It then goes through the allocation of the remaining resources. For example, for a PF scheduler, it computes \gls{pf} coefficients for all \glspl{ue}, based on the ratio of instantaneous achievable rate to historical average throughput, and then allocates \glspl{prb} to \glspl{ue} in descending order of their PF coefficients. The allocation size is determined by the \gls{mcs} and buffer status.

As shown in Fig. \ref{fig:OAI_stack}, we propose to improve this workflow by framing the resource allocation portion of the scheduler as a \gls{dapp}.
To integrate the \gls{dapp} in this context, we design an abstraction layer for scheduler implementations, which still handles \gls{harq} retransmissions with highest priority, but then delegates the resource allocation decision to a dApp. 
This is done with an \gls{api} that exposes \gls{mac} \glspl{kpi} and returns the \gls{prb} allocation for each \gls{ue} (i.e., the \texttt{compute\_target} \gls{api} shown in Fig.~\ref{fig:testing}).
We detail the \glspl{kpi} passed to the \gls{dapp} in Listing \ref{listing:datastruct}.
Once the allocations are returned, they are translated into \glspl{dci} and propagated across the stack through FAPI \cite{SCF2021FAPI}.

By exposing this abstraction, the scheduler transitions from a monolithic implementation to a modular system where allocation policies can be defined, deployed, and potentially updated independently from the core scheduler logic.
This separation of concerns introduces several advantages: it enables dynamic policy updates without modifying the underlying scheduler code; it allows for context-specific optimization strategies to be implemented as specialized \glspl{dapp}; and it enables re-use of optimization-related functions in different contexts. 

To change the logic of the scheduling \glspl{dapp} depending on context, intent, and, e.g., current traffic conditions, we design the system such that the scheduling logic becomes hot-swappable. For this reason, we select Lua as a programming language. As will be shown in Section \ref{sec:experimental-framework}, this programming language has the advantage of having extremely fast implementations \cite{luajit}, while still being a high-level interpreted language, making it easy to output for a \gls{llm} and to verify for a human. Furthermore, being primarily made for scripting in embedded environments, its luajit runtime is particularly lightweight. These advantages make it an ideal candidate to implement our scheduling \glspl{dapp}.

\begin{lstlisting}[float=t,floatplacement=t,language=json,style=mystyle-json, 
caption={Data structure passed to the scheduling \gls{dapp}}, 
label={listing:datastruct}]
typedef struct {
    float tbs;                // TBS estimate of 1 PRB
    float throughput;         // Avg Recent Goodput
    int pusch_snrx10;
    int pucch_snrx10;
    uint8_t dl_rsrp;
    uint32_t buffer_length;
    uint16_t total_pdus;      // Total PDUs transmitted
    float bler;
    uint8_t max_mcs;
    uint8_t current_mcs;
    bool ta_apply; // Presence of a timing advance update
    int16_t ta_update;        // Timing advance update value
    int cqi;
    int rssi;
    float rsrq;
    uint64_t hol_delay_us;
    bool hol_is_retransmission; // 1 if 1st pdu is retrans.
    uint16_t num_rbs_required;
    uint16_t rnti;
    uint64_t fiveQI;
} ue_metric_t;
ue_metric_t ue_metrics[UE_COUNT];
\end{lstlisting}

\vspace*{-5pt} 
\subsection{Scheduler Generation and Testing Framework}
\label{sec:testing}
\vspace{-.1cm}

\begin{figure}
    \centering
    \includegraphics[width=1.\linewidth]{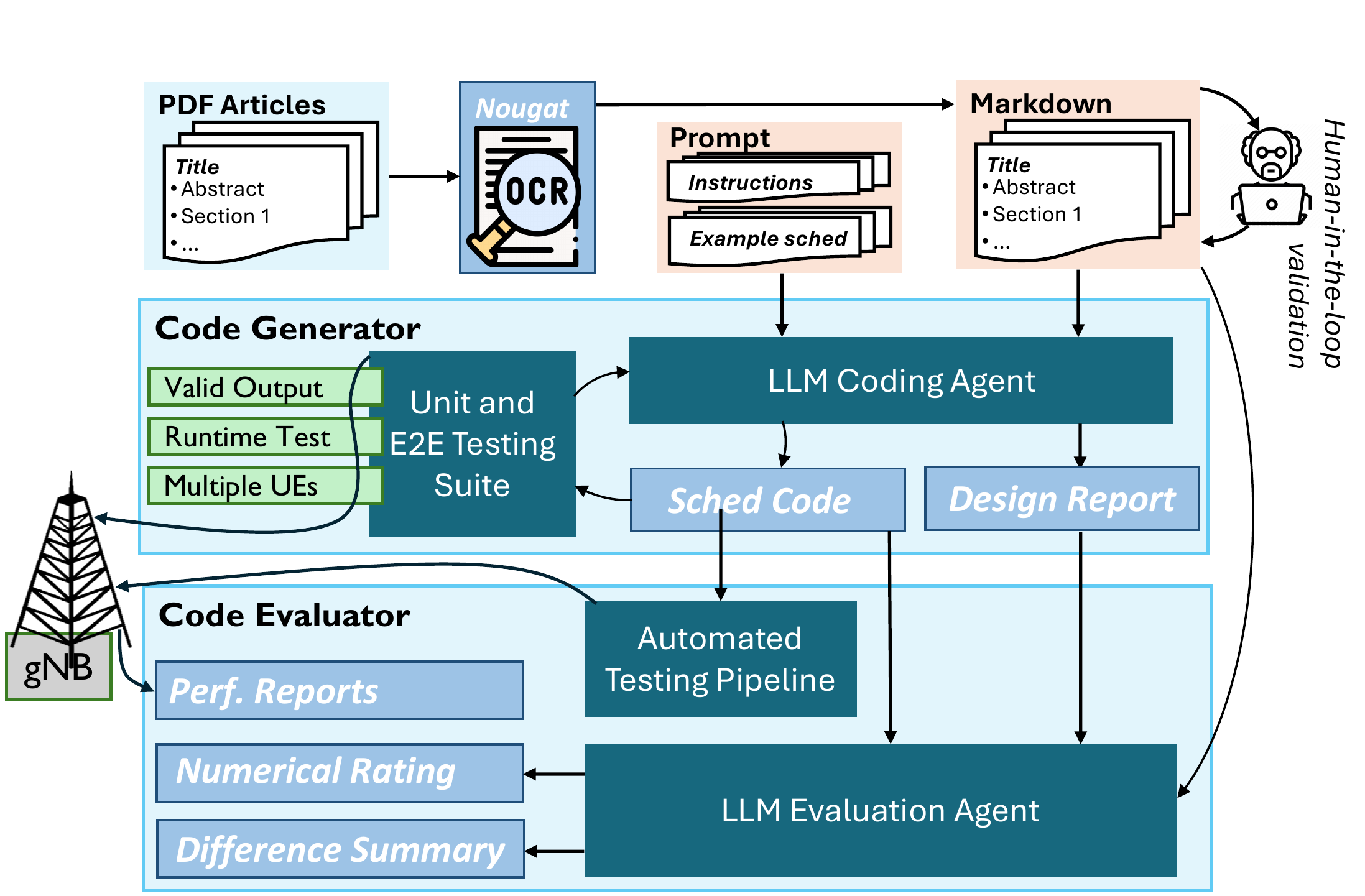}
    \setlength\abovecaptionskip{-.3cm}
    \setlength\belowcaptionskip{-.3cm}
    \caption{The \framework architecture for scheduler generation and testing.}
    \label{fig:framework}
\end{figure}

Let us now detail the architecture of the Scheduler Generation and Testing Framework. The role of this component of \framework is to take high level descriptions of schedulers (specifically, research papers) as input and to process them to obtain an executable scheduler code that is as close as possible to the source article, and that can then be deployed in the \gls{ran}.
As shown in Fig.~\ref{fig:framework}, the pipeline starts with its Data Preparation phase (top), which receives the PDF document of a research paper. Since PDF is purely Glyph-based, we use the Nougat \gls{ocr} engine \cite{blecher2023nougat} to extract the paper into structured markdown which can be consumed by a \gls{llm}. The generated Markdown can then optionally be checked by a human validator who ensures that the main equations and/or algorithms of the paper are correctly transcribed in Markdown. 
The other part of the Data Preparation pipeline consists in a prompt engineering step, in which detailed instructions and context are given to the model. 
This includes information such as the details of the data-structure passed to the scheduler (see Listing \ref{listing:datastruct}), the expected function prototype, and, if considered, the QoS requirements.

\begin{figure}
    \centering
    \includegraphics[width=1.\linewidth]{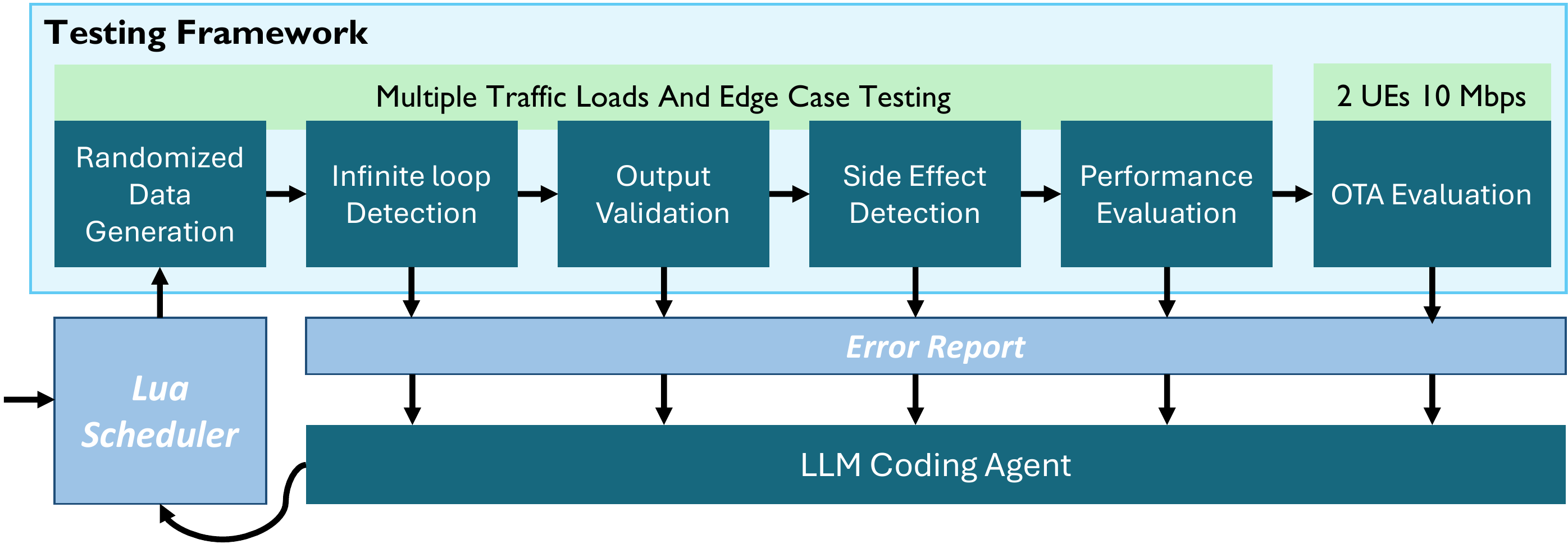}
        \setlength\abovecaptionskip{-.3cm}
    \setlength\belowcaptionskip{-.3cm}
    \caption{Architecture of the Testing suite.}
    \label{fig:testing}
\end{figure}

The output of the Data Preparation phase is then passed to the Code Generator (middle block of Fig.~\ref{fig:framework}). This pipeline always starts with a call to the \gls{llm}, which aims to generate valid code. Then, as illustrated in Fig. \ref{fig:testing}, the code goes through a series of unit tests, which do not run on the \gls{ran} but on a regular server. This tests for various usual issues such as infinite loops, respecting the maximum amount of \glspl{prb} even in edge cases, or avoiding buffer overflows, which would corrupt the memory of the rest of the \gls{du}. 
Then, if all these tests are passed successfully, we send the scheduler to the \gls{ota} \gls{ran} where we evaluate under minimal conditions whether \glspl{ue} successfully establish \gls{pdu} sessions and receive some downlink traffic. If there is a failure during any of these steps, a report is generated and passed to the \gls{llm} coding agent, which consumes it along with the code and the stack trace.
This enables the \gls{llm} to correct the code, and to pass it again to the whole testing pipeline. This loop continues until the code passes all the tests, at which point it is considered functional and sent to the Code Evaluation Pipeline along with a \gls{llm}-generated report explaining the approach to design the code. 

Finally, in the Code Evaluator (bottom part of Fig.~\ref{fig:framework}), another \gls{llm} agent receives the aforementioned design report and functional code along with the text of the original paper. This agent is prompted to evaluate the fidelity of the code compared to the paper. 
It provides a numerical rating along with a summary of the differences and similarities between the code and the paper, which the  {operator} can read as a first way to evaluate the correctness of the generated code. Furthermore, in the code evaluation phase, more tests are performed. Opposed to the tests in the code generation phase, which purely are functional, these are performance tests. We use \gls{x5g}~\cite{villa2024x5g} along with different traffic scenarios ({which are detailed in Sect. \ref{sec:num-results}}) to generate plots (see Sec.~\ref{sec:experimental-framework}) that enable us to assess the behavior of the scheduler in those different scenarios.


\vspace{-.2cm}
\subsection{\gls{ibs} Pipeline}
\vspace{-.1cm}
\label{sec:ibs}

 {We now describe the \gls{ibs} Pipeline of \framework, which builds upon the scheduler library created by our Generation and Testing pipeline (Section~\ref{sec:testing}). Having validated a comprehensive set of scheduling algorithms $\{\zeta_{code}^{(1)}, \ldots, \zeta_{code}^{(n)}\}$ with their performance profiles $\{\mathcal{M}^{(1)}, \ldots, \mathcal{M}^{(n)}\}$, we now address how to dynamically compose these validated elements to satisfy operator intents.
While individual schedulers from the literature each optimize for specific objectives (e.g., fairness, delay, throughput), real-world operator intents often require adaptive combinations of these behaviors based on runtime conditions. To address this gap, we propose a compositional framework for that can dynamically select and combine pre-tested scheduling elements—using objective functions from one algorithm, constraints from another, and grouping strategies from a third—or integrate new custom functions (operator-provided or LLM-generated) for novel requirements, balancing reliability through component reuse with the flexibility to address intents not covered by existing elements.}

\textbf{Scheduling Model.}
 {To enable flexible composition while maintaining the real-time execution constraints established in Section~\ref{sec:testing}, we will design a pipeline which implements the following function $\mathcal{G}_{IBS}$, generating the code $\zeta_{code}^{IBS}$ of the intent-based scheduler $\zeta_{IBS}$ :}
\begin{multline}
\mathcal{G}_{IBS}: \mathcal{I}, \mathcal{H}, \{\zeta_{code}^{(1)}, \ldots, \zeta_{code}^{(n)}\}, \\
\{\mathcal{M}^{(1)}, \ldots, \mathcal{M}^{(n)}\}, \{\zeta_{NL}^{(1)}, \ldots, \zeta_{NL}^{(n)}\} \mapsto \zeta_{code}^{IBS}
\end{multline}
 {
where $\mathcal{I}$ is the operator's \gls{nl} intent and $\mathcal{H}$ is an optional set of human-provided content such as additional code or algorithm descriptions.}

 {
Note that the general problem in equation \eqref{initialPB} is nonlinear. This makes it potentially slow to solve in real-time for arbitrary \gls{ibs} objectives and constraints. For this reason, we adopt the following sequential structure for $\zeta_{IBS}$.
First, $\zeta_{IBS}$ is comprised of the following grouping function, which groups multiple \glspl{ue} together based on their metrics:
\begin{equation}
    \Gamma: \mathbf{K(t)} \mapsto \{g_1, ..., g_k\}, \{B^{g_1}, ... B^{g_k} \}
\end{equation}
i.e., $\Gamma$ forms groups of \glspl{ue} $g_m$, which each can at most be allocated $B^{g_m}$ \glspl{prb}. This enables to differentiate how \glspl{ue} are scheduled based on e.g., metrics, 5QI or \gls{qos} requirements. Once the \glspl{ue} are grouped, $\zeta_{IBS}$ approximately solves different versions of the problem in equation \eqref{initialPB} for each group (these different versions are defined by $\mathcal{G}_{IBS}$ based on $\mathcal{I}$).
However, again, this problem is non-linear. For this reason, we approximate it by instead restricting the output of $\mathcal{G}_{IBS}$ to a series of linear knapsack problems. There is one series of sequential knapsacks for each group, which iteratively allocate \gls{prb} resources out of the leftovers from the previous round of knapsack. These problems each have the following structure, which we describe for any group $g_m$ at the $j^{th}$ round in the knapsack series as:
}
 {
\begin{align}
 \max_{\theta(t)} \quad & \sum_{i \in g_m} v_i^{g_m,j} \cdot\theta_i(t) \\
\text{subject to} \quad & \sum_{i \in g_m} \theta_i(t) \leq B^{g_m,j} \\
& l_i^{g_m,j} \leq \theta_i(t) \leq r_i^{g_m,j}
\end{align}
}
 {where $v_i^{g_m,j}$ is the utility coefficient for UE $i$, and $B^{g_m,j}$ is the PRB budget for that round. Note that $B^{g_m, 0}$ is initialized to $B^{g_m}$ and that the value of $B^{g_m, j+1}$ in subsequent rounds is $B^{g_m, j} - \sum\limits_{i \in g_m}\theta_i(t) $. We call $\theta^{g_m,j}(t)$ the solution of subproblem $j$ of group $g_m$, and the final allocation is $\theta_i(t) = \sum_{j} \theta_i^{g_m,j}(t)$}. This framework enables the scheduler to treat arbitrary groups of users differently, and, by composing multiple rounds of scheduling, non-linear constraints and objectives can be  {approximated through piecewise-linear decomposition, combined with the constraints limits, which can be used to selectively ensure that some leftover resources remain to leverage different objective functions in the subsequent rounds}.  { The main benefit of this approach is that each knapsack sub-problem is linear and has uniform weights, which enables optimal solutions via a simple greedy algorithm. When all items have uniform weight, the knapsack can hold exactly $k = \lfloor B^{g_m,j} \rfloor$ items. Since each item occupies the same space, maximizing total utility reduces to selecting the $k$ UEs with highest utility coefficients $v_i^{g_m,j}$, which can be done optimally in $O(n\cdot\log n)$ time via sorting.}

\begin{figure}[t]
    \centering
    \includegraphics[width=1.\linewidth]{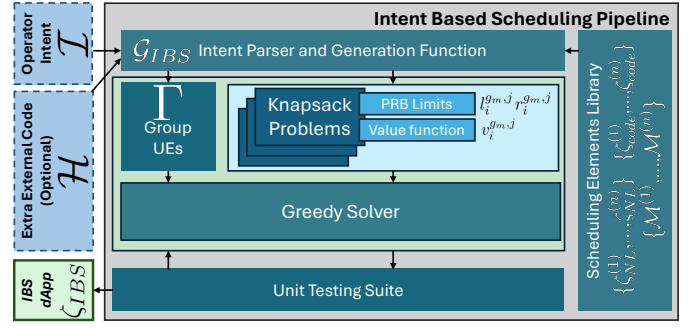}
    \setlength\belowcaptionskip{-.3cm}
    \caption{\gls{ibs} agent flow.}
    \label{fig:IBS_agent}
\end{figure}

\textbf{\gls{ibs} Lua Framework.}
We now describe the practical implementation of this solution model within \framework.  {First, $\mathcal{G}_{IBS}$ is implemented as a single-shot prompt which contains detailed instructions about the expected function signatures to return for the different elements of the solution, i.e., $\Gamma$, and the tuples of Lua functions which implement the pre-existing objective and constraints in the library.
Furthermore, while $\mathcal{G}_{IBS}$ can leverage all the code in $\{\zeta_{code}^{(1)}, \ldots, \zeta_{code}^{(n)}\}$, the prompt only contains the prototype of the different functions (i.e. the objective and limit function, which are described in greater details thereafter), which is sufficient to generate the scheduler code. Besides, we do not pass full research papers in the prompt, but instead resort to \gls{llm}-summarized versions of $\{\zeta_{NL}^{(1)}, \ldots, \zeta_{NL}^{(n)}\}$ , keeping the prompt shorter.
 Let us now detail the execution of the \gls{ibs} code, which is also depicted in Figure~\ref{fig:IBS_agent}:}
\begin{itemize}[leftmargin=*]
    \item Initially, \glspl{ue} are divided in groups  {using the Lua implementation of $\Gamma$, which is generated by $\mathcal{G}_{IBS}$}.
    \item  {$\mathcal{G}_{IBS}$ also generates a series of knapsack problems for each group. Each of those problems are specified by a }\verb|set_values|  {Lua function and a }\verb|set_limits|  {Lua function, which respectively output the values of $v_i^{g,j}$ and of $l_i^{g,j}, r_i^{g,j}$ given the \gls{mac} \glspl{kpi} (i.e., the matrix $\mathbf{K}$) for that problem. These functions are generated by the \gls{llm} which can either reuse directly the \gls{ibs} library elements or generate new code which combines the elements for new intents (for example, generating a new} \verb|set_limits|  {function which combines two pre-existing constraints).}
    
    \item From this information, the multi-round knapsack scheduler is defined and can be solved by solving each subproblem of each group sequentially.
    \item Finally, since there might be some remaining capacity (due to either not allocating all \glspl{prb} to the groups or to overprovisioning for some groups), the \framework \gls{ibs} pipeline also accepts an optional scheduler  {(as a knapsack problem)} which can distribute those extra resources.
\end{itemize}


We illustrate the simplicity of defining schedulers within this framework in Listing \ref{listing:example}. This is an example  {generated by $\mathcal{G}_{IBS}$} in which two groups are formed (i.e., $G=2$) based on the users \gls{cqi}, and where 70\% of the \glspl{prb} are given to the \glspl{ue} with the best channel. Then, each group uses a single round.
The first group allocates \glspl{prb} so that each \gls{ue} does not exceed $10$\:Mbps of throughput, and uses the VT-SH scheduler (described in Section~\ref{sec:review}), while the second group is limited to $20$\:Mbps and uses a \gls{pf} scheduler. Finally remaining resources are distributed according to \gls{pf}.

Such multi-round scheduling code can be human-written, but, as shown in Fig. \ref{fig:IBS_agent}, we also provide a \gls{llm} agent, which can accomplish that task by leveraging parts of the Testing pipeline. That agent receives as input the \gls{ibs} catalog, which describes the available \verb|get_groups|, \verb|set_limits| and \verb|set_values| function. It can also output these different functions. 
However, we highly recommend restricting to predefined \verb|set_values| functions whenever possible because they are the most error-prone, as they are typically non-intuitive, and subtle edge cases can lead to bugs such as starving some of the \glspl{ue}, which can only be avoided by extensive performance testing. 
For this reason, we argue value functions should be thoroughly tested before deployment, which would make the \gls{ibs} process much slower.
Value functions are also the most researched element of the pipeline: as will be made evident in Sec.~\ref{sec:review}, the key contribution of most schedulers from the literature is a value function, and as we will also show, even schedulers carefully crafted by researchers do not necessarily satisfy their intended targets. 
Finally, the intents for grouping and limits often map more directly into code, which makes it easier for the \gls{llm} agents to get right without extensive performance testing.


\begin{lstlisting}[float=t,floatplacement=t,language=json,style=LuaStyle, 
caption={Example \gls{ibs} scheduler. LLM-generated value functions that have been tested are in red. User-defined functions from the knapsack problem are in green}, 
label={listing:example}]
function get_schedulers()
    return { -- first scheduler
            { -- 1st sched. round for 1st scheduler
            {<@\textcolor{green!60!black}{set\_limits}@>=generate_set_limits_target_throughput(10), <@\textcolor{green!60!black}{set\_values}@>=<@\textcolor{red}{create\_VT\_SH\_values(100, 50, 1)}@>}
            }
            -- second scheduler 
            {
                -- 1st sched. round for 2nd scheduler
                {<@\textcolor{green!60!black}{set\_limits}@>=generate_set_limits_target_throughput(20), <@\textcolor{green!60!black}{set\_values}@>=<@\textcolor{red}{set\_values\_pf}@>}
            },
            -- last scheduler takes care of the remaining rbs (if no last scheduler specified, remaining RBs are unused)
            {
                {<@\textcolor{green!60!black}{set\_limits}@>=set_default_limits, 
                <@\textcolor{green!60!black}{set\_values}@>=<@\textcolor{red}{set\_values\_pf}@>}
            }
    }
end

<@\textcolor{green!60!black}{compute\_targets}@> = generate_compute_targets(get_2groups_by_cqi(0.5, 0.7), get_schedulers())
\end{lstlisting}

\vspace{-10pt} 
\section{Automated Private 5G Evaluation Framework}
\label{sec:experimental-framework}

In this section, we introduce the experimental framework which we used to prototype \framework, and the profiling of the Lua-based scheduler.

\vspace{-10pt} 
\subsection{Experimental Testbed and Framework}
\label{sec:testbed}%
We implement \framework as a modified version of \gls{oai}~\cite{kaltenberger2019openairinterface}. Most of the \glspl{kpi} discussed in Sec.~\ref{sec:dapp} and Listing \ref{listing:datastruct} are already available in the \gls{mac}. However, we find that (i) \gls{oai} does not track the \gls{hol}; and (ii) that the original throughput calculation uses an \gls{ewma} of the total amount of bytes transmitted, meaning that retransmitted data is counted multiple times. Furthermore, our experiments show that the reported throughput often has discrepancies compared with the application throughput reported by {e.g.}, iPerf. To address (i), we modify the \gls{mac} implementation such that each \gls{pdu} is now timestamped upon arrival. Then, in the scheduler, the timestamp of the first \gls{pdu} of the queue can be compared with the current time to obtain the \gls{hol}. For (ii), we resort to tracking the goodput by maintaining a circular buffer where, at each frame start, we keep track of the amount of bytes that have been sent minus the retransmitted bytes. We obtain goodput values that are only few percent apart from our measurements at the application layer, which can be attributed to the protocol overhead.


\begin{figure}
    \centering
    \includegraphics[width=0.95\linewidth]{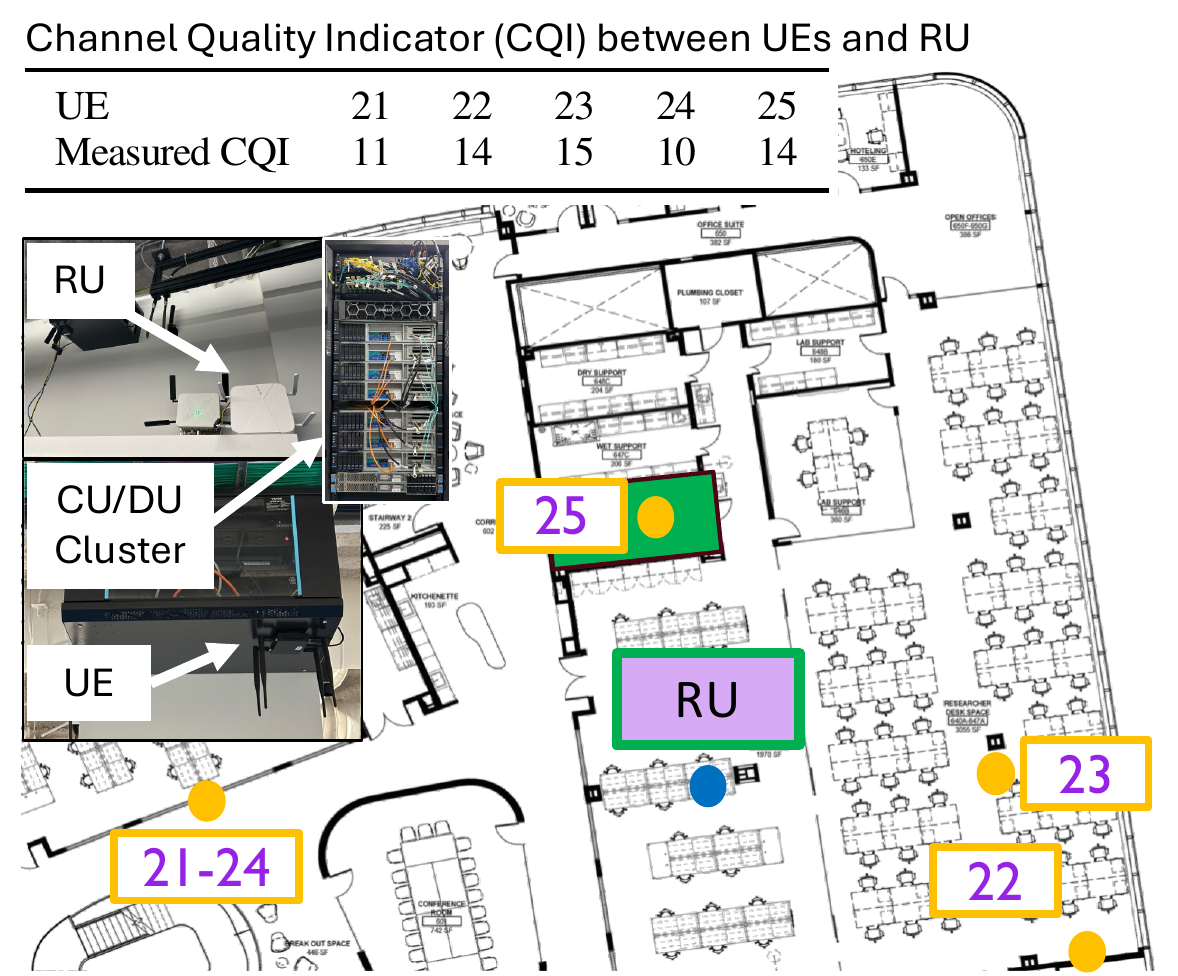}
        \setlength\abovecaptionskip{-0cm}
    \setlength\belowcaptionskip{-.6cm}
\caption{Locations of RU and UEs in our environment.}
    \label{fig:map}
\end{figure}

The experimental evaluation is carried out on the infrastructure provided by the X5G~\cite{villa2024x5g} testbed using the AutoRAN~\cite{maxenti2025autoranautomatedzerotouchopen} framework, as shown in Fig.~\ref{fig:map}.
We use a Gigabyte E251 server equipped with Intel Xeon 6240R and a Mellanox ConnectX 6. 
The \gls{ru} we use is a Foxconn RPQN, configured at 2x2 MIMO, transmitting at $20$\:MHz ($51$ \glspl{prb}) in band N77, $7DS2U$ \gls{tdd} pattern, and numerology $1$ (i.e., a slot duration of $0.5$\:ms).
Our core network is Open5GS, also running on the AutoRAN cluster. Inside the pod for the core network \gls{upf}, we deploy the server used to generate traffic.
Finally, as \glspl{ue}, we use 5 Sierra Wireless EM9293 with Qualcomm modems, positioned in the lab according to the map shown in Fig. \ref{fig:map}. Two \glspl{ue} (with IMSIs $22$ and $23$) are directly in line of sight of the radio, and have an excellent channel. \glspl{ue} $21$ and $24$ have an \gls{nlos} channel and are located in a corridor and a room with thick walls is between them and the RU. Finally, \gls{ue} 25 is in the neighboring server room and has a good channel, slightly worse than the line-of-sight \glspl{ue}. We report the average measured \gls{cqi} of those \glspl{ue} in Fig.~\ref{fig:map}.
%

Unless specified otherwise, in the remainder of this paper, we evaluate the schedulers by automatically carrying out two tests. First, all \glspl{ue} receive iPerf traffic over \gls{udp} at a constant bitrate of $24$\:Mbps. This test mainly aims at evaluating the fairness of the schedulers, as the sum of the bitrates exceeds the maximum achievable by the 20 MHz carrier which, in our tests, proved to be of $105 - 110$\:Mbps with a perfect channel. 
The second set of tests uses the mgen traffic generator. Mgen is configured with a bursty ON/OFF traffic distribution with exponentially distributed ON and OFF periods with an average duration of 2 seconds, and bursts at a rate of 2000 packets/s of size 1000 bytes. 
The goal of this test is mainly to evaluate the delay distribution obtained in bursty conditions.
Both experiments are run 10 times for a duration of 420 seconds. The metrics we shall use are hence (i) for iPerf, the throughput and fairness coefficients (Gini and Jain), (ii) for mgen, the delay distribution. In both cases, we also report the starvation rate, which indicates whether and how many \glspl{ue} were starved, which we define as a \gls{ue} either not managing to establish a \gls{pdu} session or not receiving data for more than 30 s.

\vspace{-15pt} 
\subsection{Overhead Evaluation and Comparison with \gls{oai}}

\begin{figure}[t]
\begin{subfigure}[c]{0.49\linewidth}
    \centering
    \scriptsize
    \resizebox{\linewidth}{!}{%

\begin{tikzpicture}

\definecolor{darkblue}{RGB}{31,119,180}
\definecolor{lightgray}{RGB}{240,240,240}
\definecolor{darkgray}{RGB}{130,130,130}
\definecolor{criticalred}{RGB}{214,39,40}

\begin{axis}[
width=1.1\linewidth,
height=0.7\linewidth,
title style={font=\large\bfseries},
legend cell align={left},
legend pos=south west,
legend style={font=\scriptsize, draw=darkgray, fill=white, fill opacity=0.9,at={(0.5,1)},anchor=south},
tick pos=left,
grid=both,
grid style={black, dashed},
minor grid style={lightgray!50},
xlabel={Scheduler Execution Time ($\mu$s)},
xlabel style={font=\scriptsize},
xmajorgrids,
xmin=0, xmax=500,
xtick style={color=black},
ylabel={Throughput (Mbps)},
ylabel style={font=\scriptsize},
ymajorgrids,
ymin=0, ymax=110,
ytick style={color=black}
]

\addplot [very thick, darkblue, mark=*, mark size=1, mark options={solid, fill=darkblue}, forget plot]
table {%
0 105
10 105
20 105
30 105
40 105
50 105
75 105
100 105
150 105
200 105
250 105
300 105
350 105
400 105
410 48
420 41.1
430 18.9
440 21.2
450 8.4
460 3.86
470 3.79
1000 0
};

\addplot [ultra thick, criticalred, opacity=0.8, dashed]
table {%
400 0
400 110
};

\addlegendentry{Critical Delay: 400 $\mu$s}
\end{axis}

\end{tikzpicture}
    }
    \setlength{\abovecaptionskip}{-.3cm}
    \setlength{\belowcaptionskip}{0cm}
    \caption{UE throughput as a function of the scheduler execution time.}
    \label{fig:plot_sleeps}
\end{subfigure}%
\hfill
\begin{subfigure}[c]{0.49\linewidth}
    \centering
    \scriptsize
    \begin{tabular}{ll}
    \toprule
        Algorithm & Number of \glspl{ue} \\
    \midrule
        BCQI & 1000\\
        QoS Log Rule & 350\\
        A-EXP/PF & 510\\
        Others & $\geq 600$ \\
    \bottomrule
    \end{tabular}
    \setlength{\belowcaptionskip}{0cm}
    \caption{Maximum number of \glspl{ue} that can be scheduled in less than 380 $\mu$s.}
    \label{tab:cutoffs}
\end{subfigure}
\caption{Scheduler profiling.}
\end{figure}
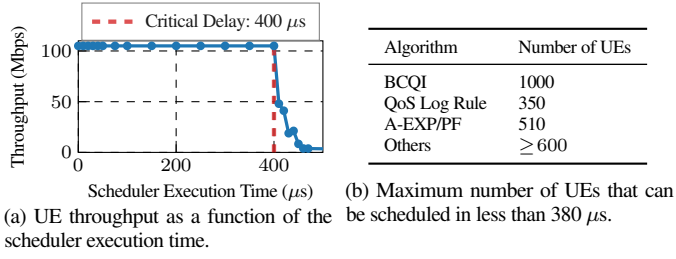

\begin{figure}[t]
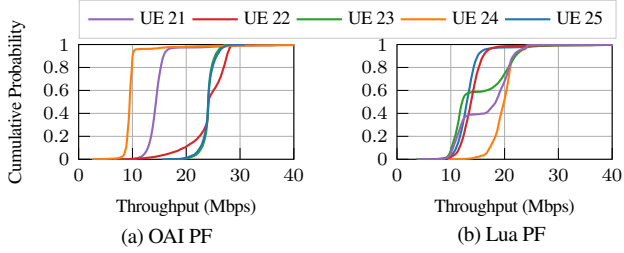

    \centering
    \begin{subfigure}[t]{0.5\linewidth}
        \centering
            \input{figures/aggregated_by_ue_smooth_oai}
        \setlength{\abovecaptionskip}{-.4cm}
        \setlength{\belowcaptionskip}{0cm}
        \caption{\gls{oai} \gls{pf}}
    \end{subfigure}%
    \hfill%
    \begin{subfigure}[t]{0.5\linewidth}
        \centering
            \input{figures/PF_smooth}
        \setlength{\abovecaptionskip}{0cm}
        \setlength{\belowcaptionskip}{0cm}
        \caption{Lua \gls{pf}}
        \label{fig:oai_comparison_lua}
    \end{subfigure}
    \caption{CDF of throughput of Lua \gls{pf} vs \gls{oai} (iPerf test).}
    \label{fig:oai_comparison}
\end{figure}

Before evaluating the scheduling algorithms themselves, we assess the overhead induced by the introduction of the Lua scheduler within \gls{oai}. 
First, we evaluate the time budget available for scheduling within the \gls{dapp} before the performance starts degrading. To do this, we connect \gls{ue} $22$ to the network and artificially increase the scheduling loop execution time using the \verb|sleep| function. For this, we use a dummy scheduler which always allocates all \glspl{prb} available.
We then push $105$\:Mbps of traffic through iPerf and iteratively increase the sleep time, measuring the performance degradation. As illustrated in Fig. \ref{fig:plot_sleeps}, the overhead is low enough that we have 400 $\mu$s of time budget to run the scheduler with our numerology 1 configuration.
In order to evaluate the number of \glspl{ue} that could be supported by the system for each scheduler, we run our unit testing suite with each scheduler with increasing numbers of \glspl{ue}. This setup enables us to isolate the scheduler from potential \gls{oai} instabilities and to scale the number of \glspl{ue} arbitrarily. We run each scheduler increasing the number of \glspl{ue} from 0 to 1000 with increments of 10 on 30 different random scheduling problems. We then report in Table \ref{tab:cutoffs} the cutoff, which is the number of \glspl{ue} for which at least one instance has a runtime larger than $380$ $\mu$s (we select this value to keep a reasonnable margin of $20$ $\mu$s).

We then assess the difference between the \gls{oai}-provided \gls{pf} scheduler (written in C) and the \framework Lua implementation. 
As shown in Fig. \ref{fig:oai_comparison}, the Lua version of the \gls{pf} algorithms achieves a much higher degree of fairness compared to the \gls{oai} scheduler, due to our modification of the throughput calculation. Indeed, the gap between the median throughput of the least and most favored \glspl{ue} is reduced by more than 45\% with the Lua \gls{pf}.

\section{Literature-Generated Schedulers}
\label{sec:review}%
In this section, we review the schedulers that are evaluated and integrated in \framework. We select papers from the wireless networking literature along with four common baselines. We focus on two classes of schedulers, which we differentiate based on support for \gls{qos} prioritization (\emph{\gls{qos}-aware}, with a focus on support for delay requirements) or not (\emph{\gls{qos}-unaware}). Table~\ref{tab:notation} summarizes the notation used in the rest of the paper.

\begin{table}[b]
    \vspace{-10pt}
    \centering
        \setlength{\abovecaptionskip}{0cm}
    \setlength{\belowcaptionskip}{0cm}
        \caption{Notation table}
    \label{tab:notation}
    \begin{tabular}{ll}
    \toprule
        Notation & Description \\
    \midrule
        $\Raveuser{i}$ & Average throughput of User $i$ \\
        $\Rinstuser{i}$ & Instantaneous achievable throughput of 1 \gls{prb}\\
        $\MCSmax$ & Maximum $\MCS$ \\
        $\Buffuser{i}$ & Length of Buffer of user $i$ \\
        $\Buffuser{RT}$ & Buffer size sum for real-time users \\
        $\Buffuser{NRT}$ & Buffer size sum for non real-time users \\
        $\MaxBuff$ & Maximum Buffer Size \\
        $\HOLdelayuser{i}$ & Head-Of-Line delay of user $i$ \\
        $\HOLdelayAvg$ & Average Head-Of-Line delay \\
        $\DelayThreshold_i$ & Maximum acceptable delay for user $i$\\
        $\PDB_i$ & Delay violation maximum probability \\
        $\mathrm{CQI}_i$ & Channel Quality Indicator of user $i$\\
    \bottomrule
    \end{tabular}

\end{table}

\begin{table*}[t]
\centering
\caption{Comparison of \gls{qos}-unaware scheduling algorithms. The first four algorithms are baselines that are not \gls{llm}-generated.}
\label{tab:scheduling_algorithms}
\scriptsize
\begin{tabularx}{\textwidth}{m{0.25cm} m{2cm} m{5.85cm} m{8.2cm}}
\toprule
\textbf{Ref.} & \textbf{Algorithm} & \textbf{Value Function} & \textbf{Parameters} \\
\midrule
\vspace{-9pt}\cite{capozzi2013downlink}  & Round Robin (RR) & Last time \gls{ue} got served & \\
\midrule
\vspace{-9pt}\cite{capozzi2013downlink} & Best CQI & \CQI & \\
\midrule
\vspace{-9pt}\cite{capozzi2013downlink} & Fair Throughput (FT) & $-\Raveuser{i}$ &\\
\midrule
\vspace{-9pt}\cite{kelly1997charging} & Proportional Fair (PF) & $\frac{\Rinstuser{i}}{\Raveuser{i}}$ & \\
\midrule
\vspace{-9pt}\cite{saglam20195g} & Lean Scheduler & $\alpha_{RRE}\log(\Rinstuser{i}/\Raveuser{i}) + \alpha_F\log(\CQIuser{i}(\slot)/\MCSmax)$ & $\alpha_{RRE}$ and $\alpha_F$: weighting factors dynamically adjusted based on $S_{lean}$ parameter such that $\alpha_{RRE} + S_{lean} \times \alpha_F = 1$. \\
\midrule
\vspace{-9pt}\cite{sadiq2010delay} & Log Rule & $b_i \log(c+a_i \Buffuser{i}) \times \Rinstuser{i}$ & $b_i$: user weight parameter typically set to $1/\mathbb{E}[\Rinstuser{i}]$, $a_i$ queue balancing weight (larger values reduce emphasis on queue), $c \geq 1$: constant affecting behavior near origin.\\
\midrule
\vspace{-8pt}\cite{mamane2021proportional} &\vspace{1pt} Proportional Fair Buffer (PFB) & $\frac{\Rinstuser{i}}{\Raveuser{i}}\times \Buffuser{i}$ & \\
\midrule
\vspace{-9pt}\cite{ramjee2006generalized} & Generalized Proportional Fair (GPF) & $\frac{[\Rinstuser{i}]^{\beta}}{[\Raveuser{i}]^{\gamma}}$ & \vspace{2pt}$\gamma$ and $\beta$ are tunable weights (inputs to the algorithm).\\
\midrule
\vspace{-9pt}\cite{afifi2021novel} & Edge User Friendly Scheduler (EUFS) & \vspace{-8pt}$
\begin{cases}
    P(\text{Sched}) = \frac{\mathrm{PF}_C}{(\mathrm{PF}_C + \sum \mathrm{FU}_C[i])} \\
    \mathrm{PF}_C = \frac{P_\mathrm{PF} \times n_{fus} \times K}{(1 - P_\mathrm{PF})}
\end{cases}
$ & 2 rounds scheduler. First applies regular \gls{pf} scheduling, then randomly decides if cell-edge users preempt resources based on $P(sched)$ value. $\mathrm{PF}_C$: parameters with default value of 0.8, $\mathrm{FU}_C[i]$: variable controlling preemption chances. Increases if no preemption, decreases if preemption is scheduled. Cell-edge users are classified based on low $\Rave$. $n_{fus}$ is the number of cell-edge users.\\
\bottomrule
\end{tabularx}
\end{table*}

\begin{table*}[t]
\centering
\scriptsize
\caption{Comparison of \gls{qos}-aware algorithms. \gls{rt} users are prioritized, \gls{nrt} are not.}
\label{tab:scheduling_algorithms_qos}
\begin{tabularx}{\textwidth}{m{0.25cm} m{2cm} m{5.85cm} m{8.2cm}}
\toprule
\textbf{Ref.} & \textbf{Algorithm} & \textbf{Criterion} & \textbf{Parameters and comments} \\
\midrule
\vspace{-7pt}\cite{sadiq2009downlink} & \vspace{2pt}QoS-Log-Rule & $b_i \times \log(c + \frac{5}{\Raveuser{i}} \HOLdelayuser{i}) \times \Rinstuser{i}$ &\vspace{2pt} Same as Log Rule in Table~\ref{tab:scheduling_algorithms}  \\ 
\midrule
\vspace{-7pt}\cite{monghal2010dynamic}&Required Activity Detection with Delay Sensitivity (RAD-DS) & $\phi(i) \times \mathrm{RA}^{\TraffType}(i) \times DS^{\TraffType}(i), \;\;\; \TraffType \in \{\mathrm{RT}, \mathrm{NRT}\}$& Criterion is used to determine which users transmit, then \gls{pf} is applied. 
$\phi(i)$: counter incremented every slot and reset to 0 when \gls{ue} is scheduled. $\mathrm{RA}^{\mathrm{RT}}(i) = \mathrm{GBR}_i / \Rinstuser{i}$ (GBR is guaranteed bitrate), 
$\mathrm{RA}^{\mathrm{NRT}}(i) = \max(0, \PRBmax - \sum \mathrm{RA}^{\mathrm{RT}} / \#{\mathrm{NRT}}) $. \\
\midrule
\vspace{-7pt}\cite{brehm2013overload} & QoS Fractional Knapsack (QoS-FK) & $16 \times \tanh(\frac{\HOLdelayuser{i}}{\DelayThresholdUser{i}}) +  2 \times \tanh(\frac{\Buffuser{i}}{\MaxBuff}) + 4 \times \tanh(\mathrm{P}_i)+  4 \times \tanh(\mathrm{L}_i) 
$ & Weights can be tweaked, we report the ones suggested by the paper. $\mathrm{P}_i$ is an operator-set priority factor and $\mathrm{L}_i$ is the normalized packet loss.\\
\midrule
\vspace{-7pt}\cite{iturralde2012resource} & \vspace{2pt}VT-SH & \vspace{-10pt}$\begin{cases} 
\exp\left(\frac{6}{\DelayThresholdUser{i}} \times \frac{V_i}{1+ \sqrt{\HOLdelayAvg}}\right) \times \frac{\Rinstuser{i}}{\Raveuser{i}} & \text{(RT)} \\
\frac{\Rinstuser{i}}{\Raveuser{i}} & \text{(NRT)} 
\end{cases}$ & 2 levels: first allocate resources between groups of users (RT/NRT) through game theory (shapley value) then schedules each group separately. Uses virtual token bucket for RT, i.e., associates each \gls{ue} with virtual queue with constant arrivals rate, and service rate based on actual transmitted data. Then calculate virtual token delay $V_i$ based on service/arrival rates.\\
\midrule
 \vspace{2pt}\cite{basukala2009performance} & \vspace{10pt}EXP/PF & $
\begin{cases}
\exp\left(\frac{\frac{-\log \PDBuser{i}}{\DelayThresholdUser{i}} \times \HOLdelayuser{i} - \HOLdelayAvg}{1 + \sqrt(\HOLdelayAvg)}\right) \times \frac{\Rinstuser{i}}{\Raveuser{i}} & \text{(RT)} \\
\frac{w}{\Buff_{\text{RT}}} \times \frac{\Rinst}{\Rave} & \text{(NRT)} 
\end{cases}
$ & \vspace{10pt} $\HOLdelayAvg$ is weighted by $\frac{-\log \DelayThresholdUser{i}}{\PDBuser{i}}$. $w$ is updated based on comparison with maximum $\HOLdelay$.\\ \midrule
\cite{rhee2003scheduling} & \vspace{10pt}Adaptative Exp/PF(A-EXP/PF) & \vspace{-2pt}$
\begin{cases} \frac{c}{\DelayThresholdUser{i}}
\exp\left(\frac{\frac{c}{\DelayThresholdUser{i}} \times \HOLdelayuser{i} - \HOLdelayAvg}{1 + \sqrt(\HOLdelayAvg)}\right) \times \frac{\Rinstuser{i}}{\Raveuser{i}} & \text{(RT)} \\
\frac{w}{\Buff_{\text{NRT}}} \times \frac{\Rinst}{\Rave} & \text{(NRT)}
\end{cases} 
$ & \vspace{4pt} c is a constant. $\HOLdelayAvg$ is weighted by $\frac{c}{\PDBuser{i}}$. $w$ is updated based on comparison of $\HOLdelayAvg$ with $c$.\\
\midrule
\vspace{-9pt}\cite{piro2011two}&2 Level Sched. w/ Frame Level Sched. (2L-FLS) & 
$l_i[t] = \Buffuser{i}[t] + \sum\limits_{n=2}^{M_i}(\Buffuser{i}[t-n+1] - \Buffuser{i}[t-n+2] - r_i[t-n+1])c_i[n]$
& Control theoretic delay guarantees. $M_i$ is the delay bound in number of slots, at each slot calculate minimum resources $l_i$ for RT \glspl{ue}, then apply \gls{pf} taking $l_i$ into account. \\ 
\midrule
\vspace{-9pt}\cite{mahfoudi2015new} &\vspace{2pt}LTTI& $\frac{-\log(\PDBuser{i})}{\DelayThresholdUser{i}} \times \frac{\Rinstuser{i}}{\Raveuser{i}} \times \exp(\DelayThresholdUser{i}/(\DelayThresholdUser{i} - \HOLdelay))$ & \\
\midrule
\vspace{-9pt}\cite{husain2020efficient} & Weighted Delay Based Packet Sched (WD-PS) & \vspace{-3pt}$\log\left(1+ -\frac{\log\PDBuser{i}}{\DelayThresholdUser{i}} \cdot \HOLdelayuser{i}\right)\cdot-\frac{\log\PDBuser{i}}{\DelayThresholdUser{i}}\cdot \HOLdelayuser{i}\cdot \frac{\Rinstuser{i}}{\Raveuser{i}}$ &  \\
\bottomrule
\end{tabularx}
\vspace{-10pt}
\end{table*}

\textbf{\gls{qos}-Unaware Schedulers}
are summarized in Table \ref{tab:scheduling_algorithms}. The first four algorithms are classic baselines, with \gls{rr}, which serves users cyclically, \gls{bcqi} which maximizes channel utilization by favoring \glspl{ue} with the best channel in all cases, \gls{ft}, which maximizes fairness by always favoring \gls{ue} with the lowest throughput and \gls{pf}, which tries to balance channel utilization and throughput fairness. We also compare with five other schedulers from the literature, mostly based on \gls{pf}.
First, the Lean Scheduler \cite{saglam20195g} tries to balance between a \gls{pf} objective and a \gls{bcqi} objective with weights that scale over time. Second, the Log Rule \cite{sadiq2010delay} strikes a balance between prioritizing instantaneous throughput and serving \glspl{ue} with larger buffers first, which can reduce the experienced latency. Third, the \gls{pf} Buffer algorithm follows a similar idea, extending \gls{pf} with a Buffer length factor. The Generalized \gls{pf} algorithm adds exponential parameters, enabling the operator to use \gls{pf} while choosing how much fairness should be prioritized over resource efficiency. Finally, the Edge User Friendly algorithm modifies \gls{pf} to assign more resources to cell-edge users.

\textbf{\gls{qos}-Aware Schedulers} are listed in Table \ref{tab:scheduling_algorithms_qos}. All these algorithms are focused on satisfying \gls{hol} latency constraints, and can be classified in two groups. First, QoS-Log-Rule, QoS-FK, LTTI, and WD-PS aim to meet a latency target for each \gls{ue} or group of \glspl{ue}.
On the other hand, RAD-DS, GT-EXP/PF, VT-SH, Adaptative EXP/PF, and 2L-FLS are restricted to two groups of \glspl{ue}, one of which has a delay target. 
The other one is considered best effort and typically dealt with using a flavor of \gls{pf}.

\section{Qualitative Evaluation}
\label{sec:qualitativeevaluation}

In this section, we evaluate the code-generation pipeline discussed in Section~\ref{sec:testing} with emphasis on code correctness and ease of spotting errors. We use Claude 3.7 Sonnet  {as it represents the state-of-the-art code generation model at the time of writing} as the \gls{llm}, as in the rest of the paper.
Note that we ask the model to use the default parameters provided by the input paper, ensuring that algorithm-specific parameters are not added to the prototype of the \verb|compute_targets| function. 
Furthermore, our prompt contains 18 instructions, along with three example schedulers (\gls{pf}, \gls{rr}, and BCQI). For the \gls{qos}-aware schedulers, we require the \gls{llm} to output a version of the scheduler which has only two classes, namely, privileged and non-privileged, which can be identified based on the 5QI.

\begin{table}[]
    \centering
        \caption{Qualitative results summary.}
    \label{tab:eval_auto}
\begin{tabular}{l c@{\hspace{0.5em}} c@{\hspace{0.5em}} c c c}
    \toprule
    \multirow{2}{*}{Scheduler} & \multicolumn{2}{c}{Rating} & \multirow{2}{*}{Iterations} & \multirow{2}{*}{Success} & \multirow{2}{*}{\shortstack{Correct\\Err. Rep.}} \\
    \cmidrule(lr){2-3}
    & $\checkmark$ & $\times$ & & & \\
    \midrule
    Lean Scheduler & 8.8/10 & N/A & 1 & 100\% & N/A \\
    Log Rule & 8.6/10 & N/A & 1 & $100$\% & N/A \\
    PFB & 8.8/10 & N/A & 1 & $100$\% & N/A \\
    GPF & 9/10 & N/A & 1 & $100$\% & N/A \\
    EUFS & 8.8/10 & N/A & 1 & $100$\% & N/A \\
    \gls{qos} log rule & 9/10 & 8/10 & 2 & $60$\% & $0$\% \\
    RAD-DS & 8/10 & 8/10 & 1.2 & $20$\% & $75$\% \\
    \gls{qos}-FK & N/A & 7.8/10 & 2 & $0$\% & $80$\% \\
    VT-SH & 8.75/10 & 8/10 & 1 & $80$\% & $100$\% \\
    EXP/PF & N/A & 8.4/10 & 2 & $0$\% & $100$\% \\
    A-EXP/PF & 8.75/10 & 8.5/10 & 1.6 & $60$\% & $100$\% \\
    2L-FLS & N/A & 6.2/10 & 1.2 & $0$\% & $100$\% \\
    LTTI & 9.1 & N/A & 2.6 & $100$\% & N/A \\
    WD-PS & N/A & 7/10 & 1.2 & $0$\% & $100$\% \\
    \bottomrule
\end{tabular}

    {\vspace{.1cm}\scriptsize The rating is outputted by the \gls{llm} and averaged separately for runs that are correct ($\checkmark$) or have an error ($\times$). The number of iterations of the generation pipeline is averaged, and the last column gives the proportion of runs where the errors are correctly detected by the \gls{llm}.}
    \vspace{-15pt}
\end{table}

For this evaluation, we run the code generation pipeline five times for each scheduler and we report the results in Table \ref{tab:eval_auto}. We observe that all \gls{qos}-unaware algorithms are successfully transcribed into Lua code after a single test iteration, and that the pipeline attributes grades close to 9/10 to all implementations. During our manual review, we find concurring results, and do not see any discrepancy between \gls{qos}-unaware papers and their generated code. This is in line with the general observation that these are the simplest algorithms we evaluate.

Conversely, \gls{qos}-aware schedulers are much harder to implement, as they are more complex algorithms. We also observe that in multiple cases, some adaptations of the original algorithm need to be made to make it work in our specific context. These occurrences typically create confusion, as will be shown by the following detail of the most common \gls{llm} errors, along with the necessary corrections:
\begin{itemize}[leftmargin=*]
    \item In RAD-DS, instead of an incrementing counter, the model often tries to use wall-clock time measurement for $\phi(i)$;
    \item In \gls{qos}-FK, \BLER is used instead of packet loss. The packet loss factor should have been ignored since it is unavailable at the \gls{mac};
    \item In EXP/PF, the number of RT users is used in lieu of $\Buffuser{RT}$;
    \item 2L-FLS is initially presented as a frame level scheduler, which confuses the \gls{llm}, as it assumes the scheduler is called every $10$ ms instead of every $0.5$ ms, making the delay constraint too tight.
    As a control-theoretic algorithm, 2L-FLS also provides deterministic guarantees that $\DelayThreshold$ will not be violated. 
    This requires (i) the scheduler to calculate the number of bytes that can be transported in one \gls{prb} and (ii) planning how many slots to give to users in the future.
    Initially, the \gls{llm} often uses a hard-coded value for (i), missing that the \gls{tbs} estimate per \gls{prb} is available as input, and also missing that the number of bytes to allocate needs to account for the control plane overhead
    (8\%).
    %
    For (ii), the \gls{llm} typically fails to notice that it would need to tweak the $M_i$ parameter to account for S and U slots, where no downlink transmission is possible.
    Finally, the history of $\Buffuser{i}$ is sometimes not stored in the \gls{llm} provided version, making it impossible to implement the control equation.
\end{itemize}
Overall, after investigating and fixing these schedulers, we find that the \gls{llm} is able to reasonably assess the quality of its responses, since the average rating of error-free results is above 8.75 for all schedulers but RAD-DS, while it is below that value for code with error, with significantly lower values for 2L-FLS, i.e., the most error-prone case. Furthermore, the rate of correct error reporting is close or equal to 100\% for almost all schedulers, meaning the system can successfully guide an operator to quickly assess and fix its errors, reducing the implementation time of schedulers from hours to minutes. 

\vspace{-5pt} 
\section{Experimental Evaluation}
\label{sec:num-results}
In this section, we present numerical results related to the evaluation of the literature-generated schedulers.

\vspace{-10pt} 
\subsection{\gls{qos}-Unaware Schedulers}
\label{sec:qos_unaware}
We start by evaluating \gls{qos}-unaware schedulers. We use the parameters suggested from the papers, except for GPF, for which we use multiple combinations of $\beta$ and $\gamma$ (i.e., $\beta = 0.6, \gamma = 0.7$, which is the default from the paper, $\beta = 0.2, \gamma=0.7$, for increased fairness, and $\beta=0.6, \gamma=0.2$, for decreased fairness).

\begin{figure}[t]
    \centering
    \begin{subfigure}[t]{0.49\linewidth}
        \centering
\begin{tikzpicture}

\definecolor{crimson2143940}{RGB}{214,39,40}
\definecolor{darkgrey176}{RGB}{176,176,176}
\definecolor{darkorange25512714}{RGB}{255,127,14}
\definecolor{forestgreen4416044}{RGB}{44,160,44}
\definecolor{lightgrey204}{RGB}{204,204,204}
\definecolor{mediumpurple148103189}{RGB}{148,103,189}
\definecolor{steelblue31119180}{RGB}{31,119,180}

\begin{axis}[
width=\linewidth,
height=0.8\linewidth,
legend cell align={left},
legend style={fill opacity=0.8, draw opacity=1, text opacity=1, draw=lightgrey204,at={(1.25,1.05)},anchor=south,font=\scriptsize},
legend columns=6,
tick pos=left,
x grid style={darkgrey176},
xmin=-0.83, xmax=10.83,
xtick style={color=black},
xtick={0,1,2,3,4,5,6,7,8,9,10},
xticklabel style={rotate=45.0,anchor=east,font=\tiny},
xticklabels={
  RR,
  BCQI,
  FT,
  PF,
  Lean Sched,
  Log Rule,
  PFB,
  {GPF \(\displaystyle 0.6\), \(\displaystyle 0.7\)},
  {GPF \(\displaystyle 0.2\), \(\displaystyle 0.7\)},
  {GPF \(\displaystyle 0.6\), \(\displaystyle 0.2\)},
  EUFS
},
y grid style={darkgrey176},
ylabel={Starvation Rate},
ymin=0, ymax=1.0,
ytick style={color=black},
ymajorgrids,
yminorgrids,
xlabel style={font=\scriptsize},
ylabel style={font=\scriptsize}
]

\addlegendimage{empty legend}
\addlegendentry{Number of starved UEs:}

\draw[draw=black,fill=steelblue31119180] (axis cs:-0.3,0) rectangle (axis cs:0.3,0);
\addlegendimage{ybar,ybar legend,draw=black,fill=steelblue31119180}
\addlegendentry{1}

\draw[draw=black,fill=steelblue31119180] (axis cs:0.7,0) rectangle (axis cs:1.3,0.1);
\draw[draw=black,fill=steelblue31119180] (axis cs:1.7,0) rectangle (axis cs:2.3,0);
\draw[draw=black,fill=steelblue31119180] (axis cs:2.7,0) rectangle (axis cs:3.3,0);
\draw[draw=black,fill=steelblue31119180] (axis cs:3.7,0) rectangle (axis cs:4.3,0);
\draw[draw=black,fill=steelblue31119180] (axis cs:4.7,0) rectangle (axis cs:5.3,0);
\draw[draw=black,fill=steelblue31119180] (axis cs:5.7,0) rectangle (axis cs:6.3,0.1);
\draw[draw=black,fill=steelblue31119180] (axis cs:6.7,0) rectangle (axis cs:7.3,0.1);
\draw[draw=black,fill=steelblue31119180] (axis cs:7.7,0) rectangle (axis cs:8.3,0);
\draw[draw=black,fill=steelblue31119180] (axis cs:8.7,0) rectangle (axis cs:9.3,0);
\draw[draw=black,fill=steelblue31119180] (axis cs:9.7,0) rectangle (axis cs:10.3,0);
\draw[draw=black,fill=darkorange25512714] (axis cs:-0.3,0) rectangle (axis cs:0.3,0);
\addlegendimage{ybar,ybar legend,draw=black,fill=darkorange25512714}
\addlegendentry{2}

\draw[draw=black,fill=darkorange25512714] (axis cs:0.7,0.1) rectangle (axis cs:1.3,1);
\draw[draw=black,fill=darkorange25512714] (axis cs:1.7,0) rectangle (axis cs:2.3,0);
\draw[draw=black,fill=darkorange25512714] (axis cs:2.7,0) rectangle (axis cs:3.3,0);
\draw[draw=black,fill=darkorange25512714] (axis cs:3.7,0) rectangle (axis cs:4.3,0.7);
\draw[draw=black,fill=darkorange25512714] (axis cs:4.7,0) rectangle (axis cs:5.3,0);
\draw[draw=black,fill=darkorange25512714] (axis cs:5.7,0.1) rectangle (axis cs:6.3,0.1);
\draw[draw=black,fill=darkorange25512714] (axis cs:6.7,0.1) rectangle (axis cs:7.3,0.1);
\draw[draw=black,fill=darkorange25512714] (axis cs:7.7,0) rectangle (axis cs:8.3,0);
\draw[draw=black,fill=darkorange25512714] (axis cs:8.7,0) rectangle (axis cs:9.3,0);
\draw[draw=black,fill=darkorange25512714] (axis cs:9.7,0) rectangle (axis cs:10.3,0);
\draw[draw=black,fill=forestgreen4416044] (axis cs:-0.3,0) rectangle (axis cs:0.3,0);
\addlegendimage{ybar,ybar legend,draw=black,fill=forestgreen4416044}
\addlegendentry{3}

\draw[draw=black,fill=forestgreen4416044] (axis cs:0.7,1) rectangle (axis cs:1.3,1);
\draw[draw=black,fill=forestgreen4416044] (axis cs:1.7,0) rectangle (axis cs:2.3,0);
\draw[draw=black,fill=forestgreen4416044] (axis cs:2.7,0) rectangle (axis cs:3.3,0);
\draw[draw=black,fill=forestgreen4416044] (axis cs:3.7,0.7) rectangle (axis cs:4.3,0.7);
\draw[draw=black,fill=forestgreen4416044] (axis cs:4.7,0) rectangle (axis cs:5.3,0);
\draw[draw=black,fill=forestgreen4416044] (axis cs:5.7,0.1) rectangle (axis cs:6.3,0.1);
\draw[draw=black,fill=forestgreen4416044] (axis cs:6.7,0.1) rectangle (axis cs:7.3,0.1);
\draw[draw=black,fill=forestgreen4416044] (axis cs:7.7,0) rectangle (axis cs:8.3,0);
\draw[draw=black,fill=forestgreen4416044] (axis cs:8.7,0) rectangle (axis cs:9.3,0);
\draw[draw=black,fill=forestgreen4416044] (axis cs:9.7,0) rectangle (axis cs:10.3,0);
\draw[draw=black,fill=crimson2143940] (axis cs:-0.3,0) rectangle (axis cs:0.3,0);
\addlegendimage{ybar,ybar legend,draw=black,fill=crimson2143940}
\addlegendentry{4}

\draw[draw=black,fill=crimson2143940] (axis cs:0.7,1) rectangle (axis cs:1.3,1);
\draw[draw=black,fill=crimson2143940] (axis cs:1.7,0) rectangle (axis cs:2.3,0);
\draw[draw=black,fill=crimson2143940] (axis cs:2.7,0) rectangle (axis cs:3.3,0);
\draw[draw=black,fill=crimson2143940] (axis cs:3.7,0.7) rectangle (axis cs:4.3,0.7);
\draw[draw=black,fill=crimson2143940] (axis cs:4.7,0) rectangle (axis cs:5.3,0);
\draw[draw=black,fill=crimson2143940] (axis cs:5.7,0.1) rectangle (axis cs:6.3,0.1);
\draw[draw=black,fill=crimson2143940] (axis cs:6.7,0.1) rectangle (axis cs:7.3,0.1);
\draw[draw=black,fill=crimson2143940] (axis cs:7.7,0) rectangle (axis cs:8.3,0);
\draw[draw=black,fill=crimson2143940] (axis cs:8.7,0) rectangle (axis cs:9.3,0);
\draw[draw=black,fill=crimson2143940] (axis cs:9.7,0) rectangle (axis cs:10.3,0);
\draw[draw=black,fill=mediumpurple148103189] (axis cs:-0.3,0) rectangle (axis cs:0.3,0);
\addlegendimage{ybar,ybar legend,draw=black,fill=mediumpurple148103189}
\addlegendentry{5}

\draw[draw=black,fill=mediumpurple148103189] (axis cs:0.7,1) rectangle (axis cs:1.3,1);
\draw[draw=black,fill=mediumpurple148103189] (axis cs:1.7,0) rectangle (axis cs:2.3,0);
\draw[draw=black,fill=mediumpurple148103189] (axis cs:2.7,0) rectangle (axis cs:3.3,0);
\draw[draw=black,fill=mediumpurple148103189] (axis cs:3.7,0.7) rectangle (axis cs:4.3,0.7);
\draw[draw=black,fill=mediumpurple148103189] (axis cs:4.7,0) rectangle (axis cs:5.3,0);
\draw[draw=black,fill=mediumpurple148103189] (axis cs:5.7,0.1) rectangle (axis cs:6.3,0.1);
\draw[draw=black,fill=mediumpurple148103189] (axis cs:6.7,0.1) rectangle (axis cs:7.3,0.1);
\draw[draw=black,fill=mediumpurple148103189] (axis cs:7.7,0) rectangle (axis cs:8.3,0);
\draw[draw=black,fill=mediumpurple148103189] (axis cs:8.7,0) rectangle (axis cs:9.3,0);
\draw[draw=black,fill=mediumpurple148103189] (axis cs:9.7,0) rectangle (axis cs:10.3,0);
\end{axis}

\end{tikzpicture}
                        \setlength{\abovecaptionskip}{-.45cm}
        \setlength{\belowcaptionskip}{0cm}
        \caption{Iperf}
        \label{fig:starving-rates-number}
    \end{subfigure}%
    \hfill%
    \begin{subfigure}[t]{0.49\linewidth}
        \centering
\begin{tikzpicture}

\definecolor{crimson2143940}{RGB}{214,39,40}
\definecolor{darkgrey176}{RGB}{176,176,176}
\definecolor{darkorange25512714}{RGB}{255,127,14}
\definecolor{forestgreen4416044}{RGB}{44,160,44}
\definecolor{lightgrey204}{RGB}{204,204,204}
\definecolor{mediumpurple148103189}{RGB}{148,103,189}
\definecolor{steelblue31119180}{RGB}{31,119,180}

\begin{axis}[
width=\linewidth,
height=0.8\linewidth,
legend cell align={left},
legend style={
  fill opacity=0.8,
  draw opacity=1,
  text opacity=1,
  at={(0.03,0.97)},
  anchor=north west,
  draw=lightgrey204
},
tick pos=left,
x grid style={darkgrey176},
xmin=-0.94, xmax=10.94,
xtick style={color=black},
xtick={0,1,2,3,4,5,6,7,8,9,10},
xticklabel style={rotate=45.0,anchor=east,font=\tiny},
xticklabels={
  RR,
  BCQI,
  FT,
  PF,
  Lean Sched,
  Log Rule,
  PFB,
  {GPF \(\displaystyle 0.6\), \(\displaystyle 0.7\)},
  {GPF \(\displaystyle 0.2\), \(\displaystyle 0.7\)},
  {GPF \(\displaystyle 0.6\), \(\displaystyle 0.2\)},
  EUFS
},
y grid style={darkgrey176},
ylabel={Starvation Rate},
ymajorgrids,
ymin=0, ymax=1,
ytick style={color=black},
ymajorgrids,
yminorgrids,
xlabel style={font=\scriptsize},
ylabel style={font=\scriptsize}
]
\draw[draw=black,fill=steelblue31119180,fill opacity=0.8] (axis cs:-0.4,0) rectangle (axis cs:0.4,0);
\addlegendimage{ybar,ybar legend,draw=black,fill=steelblue31119180,fill opacity=0.8}

\draw[draw=black,fill=steelblue31119180,fill opacity=0.8] (axis cs:0.6,0) rectangle (axis cs:1.4,0.111111111111111);
\draw[draw=black,fill=steelblue31119180,fill opacity=0.8] (axis cs:1.6,0) rectangle (axis cs:2.4,0);
\draw[draw=black,fill=steelblue31119180,fill opacity=0.8] (axis cs:2.6,0) rectangle (axis cs:3.4,0);
\draw[draw=black,fill=steelblue31119180,fill opacity=0.8] (axis cs:3.6,0) rectangle (axis cs:4.4,0.5);
\draw[draw=black,fill=steelblue31119180,fill opacity=0.8] (axis cs:4.6,0) rectangle (axis cs:5.4,0);
\draw[draw=black,fill=steelblue31119180,fill opacity=0.8] (axis cs:5.6,0) rectangle (axis cs:6.4,0);
\draw[draw=black,fill=steelblue31119180,fill opacity=0.8] (axis cs:6.6,0) rectangle (axis cs:7.4,0);
\draw[draw=black,fill=steelblue31119180,fill opacity=0.8] (axis cs:7.6,0) rectangle (axis cs:8.4,0);
\draw[draw=black,fill=steelblue31119180,fill opacity=0.8] (axis cs:8.6,0) rectangle (axis cs:9.4,0);
\draw[draw=black,fill=steelblue31119180,fill opacity=0.8] (axis cs:9.6,0) rectangle (axis cs:10.4,0.1);
\draw[draw=black,fill=darkorange25512714,fill opacity=0.8] (axis cs:-0.4,0) rectangle (axis cs:0.4,0);
\addlegendimage{ybar,ybar legend,draw=black,fill=darkorange25512714,fill opacity=0.8}

\draw[draw=black,fill=darkorange25512714,fill opacity=0.8] (axis cs:0.6,0.111111111111111) rectangle (axis cs:1.4,1);
\draw[draw=black,fill=darkorange25512714,fill opacity=0.8] (axis cs:1.6,0) rectangle (axis cs:2.4,0);
\draw[draw=black,fill=darkorange25512714,fill opacity=0.8] (axis cs:2.6,0) rectangle (axis cs:3.4,0);
\draw[draw=black,fill=darkorange25512714,fill opacity=0.8] (axis cs:3.6,0.5) rectangle (axis cs:4.4,0.9);
\draw[draw=black,fill=darkorange25512714,fill opacity=0.8] (axis cs:4.6,0) rectangle (axis cs:5.4,0);
\draw[draw=black,fill=darkorange25512714,fill opacity=0.8] (axis cs:5.6,0) rectangle (axis cs:6.4,0);
\draw[draw=black,fill=darkorange25512714,fill opacity=0.8] (axis cs:6.6,0) rectangle (axis cs:7.4,0);
\draw[draw=black,fill=darkorange25512714,fill opacity=0.8] (axis cs:7.6,0) rectangle (axis cs:8.4,0);
\draw[draw=black,fill=darkorange25512714,fill opacity=0.8] (axis cs:8.6,0) rectangle (axis cs:9.4,0);
\draw[draw=black,fill=darkorange25512714,fill opacity=0.8] (axis cs:9.6,0.1) rectangle (axis cs:10.4,0.1);
\draw[draw=black,fill=forestgreen4416044,fill opacity=0.8] (axis cs:-0.4,0) rectangle (axis cs:0.4,0);
\addlegendimage{ybar,ybar legend,draw=black,fill=forestgreen4416044,fill opacity=0.8}

\draw[draw=black,fill=forestgreen4416044,fill opacity=0.8] (axis cs:0.6,1) rectangle (axis cs:1.4,1);
\draw[draw=black,fill=forestgreen4416044,fill opacity=0.8] (axis cs:1.6,0) rectangle (axis cs:2.4,0);
\draw[draw=black,fill=forestgreen4416044,fill opacity=0.8] (axis cs:2.6,0) rectangle (axis cs:3.4,0);
\draw[draw=black,fill=forestgreen4416044,fill opacity=0.8] (axis cs:3.6,0.9) rectangle (axis cs:4.4,0.9);
\draw[draw=black,fill=forestgreen4416044,fill opacity=0.8] (axis cs:4.6,0) rectangle (axis cs:5.4,0);
\draw[draw=black,fill=forestgreen4416044,fill opacity=0.8] (axis cs:5.6,0) rectangle (axis cs:6.4,0);
\draw[draw=black,fill=forestgreen4416044,fill opacity=0.8] (axis cs:6.6,0) rectangle (axis cs:7.4,0);
\draw[draw=black,fill=forestgreen4416044,fill opacity=0.8] (axis cs:7.6,0) rectangle (axis cs:8.4,0);
\draw[draw=black,fill=forestgreen4416044,fill opacity=0.8] (axis cs:8.6,0) rectangle (axis cs:9.4,0);
\draw[draw=black,fill=forestgreen4416044,fill opacity=0.8] (axis cs:9.6,0.1) rectangle (axis cs:10.4,0.1);
\draw[draw=black,fill=crimson2143940,fill opacity=0.8] (axis cs:-0.4,0) rectangle (axis cs:0.4,0);
\addlegendimage{ybar,ybar legend,draw=black,fill=crimson2143940,fill opacity=0.8}

\draw[draw=black,fill=crimson2143940,fill opacity=0.8] (axis cs:0.6,1) rectangle (axis cs:1.4,1);
\draw[draw=black,fill=crimson2143940,fill opacity=0.8] (axis cs:1.6,0) rectangle (axis cs:2.4,0);
\draw[draw=black,fill=crimson2143940,fill opacity=0.8] (axis cs:2.6,0) rectangle (axis cs:3.4,0);
\draw[draw=black,fill=crimson2143940,fill opacity=0.8] (axis cs:3.6,0.9) rectangle (axis cs:4.4,0.9);
\draw[draw=black,fill=crimson2143940,fill opacity=0.8] (axis cs:4.6,0) rectangle (axis cs:5.4,0);
\draw[draw=black,fill=crimson2143940,fill opacity=0.8] (axis cs:5.6,0) rectangle (axis cs:6.4,0);
\draw[draw=black,fill=crimson2143940,fill opacity=0.8] (axis cs:6.6,0) rectangle (axis cs:7.4,0);
\draw[draw=black,fill=crimson2143940,fill opacity=0.8] (axis cs:7.6,0) rectangle (axis cs:8.4,0);
\draw[draw=black,fill=crimson2143940,fill opacity=0.8] (axis cs:8.6,0) rectangle (axis cs:9.4,0);
\draw[draw=black,fill=crimson2143940,fill opacity=0.8] (axis cs:9.6,0.1) rectangle (axis cs:10.4,0.1);
\draw[draw=black,fill=mediumpurple148103189,fill opacity=0.8] (axis cs:-0.4,0) rectangle (axis cs:0.4,0.1);
\addlegendimage{ybar,ybar legend,draw=black,fill=mediumpurple148103189,fill opacity=0.8}

\draw[draw=black,fill=mediumpurple148103189,fill opacity=0.8] (axis cs:0.6,1) rectangle (axis cs:1.4,1);
\draw[draw=black,fill=mediumpurple148103189,fill opacity=0.8] (axis cs:1.6,0) rectangle (axis cs:2.4,0);
\draw[draw=black,fill=mediumpurple148103189,fill opacity=0.8] (axis cs:2.6,0) rectangle (axis cs:3.4,0);
\draw[draw=black,fill=mediumpurple148103189,fill opacity=0.8] (axis cs:3.6,0.9) rectangle (axis cs:4.4,1);
\draw[draw=black,fill=mediumpurple148103189,fill opacity=0.8] (axis cs:4.6,0) rectangle (axis cs:5.4,0);
\draw[draw=black,fill=mediumpurple148103189,fill opacity=0.8] (axis cs:5.6,0) rectangle (axis cs:6.4,0);
\draw[draw=black,fill=mediumpurple148103189,fill opacity=0.8] (axis cs:6.6,0) rectangle (axis cs:7.4,0.2);
\draw[draw=black,fill=mediumpurple148103189,fill opacity=0.8] (axis cs:7.6,0) rectangle (axis cs:8.4,0.125);
\draw[draw=black,fill=mediumpurple148103189,fill opacity=0.8] (axis cs:8.6,0) rectangle (axis cs:9.4,0);
\draw[draw=black,fill=mediumpurple148103189,fill opacity=0.8] (axis cs:9.6,0.1) rectangle (axis cs:10.4,0.1);
\end{axis}

\end{tikzpicture}
                        \setlength{\abovecaptionskip}{0cm}
        \setlength{\belowcaptionskip}{0cm}
        \caption{Bursty Traffic mgen}
        \label{fig:starvation-rates-ftp}
    \end{subfigure}
    \caption{Starving rate of QoS-unaware schedulers.}
    \label{fig:disconnection_qos_unaware}
\end{figure}
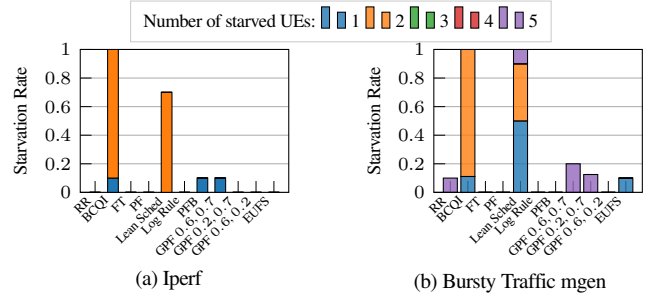

The first metric we measure is the starving rate in both scenarios, depicted in Fig.~\ref{fig:disconnection_qos_unaware}. We observe that for this set of schedulers, the only one which always starves at least one \gls{ue} is BCQI, which is unsurprising since \glspl{ue} $21$ and $24$ never have a better \gls{cqi} than \glspl{ue} $22$, $23$ and $25$. Lean Sched also significantly struggles with respectively 70\% and 100\% of starved runs for the iPerf and Bursty scenarios. This first result suggests that those two algorithms should be avoided in most practical scenarios. Furthermore, since \gls{bcqi} only has starved runs, we exclude it from our further fairness analysis, which requires schedulers to handle the five \glspl{ue} properly.   

\begin{figure}[t]
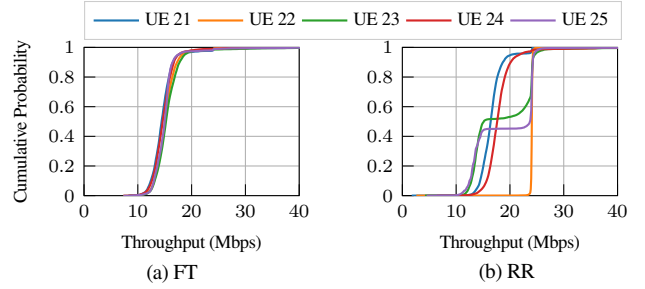

    \centering
    \begin{subfigure}[t]{0.5\linewidth}
        \centering
            \input{figures/FT_smooth}
                \setlength{\abovecaptionskip}{-.4cm}
        \setlength{\belowcaptionskip}{0cm}
        \caption{\gls{ft}}
    \end{subfigure}%
    \hfill%
    \begin{subfigure}[t]{0.5\linewidth}
        \centering
            \input{figures/RR_smooth}
                \setlength{\abovecaptionskip}{0cm}
        \setlength{\belowcaptionskip}{0cm}
        \caption{\gls{rr}}
    \end{subfigure}
    \caption{Throughput CDF for FT and RR baselines.}
    \label{fig:baselines}
    \vspace{-10pt}
\end{figure}

We then analyze the \glspl{cdf} of the throughput of the baselines, which are in Fig. \ref{fig:baselines} for \gls{rr} and \gls{ft} and in Fig. \ref{fig:oai_comparison_lua} for \gls{pf}. We observe that as expected, \gls{ft} achieves near-perfect fairness, \gls{pf} is skewed towards favoring better channels but allocates significant portions of the resources to \glspl{ue} with lower channel quality, and \gls{rr} is slightly more unfair than \gls{pf}, as commonly observed in the literature: in our case, the median throughput of the three \glspl{ue} with good channel reach at least $22$ Mbps for \gls{rr} while in \gls{pf} they do not go above $20$ Mbps, since more resources are used to give a higher throughput to \glspl{ue} with the poorer channel. 

%
\begin{figure}[htbp]
    \vspace*{-10pt}
    \centering
    \scriptsize
    \resizebox{\linewidth}{!}{%
\begin{tikzpicture}

\definecolor{darkgrey176}{RGB}{176,176,176}
\definecolor{firebrick}{RGB}{178,34,34}
\definecolor{steelblue}{RGB}{70,130,180}

\begin{axis}[
width=\linewidth,
height=0.4\linewidth,
tick pos=left,
x grid style={darkgrey176},
xmin=-0.835, xmax=9.835,
xtick style={color=black},
xtick={0,1,2,3,4,5,6,7,8,9},
xticklabel style={rotate=45.0,anchor=east},
xticklabels={
  RR,
  FT,
  PF,
  Lean Sched,
  Log Rule,
  PFB,
  {GPF \(\displaystyle 0.6\), \(\displaystyle 0.7\)},
  {GPF \(\displaystyle 0.2\), \(\displaystyle 0.7\)},
  {GPF \(\displaystyle 0.6\), \(\displaystyle 0.2\)},
  EUFS
},
y grid style={darkgrey176},
ylabel=\textcolor{steelblue}{Jain Index, $\uparrow$ = fairer},
ymin=0, ymax=1.0,
ytick style={color=black},
xlabel style={font=\scriptsize},
ylabel style={font=\scriptsize}
]
\draw[draw=black,fill=steelblue,fill opacity=0.8] (axis cs:-0.35,0) rectangle (axis cs:0,0.95198693187796);
\addlegendimage{ybar,ybar legend,draw=black,fill=steelblue,fill opacity=0.8}

\draw[draw=black,fill=steelblue,fill opacity=0.8] (axis cs:0.65,0) rectangle (axis cs:1,0.99901619028576);
\draw[draw=black,fill=steelblue,fill opacity=0.8] (axis cs:1.65,0) rectangle (axis cs:2,0.946880560074304);
\draw[draw=black,fill=steelblue,fill opacity=0.8] (axis cs:2.65,0) rectangle (axis cs:3,0.955567437781418);
\draw[draw=black,fill=steelblue,fill opacity=0.8] (axis cs:3.65,0) rectangle (axis cs:4,0.953491434748863);
\draw[draw=black,fill=steelblue,fill opacity=0.8] (axis cs:4.65,0) rectangle (axis cs:5,0.955074773264407);
\draw[draw=black,fill=steelblue,fill opacity=0.8] (axis cs:5.65,0) rectangle (axis cs:6,0.959026511178614);
\draw[draw=black,fill=steelblue,fill opacity=0.8] (axis cs:6.65,0) rectangle (axis cs:7,0.994851857546326);
\draw[draw=black,fill=steelblue,fill opacity=0.8] (axis cs:7.65,0) rectangle (axis cs:8,0.732680792780498);
\draw[draw=black,fill=steelblue,fill opacity=0.8] (axis cs:8.65,0) rectangle (axis cs:9,0.934156458558888);
\end{axis}

\begin{axis}[
width=\linewidth,
height=0.4\linewidth,
axis y line*=right,
x grid style={darkgrey176},
xmin=-0.835, xmax=9.835,
xtick style={color=black},
xtick=\empty,
y grid style={darkgrey176},
ylabel=\textcolor{firebrick}{Gini Index, $\downarrow$ = fairer},
ymin=0, ymax=0.397233173215185,
ytick pos=right,
ytick style={color=black},
yticklabel style={anchor=west},
xlabel style={font=\scriptsize},
ylabel style={font=\scriptsize}
]
\draw[draw=black,fill=firebrick,fill opacity=0.8] (axis cs:2.77555756156289e-17,0) rectangle (axis cs:0.35,0.121253958687122);
\addlegendimage{ybar,ybar legend,draw=black,fill=firebrick,fill opacity=0.8}

\draw[draw=black,fill=firebrick,fill opacity=0.8] (axis cs:1,0) rectangle (axis cs:1.35,0.0165639450521487);
\draw[draw=black,fill=firebrick,fill opacity=0.8] (axis cs:2,0) rectangle (axis cs:2.35,0.126970815156285);
\draw[draw=black,fill=firebrick,fill opacity=0.8] (axis cs:3,0) rectangle (axis cs:3.35,0.117698034341185);
\draw[draw=black,fill=firebrick,fill opacity=0.8] (axis cs:4,0) rectangle (axis cs:4.35,0.120221474551158);
\draw[draw=black,fill=firebrick,fill opacity=0.8] (axis cs:5,0) rectangle (axis cs:5.35,0.118480540728007);
\draw[draw=black,fill=firebrick,fill opacity=0.8] (axis cs:6,0) rectangle (axis cs:6.35,0.11191661192655);
\draw[draw=black,fill=firebrick,fill opacity=0.8] (axis cs:7,0) rectangle (axis cs:7.35,0.0372322334320655);
\draw[draw=black,fill=firebrick,fill opacity=0.8] (axis cs:8,0) rectangle (axis cs:8.35,0.331027644345988);
\draw[draw=black,fill=firebrick,fill opacity=0.8] (axis cs:9,0) rectangle (axis cs:9.35,0.139831254301263);
\end{axis}

\end{tikzpicture}
    }
    \vspace{-15pt}
    \caption{Fairness indices for \gls{qos}-unaware.}
    \label{fig:gini_unaware}
\end{figure}
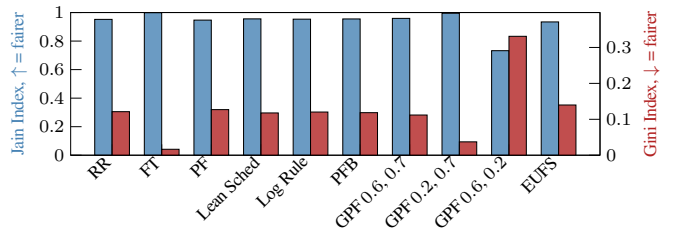

\begin{figure}[htbp]
    \vspace*{-10pt}
    \centering
    \scriptsize
    \resizebox{\linewidth}{!}{%
\begin{tikzpicture}

\definecolor{darkgrey176}{RGB}{176,176,176}
\definecolor{darkslategrey38}{RGB}{38,38,38}

\begin{axis}[
colorbar,
colorbar style={ylabel={RMSE}},
colormap={mymap}{[1pt]
  rgb(0pt)=(0.0313725490196078,0.113725490196078,0.345098039215686);
  rgb(1pt)=(0.145098039215686,0.203921568627451,0.580392156862745);
  rgb(2pt)=(0.133333333333333,0.368627450980392,0.658823529411765);
  rgb(3pt)=(0.113725490196078,0.568627450980392,0.752941176470588);
  rgb(4pt)=(0.254901960784314,0.713725490196078,0.768627450980392);
  rgb(5pt)=(0.498039215686275,0.803921568627451,0.733333333333333);
  rgb(6pt)=(0.780392156862745,0.913725490196078,0.705882352941177);
  rgb(7pt)=(0.929411764705882,0.972549019607843,0.694117647058824);
  rgb(8pt)=(1,1,0.850980392156863)
},
point meta max=5.68652342745627,
point meta min=0.247169190533907,
tick align=outside,
tick pos=left,
x grid style={darkgrey176},
xmin=0, xmax=10,
xtick style={color=black},
xtick={0.5,1.5,2.5,3.5,4.5,5.5,6.5,7.5,8.5,9.5},
xticklabel style={rotate=45.0,anchor=east},
xticklabels={
  RR,
  FT,
  PF,
  Lean Sched,
  Log Rule,
  PFB,
  {GPF \(\displaystyle 0.6\), \(\displaystyle 0.7\)},
  {GPF \(\displaystyle 0.2\), \(\displaystyle 0.7\)},
  {GPF \(\displaystyle 0.6\), \(\displaystyle 0.2\)},
  EUFS
},
y dir=reverse,
y grid style={darkgrey176},
ymin=0, ymax=10,
ytick style={color=black},
ytick={0.5,1.5,2.5,3.5,4.5,5.5,6.5,7.5,8.5,9.5},
yticklabels={
  RR,
  FT,
  PF,
  Lean Sched,
  Log Rule,
  PFB,
  {GPF \(\displaystyle 0.6\), \(\displaystyle 0.7\)},
  {GPF \(\displaystyle 0.2\), \(\displaystyle 0.7\)},
  {GPF \(\displaystyle 0.6\), \(\displaystyle 0.2\)},
  EUFS
}
]
\addplot graphics [includegraphics cmd=\pgfimage,xmin=0, xmax=10, ymin=10, ymax=0] {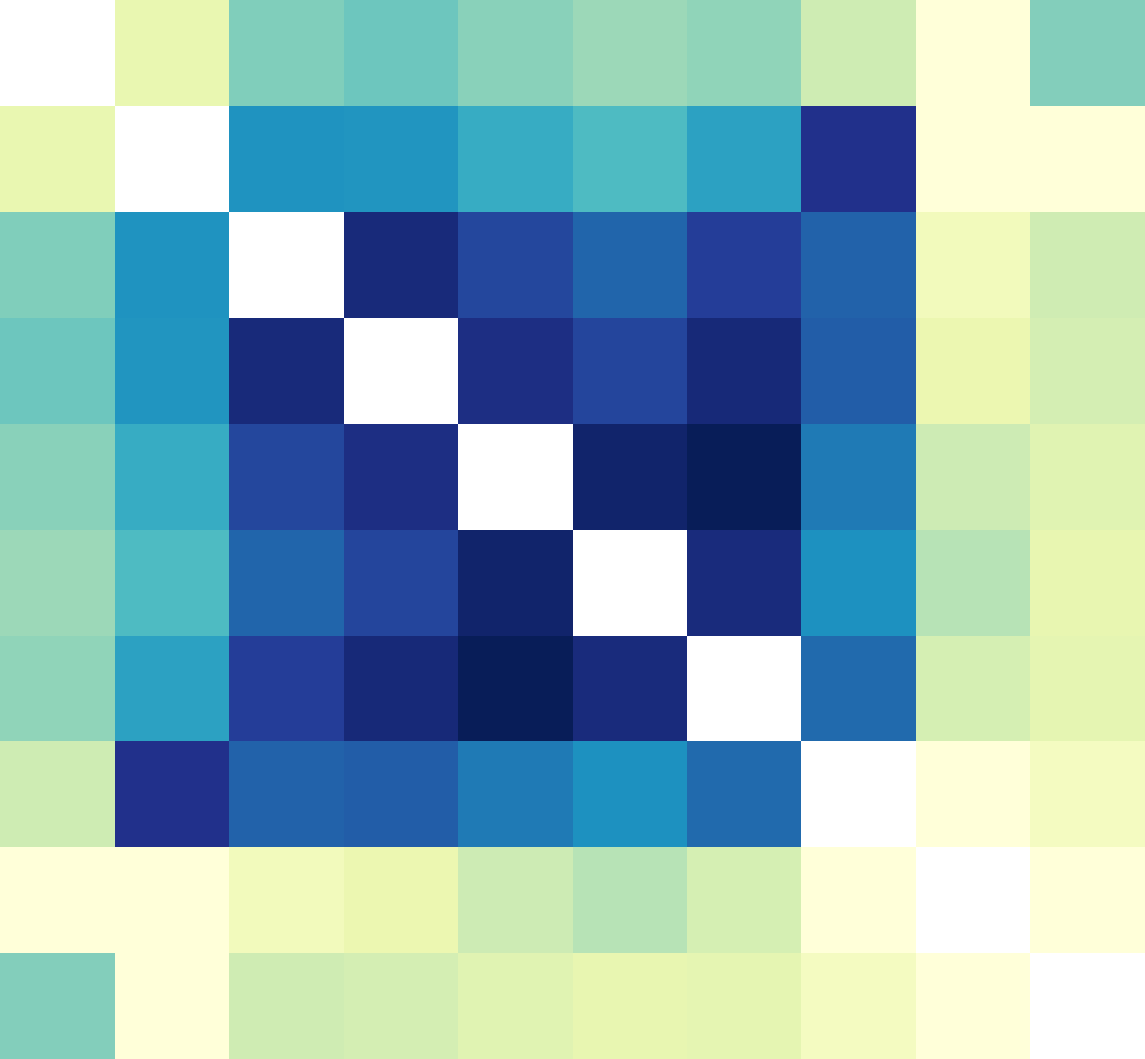};
\draw (axis cs:1.5,0.5) node[
  scale=0.8,
  text=darkslategrey38,
  rotate=0.0
]{4.94};
\draw (axis cs:2.5,0.5) node[
  scale=0.8,
  text=darkslategrey38,
  rotate=0.0
]{3.66};
\draw (axis cs:3.5,0.5) node[
  scale=0.8,
  text=darkslategrey38,
  rotate=0.0
]{3.44};
\draw (axis cs:4.5,0.5) node[
  scale=0.8,
  text=darkslategrey38,
  rotate=0.0
]{3.74};
\draw (axis cs:5.5,0.5) node[
  scale=0.8,
  text=darkslategrey38,
  rotate=0.0
]{3.90};
\draw (axis cs:6.5,0.5) node[
  scale=0.8,
  text=darkslategrey38,
  rotate=0.0
]{3.81};
\draw (axis cs:7.5,0.5) node[
  scale=0.8,
  text=darkslategrey38,
  rotate=0.0
]{4.45};
\draw (axis cs:8.5,0.5) node[
  scale=0.8,
  text=darkslategrey38,
  rotate=0.0
]{6.78};
\draw (axis cs:9.5,0.5) node[
  scale=0.8,
  text=darkslategrey38,
  rotate=0.0
]{3.68};
\draw (axis cs:0.5,1.5) node[
  scale=0.8,
  text=darkslategrey38,
  rotate=0.0
]{4.94};
\draw (axis cs:2.5,1.5) node[
  scale=0.8,
  text=white,
  rotate=0.0
]{2.32};
\draw (axis cs:3.5,1.5) node[
  scale=0.8,
  text=white,
  rotate=0.0
]{2.36};
\draw (axis cs:4.5,1.5) node[
  scale=0.8,
  text=white,
  rotate=0.0
]{2.79};
\draw (axis cs:5.5,1.5) node[
  scale=0.8,
  text=darkslategrey38,
  rotate=0.0
]{3.11};
\draw (axis cs:6.5,1.5) node[
  scale=0.8,
  text=white,
  rotate=0.0
]{2.57};
\draw (axis cs:7.5,1.5) node[
  scale=0.8,
  text=white,
  rotate=0.0
]{0.84};
\draw (axis cs:8.5,1.5) node[
  scale=0.8,
  text=darkslategrey38,
  rotate=0.0
]{6.76};
\draw (axis cs:9.5,1.5) node[
  scale=0.8,
  text=darkslategrey38,
  rotate=0.0
]{5.69};
\draw (axis cs:0.5,2.5) node[
  scale=0.8,
  text=darkslategrey38,
  rotate=0.0
]{3.66};
\draw (axis cs:1.5,2.5) node[
  scale=0.8,
  text=white,
  rotate=0.0
]{2.32};
\draw (axis cs:3.5,2.5) node[
  scale=0.8,
  text=white,
  rotate=0.0
]{0.65};
\draw (axis cs:4.5,2.5) node[
  scale=0.8,
  text=white,
  rotate=0.0
]{1.23};
\draw (axis cs:5.5,2.5) node[
  scale=0.8,
  text=white,
  rotate=0.0
]{1.70};
\draw (axis cs:6.5,2.5) node[
  scale=0.8,
  text=white,
  rotate=0.0
]{1.09};
\draw (axis cs:7.5,2.5) node[
  scale=0.8,
  text=white,
  rotate=0.0
]{1.66};
\draw (axis cs:8.5,2.5) node[
  scale=0.8,
  text=darkslategrey38,
  rotate=0.0
]{5.18};
\draw (axis cs:9.5,2.5) node[
  scale=0.8,
  text=darkslategrey38,
  rotate=0.0
]{4.47};
\draw (axis cs:0.5,3.5) node[
  scale=0.8,
  text=darkslategrey38,
  rotate=0.0
]{3.44};
\draw (axis cs:1.5,3.5) node[
  scale=0.8,
  text=white,
  rotate=0.0
]{2.36};
\draw (axis cs:2.5,3.5) node[
  scale=0.8,
  text=white,
  rotate=0.0
]{0.65};
\draw (axis cs:4.5,3.5) node[
  scale=0.8,
  text=white,
  rotate=0.0
]{0.75};
\draw (axis cs:5.5,3.5) node[
  scale=0.8,
  text=white,
  rotate=0.0
]{1.21};
\draw (axis cs:6.5,3.5) node[
  scale=0.8,
  text=white,
  rotate=0.0
]{0.62};
\draw (axis cs:7.5,3.5) node[
  scale=0.8,
  text=white,
  rotate=0.0
]{1.60};
\draw (axis cs:8.5,3.5) node[
  scale=0.8,
  text=darkslategrey38,
  rotate=0.0
]{4.97};
\draw (axis cs:9.5,3.5) node[
  scale=0.8,
  text=darkslategrey38,
  rotate=0.0
]{4.54};
\draw (axis cs:0.5,4.5) node[
  scale=0.8,
  text=darkslategrey38,
  rotate=0.0
]{3.74};
\draw (axis cs:1.5,4.5) node[
  scale=0.8,
  text=white,
  rotate=0.0
]{2.79};
\draw (axis cs:2.5,4.5) node[
  scale=0.8,
  text=white,
  rotate=0.0
]{1.23};
\draw (axis cs:3.5,4.5) node[
  scale=0.8,
  text=white,
  rotate=0.0
]{0.75};
\draw (axis cs:5.5,4.5) node[
  scale=0.8,
  text=white,
  rotate=0.0
]{0.48};
\draw (axis cs:6.5,4.5) node[
  scale=0.8,
  text=white,
  rotate=0.0
]{0.25};
\draw (axis cs:7.5,4.5) node[
  scale=0.8,
  text=white,
  rotate=0.0
]{1.98};
\draw (axis cs:8.5,4.5) node[
  scale=0.8,
  text=darkslategrey38,
  rotate=0.0
]{4.42};
\draw (axis cs:9.5,4.5) node[
  scale=0.8,
  text=darkslategrey38,
  rotate=0.0
]{4.76};
\draw (axis cs:0.5,5.5) node[
  scale=0.8,
  text=darkslategrey38,
  rotate=0.0
]{3.90};
\draw (axis cs:1.5,5.5) node[
  scale=0.8,
  text=darkslategrey38,
  rotate=0.0
]{3.11};
\draw (axis cs:2.5,5.5) node[
  scale=0.8,
  text=white,
  rotate=0.0
]{1.70};
\draw (axis cs:3.5,5.5) node[
  scale=0.8,
  text=white,
  rotate=0.0
]{1.21};
\draw (axis cs:4.5,5.5) node[
  scale=0.8,
  text=white,
  rotate=0.0
]{0.48};
\draw (axis cs:6.5,5.5) node[
  scale=0.8,
  text=white,
  rotate=0.0
]{0.67};
\draw (axis cs:7.5,5.5) node[
  scale=0.8,
  text=white,
  rotate=0.0
]{2.30};
\draw (axis cs:8.5,5.5) node[
  scale=0.8,
  text=darkslategrey38,
  rotate=0.0
]{4.17};
\draw (axis cs:9.5,5.5) node[
  scale=0.8,
  text=darkslategrey38,
  rotate=0.0
]{4.91};
\draw (axis cs:0.5,6.5) node[
  scale=0.8,
  text=darkslategrey38,
  rotate=0.0
]{3.81};
\draw (axis cs:1.5,6.5) node[
  scale=0.8,
  text=white,
  rotate=0.0
]{2.57};
\draw (axis cs:2.5,6.5) node[
  scale=0.8,
  text=white,
  rotate=0.0
]{1.09};
\draw (axis cs:3.5,6.5) node[
  scale=0.8,
  text=white,
  rotate=0.0
]{0.62};
\draw (axis cs:4.5,6.5) node[
  scale=0.8,
  text=white,
  rotate=0.0
]{0.25};
\draw (axis cs:5.5,6.5) node[
  scale=0.8,
  text=white,
  rotate=0.0
]{0.67};
\draw (axis cs:7.5,6.5) node[
  scale=0.8,
  text=white,
  rotate=0.0
]{1.76};
\draw (axis cs:8.5,6.5) node[
  scale=0.8,
  text=darkslategrey38,
  rotate=0.0
]{4.58};
\draw (axis cs:9.5,6.5) node[
  scale=0.8,
  text=darkslategrey38,
  rotate=0.0
]{4.84};
\draw (axis cs:0.5,7.5) node[
  scale=0.8,
  text=darkslategrey38,
  rotate=0.0
]{4.45};
\draw (axis cs:1.5,7.5) node[
  scale=0.8,
  text=white,
  rotate=0.0
]{0.84};
\draw (axis cs:2.5,7.5) node[
  scale=0.8,
  text=white,
  rotate=0.0
]{1.66};
\draw (axis cs:3.5,7.5) node[
  scale=0.8,
  text=white,
  rotate=0.0
]{1.60};
\draw (axis cs:4.5,7.5) node[
  scale=0.8,
  text=white,
  rotate=0.0
]{1.98};
\draw (axis cs:5.5,7.5) node[
  scale=0.8,
  text=white,
  rotate=0.0
]{2.30};
\draw (axis cs:6.5,7.5) node[
  scale=0.8,
  text=white,
  rotate=0.0
]{1.76};
\draw (axis cs:8.5,7.5) node[
  scale=0.8,
  text=darkslategrey38,
  rotate=0.0
]{6.09};
\draw (axis cs:9.5,7.5) node[
  scale=0.8,
  text=darkslategrey38,
  rotate=0.0
]{5.27};
\draw (axis cs:0.5,8.5) node[
  scale=0.8,
  text=darkslategrey38,
  rotate=0.0
]{6.78};
\draw (axis cs:1.5,8.5) node[
  scale=0.8,
  text=darkslategrey38,
  rotate=0.0
]{6.76};
\draw (axis cs:2.5,8.5) node[
  scale=0.8,
  text=darkslategrey38,
  rotate=0.0
]{5.18};
\draw (axis cs:3.5,8.5) node[
  scale=0.8,
  text=darkslategrey38,
  rotate=0.0
]{4.97};
\draw (axis cs:4.5,8.5) node[
  scale=0.8,
  text=darkslategrey38,
  rotate=0.0
]{4.42};
\draw (axis cs:5.5,8.5) node[
  scale=0.8,
  text=darkslategrey38,
  rotate=0.0
]{4.17};
\draw (axis cs:6.5,8.5) node[
  scale=0.8,
  text=darkslategrey38,
  rotate=0.0
]{4.58};
\draw (axis cs:7.5,8.5) node[
  scale=0.8,
  text=darkslategrey38,
  rotate=0.0
]{6.09};
\draw (axis cs:9.5,8.5) node[
  scale=0.8,
  text=darkslategrey38,
  rotate=0.0
]{7.67};
\draw (axis cs:0.5,9.5) node[
  scale=0.8,
  text=darkslategrey38,
  rotate=0.0
]{3.68};
\draw (axis cs:1.5,9.5) node[
  scale=0.8,
  text=darkslategrey38,
  rotate=0.0
]{5.69};
\draw (axis cs:2.5,9.5) node[
  scale=0.8,
  text=darkslategrey38,
  rotate=0.0
]{4.47};
\draw (axis cs:3.5,9.5) node[
  scale=0.8,
  text=darkslategrey38,
  rotate=0.0
]{4.54};
\draw (axis cs:4.5,9.5) node[
  scale=0.8,
  text=darkslategrey38,
  rotate=0.0
]{4.76};
\draw (axis cs:5.5,9.5) node[
  scale=0.8,
  text=darkslategrey38,
  rotate=0.0
]{4.91};
\draw (axis cs:6.5,9.5) node[
  scale=0.8,
  text=darkslategrey38,
  rotate=0.0
]{4.84};
\draw (axis cs:7.5,9.5) node[
  scale=0.8,
  text=darkslategrey38,
  rotate=0.0
]{5.27};
\draw (axis cs:8.5,9.5) node[
  scale=0.8,
  text=darkslategrey38,
  rotate=0.0
]{7.67};
\end{axis}

\end{tikzpicture}
    }
    \vspace{-15pt}
    \caption{RMSE of average throughput per UE between QoS-unaware schedulers. Tested with iPerf at $24$\:Mbps.}
    \label{fig:similarity}
\end{figure}
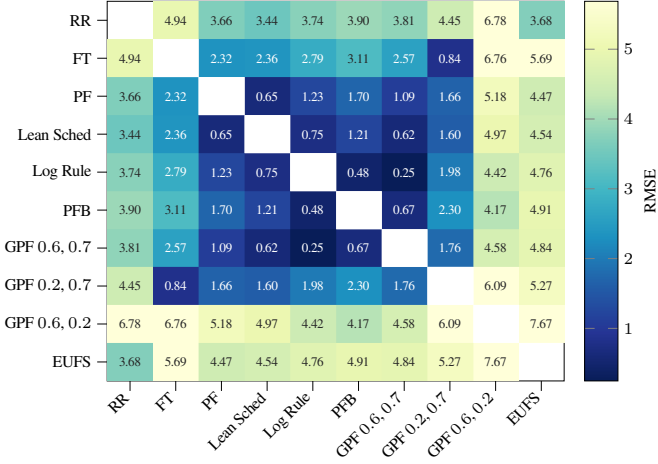

Next, we analyze the similarity of algorithms (both baselines and from the literature) between one another, focusing only on runs without starvation in the iPerf scenario.
%
To do this, we stack the average throughput of each \gls{ue} in vectors, and then compute the \gls{rmse} between each average throughput vector. 
The results, depicted in Fig. \ref{fig:similarity}, reveal that most schedulers from the literature have a similar behavior to \gls{pf}, which also translates in similar fairness indices in Fig. \ref{fig:gini_unaware}. We also see that GPF's fairness behaves as expected with respect to its parameters, making it a good algorithm for tuning the behavior of the scheduler between the extreme fairness of FT (which is approached by using $\beta=0.6$ and $\gamma=0.2$), and the higher resource utilization that can be achieved with $\beta = 0.6$ and $\gamma=0.2$, at the cost of fairness.

Finally we observe that EUFS is the only algorithm that significantly differ from the rest of the tested algorithms, despite being also based on a modification of \gls{pf}. Its Gini and Jain indices (Fig. \ref{fig:gini_unaware}) indicate that it is slightly less fair than \gls{pf}. As reported in Table \ref{tab:eufs}, this is due to \gls{ue} 21 receiving less resources despite EUFS supposedly being designed to favor cell-edge \glspl{ue} such as \gls{ue} 21. We argue this is due to using $\Raveuser{i}$ to identify cell-edge users, which may create a feedback loop, favoring some \glspl{ue} momentarily. This then lowers the $\Raveuser{i}$ of the other \glspl{ue}, which makes them favored by the algorithm at later rounds, despite their good channel conditions. We argue a more stable and suitable way to identify cell-edge users would be to use directly channel \glspl{kpi} such as the \gls{cqi}. This insight shows the importance of such \gls{ota} analysis, which enables us to uncover non-trivial feedback loops in the algorithm's behavior.

\begin{table}[t]
    \centering
        \caption{Average throughput (Mbps) of each \gls{ue} for \gls{pf} and EUFS.} 
    \label{tab:eufs}
    \begin{tabular}{llllll}
    \toprule
    Algorithm & UE 21 & UE 22 & UE 23 & UE 24 & UE 25\\
    \midrule
    EUFS     &  11.07      & 21.85      & 22.45      & 15.67      & 22.27\\
    \gls{pf}     & 13.25      & 19.84      & 15.27      & 14.00      & 16.68 \\
    \bottomrule
    \end{tabular}
    \vspace{-10pt}
\end{table}

We now study the packet latency in the bursty scenario, which, contrary to the iPerf throughput, varies significantly between schedulers.
Notably, \gls{pf}'s latency profile is one of the worst, with significant portions of packets delivered in up to a second. Indeed, in \gls{pf}, the \glspl{ue} with the worst channel have a lower $\Rinst$, and hence only get served when their $\Rave$ is low enough, which translates into buffering times.
Other \gls{pf}-based algorithms have a similarly poor delay profile, such as EUFS and GPF.
On the other hand, while it did not have a negative impact on throughput in the first experiment, the buffer awareness of PFB and Log Rule successfully balances latency, making it a tradeoff-free improvement over \gls{pf}.
\gls{rr} also has a balanced latency profile, due to always regularly giving resources to all \glspl{ue}, which avoids excessive buffering.

\begin{figure}[htbp]
    \centering
    \vspace{-7pt}
    \includegraphics[width=\linewidth]{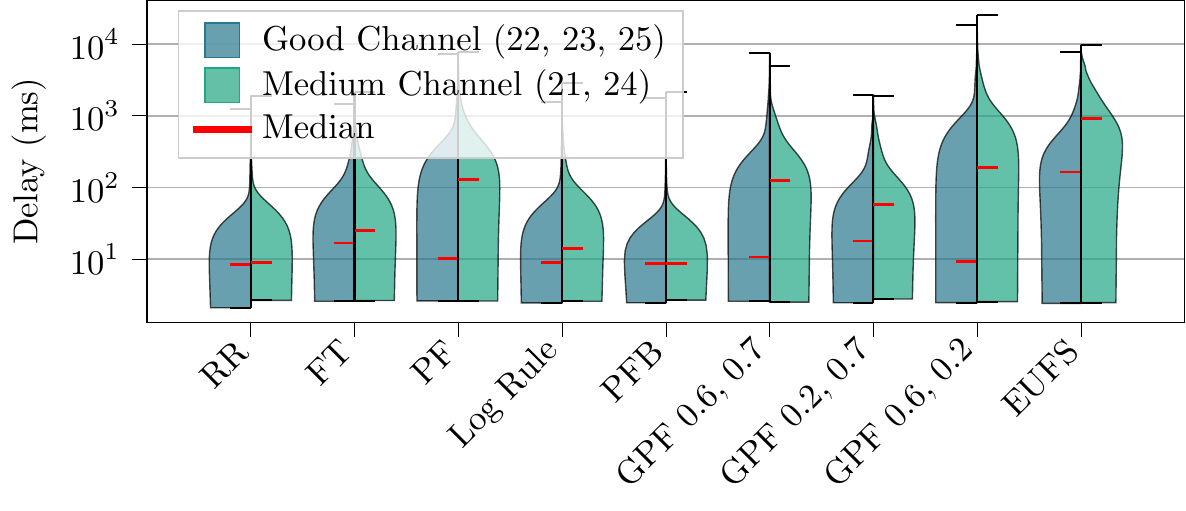}
    \vspace*{-19pt}
    \caption{Delay for of QoS-Unaware schedulers (bursty traffic).}
    \label{fig:latency_unaware}
\end{figure}

\vspace{-15pt} 
\subsection{\gls{qos}-Aware Schedulers}
\label{sec:qos_aware}
For this class of schedulers, we define two \gls{qos} classes, which the scheduler is aware of via the \FQI~indicator. The privileged class is comprised of \glspl{ue} 21 and 24. Their requirements is set at $50$ ms of target delay for $99\%$ of packets. The rest of \glspl{ue} are non-priviledged. For schedulers that take a delay target for each \gls{ue} as input (\textit{i.e.}, WD-PS, LTTI and QoS-FK), we set the non-priviledged target to $500$ ms for $97\%$ of packets. For the other schedulers, these \glspl{ue} are just part of the best-effort class.
\begin{figure}[htbp]
    \centering
    \vspace{-8pt}
    \begin{subfigure}[t]{0.49\linewidth}
        \centering
\begin{tikzpicture}

\definecolor{crimson2143940}{RGB}{214,39,40}
\definecolor{darkgrey176}{RGB}{176,176,176}
\definecolor{darkorange25512714}{RGB}{255,127,14}
\definecolor{forestgreen4416044}{RGB}{44,160,44}
\definecolor{lightgrey204}{RGB}{204,204,204}
\definecolor{mediumpurple148103189}{RGB}{148,103,189}
\definecolor{steelblue31119180}{RGB}{31,119,180}

\begin{axis}[
width=\linewidth,
height=0.8\linewidth,
legend cell align={left},
legend style={fill opacity=0.8, draw opacity=1, text opacity=1, draw=lightgrey204,at={(1.25,1.05)},anchor=south,font=\scriptsize},
legend columns=6,
tick pos=left,
x grid style={darkgrey176},
xmin=-0.73, xmax=8.73,
xtick style={color=black},
xtick={0,1,2,3,4,5,6,7,8},
xticklabel style={rotate=45.0,anchor=east,font=\tiny},
xticklabels={QoS Log Rule,RAD-DS,QoS-FK,VT-SH,EXP/PF,A-EXP/PF,2L-FLS,LTTI,WD-PS},
y grid style={darkgrey176},
ymajorgrids,
ylabel={Starvation Rate},
ymin=0, ymax=1,
ytick style={color=black},
ymajorgrids,
yminorgrids,
xlabel style={font=\scriptsize},
ylabel style={font=\scriptsize}
]

\addlegendimage{empty legend}
\addlegendentry{Number of starved UEs:}

\draw[draw=black,fill=steelblue31119180] (axis cs:-0.3,0) rectangle (axis cs:0.3,0);
\addlegendimage{ybar,ybar legend,draw=black,fill=steelblue31119180}
\addlegendentry{1}

\draw[draw=black,fill=steelblue31119180] (axis cs:0.7,0) rectangle (axis cs:1.3,0);
\draw[draw=black,fill=steelblue31119180] (axis cs:1.7,0) rectangle (axis cs:2.3,0.3);
\draw[draw=black,fill=steelblue31119180] (axis cs:2.7,0) rectangle (axis cs:3.3,0);
\draw[draw=black,fill=steelblue31119180] (axis cs:3.7,0) rectangle (axis cs:4.3,0);
\draw[draw=black,fill=steelblue31119180] (axis cs:4.7,0) rectangle (axis cs:5.3,0.9);
\draw[draw=black,fill=steelblue31119180] (axis cs:5.7,0) rectangle (axis cs:6.3,0);
\draw[draw=black,fill=steelblue31119180] (axis cs:6.7,0) rectangle (axis cs:7.3,0.444444444444444);
\draw[draw=black,fill=steelblue31119180] (axis cs:7.7,0) rectangle (axis cs:8.3,0);
\draw[draw=black,fill=darkorange25512714] (axis cs:-0.3,0) rectangle (axis cs:0.3,0.9);
\addlegendimage{ybar,ybar legend,draw=black,fill=darkorange25512714}
\addlegendentry{2}

\draw[draw=black,fill=darkorange25512714] (axis cs:0.7,0) rectangle (axis cs:1.3,0);
\draw[draw=black,fill=darkorange25512714] (axis cs:1.7,0.3) rectangle (axis cs:2.3,0.3);
\draw[draw=black,fill=darkorange25512714] (axis cs:2.7,0) rectangle (axis cs:3.3,0);
\draw[draw=black,fill=darkorange25512714] (axis cs:3.7,0) rectangle (axis cs:4.3,0.1);
\draw[draw=black,fill=darkorange25512714] (axis cs:4.7,0.9) rectangle (axis cs:5.3,1);
\draw[draw=black,fill=darkorange25512714] (axis cs:5.7,0) rectangle (axis cs:6.3,0);
\draw[draw=black,fill=darkorange25512714] (axis cs:6.7,0.444444444444444) rectangle (axis cs:7.3,0.888888888888889);
\draw[draw=black,fill=darkorange25512714] (axis cs:7.7,0) rectangle (axis cs:8.3,0);
\draw[draw=black,fill=forestgreen4416044] (axis cs:-0.3,0.9) rectangle (axis cs:0.3,1);
\addlegendimage{ybar,ybar legend,draw=black,fill=forestgreen4416044}
\addlegendentry{3}

\draw[draw=black,fill=forestgreen4416044] (axis cs:0.7,0) rectangle (axis cs:1.3,0);
\draw[draw=black,fill=forestgreen4416044] (axis cs:1.7,0.3) rectangle (axis cs:2.3,0.3);
\draw[draw=black,fill=forestgreen4416044] (axis cs:2.7,0) rectangle (axis cs:3.3,0);
\draw[draw=black,fill=forestgreen4416044] (axis cs:3.7,0.1) rectangle (axis cs:4.3,0.2);
\draw[draw=black,fill=forestgreen4416044] (axis cs:4.7,1) rectangle (axis cs:5.3,1);
\draw[draw=black,fill=forestgreen4416044] (axis cs:5.7,0) rectangle (axis cs:6.3,0);
\draw[draw=black,fill=forestgreen4416044] (axis cs:6.7,0.888888888888889) rectangle (axis cs:7.3,1);
\draw[draw=black,fill=forestgreen4416044] (axis cs:7.7,0) rectangle (axis cs:8.3,0);
\draw[draw=black,fill=crimson2143940] (axis cs:-0.3,1) rectangle (axis cs:0.3,1);
\addlegendimage{ybar,ybar legend,draw=black,fill=crimson2143940}
\addlegendentry{4}

\draw[draw=black,fill=crimson2143940] (axis cs:0.7,0) rectangle (axis cs:1.3,0);
\draw[draw=black,fill=crimson2143940] (axis cs:1.7,0.3) rectangle (axis cs:2.3,0.3);
\draw[draw=black,fill=crimson2143940] (axis cs:2.7,0) rectangle (axis cs:3.3,0);
\draw[draw=black,fill=crimson2143940] (axis cs:3.7,0.2) rectangle (axis cs:4.3,0.2);
\draw[draw=black,fill=crimson2143940] (axis cs:4.7,1) rectangle (axis cs:5.3,1);
\draw[draw=black,fill=crimson2143940] (axis cs:5.7,0) rectangle (axis cs:6.3,0);
\draw[draw=black,fill=crimson2143940] (axis cs:6.7,1) rectangle (axis cs:7.3,1);
\draw[draw=black,fill=crimson2143940] (axis cs:7.7,0) rectangle (axis cs:8.3,0);
\draw[draw=black,fill=mediumpurple148103189] (axis cs:-0.3,1) rectangle (axis cs:0.3,1);
\addlegendimage{ybar,ybar legend,draw=black,fill=mediumpurple148103189}
\addlegendentry{5}

\draw[draw=black,fill=mediumpurple148103189] (axis cs:0.7,0) rectangle (axis cs:1.3,0);
\draw[draw=black,fill=mediumpurple148103189] (axis cs:1.7,0.3) rectangle (axis cs:2.3,0.3);
\draw[draw=black,fill=mediumpurple148103189] (axis cs:2.7,0) rectangle (axis cs:3.3,0);
\draw[draw=black,fill=mediumpurple148103189] (axis cs:3.7,0.2) rectangle (axis cs:4.3,0.2);
\draw[draw=black,fill=mediumpurple148103189] (axis cs:4.7,1) rectangle (axis cs:5.3,1);
\draw[draw=black,fill=mediumpurple148103189] (axis cs:5.7,0) rectangle (axis cs:6.3,0);
\draw[draw=black,fill=mediumpurple148103189] (axis cs:6.7,1) rectangle (axis cs:7.3,1);
\draw[draw=black,fill=mediumpurple148103189] (axis cs:7.7,0) rectangle (axis cs:8.3,0);
\end{axis}

\end{tikzpicture}
        \setlength{\abovecaptionskip}{-.45cm}
        \setlength{\belowcaptionskip}{0cm}
        \vspace{-7pt}
        \caption{iPerf}
        \label{fig:starving-rates-number}
    \end{subfigure}%
    \hfill%
    \begin{subfigure}[t]{0.49\linewidth}
        \centering
\begin{tikzpicture}

\definecolor{crimson2143940}{RGB}{214,39,40}
\definecolor{darkgrey176}{RGB}{176,176,176}
\definecolor{darkorange25512714}{RGB}{255,127,14}
\definecolor{forestgreen4416044}{RGB}{44,160,44}
\definecolor{lightgrey204}{RGB}{204,204,204}
\definecolor{mediumpurple148103189}{RGB}{148,103,189}
\definecolor{steelblue31119180}{RGB}{31,119,180}

\begin{axis}[
width=\linewidth,
height=0.8\linewidth,
legend cell align={left},
legend cell align={left},
legend style={
  fill opacity=0.8,
  draw opacity=1,
  text opacity=1,
  at={(0.03,0.97)},
  anchor=north west,
  draw=lightgrey204
},
tick pos=left,
x grid style={darkgrey176},
xmin=-0.84, xmax=8.84,
xtick style={color=black},
xtick={0,1,2,3,4,5,6,7,8},
xticklabel style={rotate=45.0,anchor=east,font=\tiny},
xticklabels={QoS Log Rule,RAD-DS,QoS-FK,VT-SH,EXP/PF,A-EXP/PF,2L-FLS,LTTI,WD-PS},
y grid style={darkgrey176},
ylabel={Starvation Rate},
ymajorgrids,
ymin=0, ymax=1,
ytick style={color=black},
ymajorgrids,
yminorgrids,
xlabel style={font=\scriptsize},
ylabel style={font=\scriptsize}
]
\draw[draw=black,fill=steelblue31119180,fill opacity=0.8] (axis cs:-0.4,0) rectangle (axis cs:0.4,0);
\addlegendimage{ybar,ybar legend,draw=black,fill=steelblue31119180,fill opacity=0.8}

\draw[draw=black,fill=steelblue31119180,fill opacity=0.8] (axis cs:0.6,0) rectangle (axis cs:1.4,0);
\draw[draw=black,fill=steelblue31119180,fill opacity=0.8] (axis cs:1.6,0) rectangle (axis cs:2.4,0.1);
\draw[draw=black,fill=steelblue31119180,fill opacity=0.8] (axis cs:2.6,0) rectangle (axis cs:3.4,0);
\draw[draw=black,fill=steelblue31119180,fill opacity=0.8] (axis cs:3.6,0) rectangle (axis cs:4.4,0);
\draw[draw=black,fill=steelblue31119180,fill opacity=0.8] (axis cs:4.6,0) rectangle (axis cs:5.4,0.9);
\draw[draw=black,fill=steelblue31119180,fill opacity=0.8] (axis cs:5.6,0) rectangle (axis cs:6.4,0);
\draw[draw=black,fill=steelblue31119180,fill opacity=0.8] (axis cs:6.6,0) rectangle (axis cs:7.4,0);
\draw[draw=black,fill=steelblue31119180,fill opacity=0.8] (axis cs:7.6,0) rectangle (axis cs:8.4,0.2);
\draw[draw=black,fill=darkorange25512714,fill opacity=0.8] (axis cs:-0.4,0) rectangle (axis cs:0.4,0.1);
\addlegendimage{ybar,ybar legend,draw=black,fill=darkorange25512714,fill opacity=0.8}

\draw[draw=black,fill=darkorange25512714,fill opacity=0.8] (axis cs:0.6,0) rectangle (axis cs:1.4,0);
\draw[draw=black,fill=darkorange25512714,fill opacity=0.8] (axis cs:1.6,0.1) rectangle (axis cs:2.4,0.2);
\draw[draw=black,fill=darkorange25512714,fill opacity=0.8] (axis cs:2.6,0) rectangle (axis cs:3.4,0);
\draw[draw=black,fill=darkorange25512714,fill opacity=0.8] (axis cs:3.6,0) rectangle (axis cs:4.4,0);
\draw[draw=black,fill=darkorange25512714,fill opacity=0.8] (axis cs:4.6,0.9) rectangle (axis cs:5.4,1);
\draw[draw=black,fill=darkorange25512714,fill opacity=0.8] (axis cs:5.6,0) rectangle (axis cs:6.4,0);
\draw[draw=black,fill=darkorange25512714,fill opacity=0.8] (axis cs:6.6,0) rectangle (axis cs:7.4,0);
\draw[draw=black,fill=darkorange25512714,fill opacity=0.8] (axis cs:7.6,0.2) rectangle (axis cs:8.4,0.2);
\draw[draw=black,fill=forestgreen4416044,fill opacity=0.8] (axis cs:-0.4,0.1) rectangle (axis cs:0.4,0.9);
\addlegendimage{ybar,ybar legend,draw=black,fill=forestgreen4416044,fill opacity=0.8}

\draw[draw=black,fill=forestgreen4416044,fill opacity=0.8] (axis cs:0.6,0) rectangle (axis cs:1.4,0);
\draw[draw=black,fill=forestgreen4416044,fill opacity=0.8] (axis cs:1.6,0.2) rectangle (axis cs:2.4,0.4);
\draw[draw=black,fill=forestgreen4416044,fill opacity=0.8] (axis cs:2.6,0) rectangle (axis cs:3.4,0);
\draw[draw=black,fill=forestgreen4416044,fill opacity=0.8] (axis cs:3.6,0) rectangle (axis cs:4.4,0);
\draw[draw=black,fill=forestgreen4416044,fill opacity=0.8] (axis cs:4.6,1) rectangle (axis cs:5.4,1);
\draw[draw=black,fill=forestgreen4416044,fill opacity=0.8] (axis cs:5.6,0) rectangle (axis cs:6.4,0);
\draw[draw=black,fill=forestgreen4416044,fill opacity=0.8] (axis cs:6.6,0) rectangle (axis cs:7.4,0);
\draw[draw=black,fill=forestgreen4416044,fill opacity=0.8] (axis cs:7.6,0.2) rectangle (axis cs:8.4,0.2);
\draw[draw=black,fill=crimson2143940,fill opacity=0.8] (axis cs:-0.4,0.9) rectangle (axis cs:0.4,0.9);
\addlegendimage{ybar,ybar legend,draw=black,fill=crimson2143940,fill opacity=0.8}

\draw[draw=black,fill=crimson2143940,fill opacity=0.8] (axis cs:0.6,0) rectangle (axis cs:1.4,0);
\draw[draw=black,fill=crimson2143940,fill opacity=0.8] (axis cs:1.6,0.4) rectangle (axis cs:2.4,0.7);
\draw[draw=black,fill=crimson2143940,fill opacity=0.8] (axis cs:2.6,0) rectangle (axis cs:3.4,0);
\draw[draw=black,fill=crimson2143940,fill opacity=0.8] (axis cs:3.6,0) rectangle (axis cs:4.4,0);
\draw[draw=black,fill=crimson2143940,fill opacity=0.8] (axis cs:4.6,1) rectangle (axis cs:5.4,1);
\draw[draw=black,fill=crimson2143940,fill opacity=0.8] (axis cs:5.6,0) rectangle (axis cs:6.4,0);
\draw[draw=black,fill=crimson2143940,fill opacity=0.8] (axis cs:6.6,0) rectangle (axis cs:7.4,0);
\draw[draw=black,fill=crimson2143940,fill opacity=0.8] (axis cs:7.6,0.2) rectangle (axis cs:8.4,0.2);
\draw[draw=black,fill=mediumpurple148103189,fill opacity=0.8] (axis cs:-0.4,0.9) rectangle (axis cs:0.4,1);
\addlegendimage{ybar,ybar legend,draw=black,fill=mediumpurple148103189,fill opacity=0.8}

\draw[draw=black,fill=mediumpurple148103189,fill opacity=0.8] (axis cs:0.6,0) rectangle (axis cs:1.4,0);
\draw[draw=black,fill=mediumpurple148103189,fill opacity=0.8] (axis cs:1.6,0.7) rectangle (axis cs:2.4,0.8);
\draw[draw=black,fill=mediumpurple148103189,fill opacity=0.8] (axis cs:2.6,0) rectangle (axis cs:3.4,0.1);
\draw[draw=black,fill=mediumpurple148103189,fill opacity=0.8] (axis cs:3.6,0) rectangle (axis cs:4.4,0);
\draw[draw=black,fill=mediumpurple148103189,fill opacity=0.8] (axis cs:4.6,1) rectangle (axis cs:5.4,1);
\draw[draw=black,fill=mediumpurple148103189,fill opacity=0.8] (axis cs:5.6,0) rectangle (axis cs:6.4,0);
\draw[draw=black,fill=mediumpurple148103189,fill opacity=0.8] (axis cs:6.6,0) rectangle (axis cs:7.4,0);
\draw[draw=black,fill=mediumpurple148103189,fill opacity=0.8] (axis cs:7.6,0.2) rectangle (axis cs:8.4,0.2);
\end{axis}

\end{tikzpicture}
        \setlength{\abovecaptionskip}{0cm}
        \setlength{\belowcaptionskip}{0cm}
        \vspace{-7pt}
        \caption{Bursty Traffic}
        \label{fig:starvation-rates-ftp}
    \end{subfigure}
    \vspace{-5pt}
    \caption{Starving rate of QoS-aware schedulers.}
    \label{fig:disconnection_qos_aware}
    \vspace{-3pt}
\end{figure}
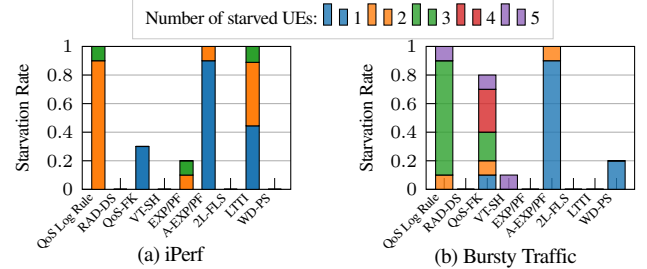
We plot starvation rates in Fig. \ref{fig:disconnection_qos_aware}, and we observe that, in the iPerf scenario \gls{qos} Log Rule, A-EXP/PF and LTTI also have $100\%$ starvation rate. However, a further analysis shows a fundamental difference: A-EXP/PF always starves one of the priviledged \glspl{ue}, while the other two algorithms always starve non-priviledged \glspl{ue}, which means, in these cases, that starvation is due to  aggressive prioritization which is not necessarily a flaw. We observe the same trend with the Bursty scenario, the main difference being that QoS-FK also starves 80\% of the runs, with both RT and NRT \glspl{ue} starved.
\begin{figure}[htbp]
    \vspace{-10pt}
    \centering
    \scriptsize
    \resizebox{\linewidth}{!}{%
\begin{tikzpicture}
\definecolor{darkgrey176}{RGB}{176,176,176}
\definecolor{lightgrey204}{RGB}{204,204,204}
\definecolor{mediumseagreen34167132}{RGB}{34,167,132}
\definecolor{teal41120142}{RGB}{41,120,142}
\begin{axis}[
width=\linewidth,
height=0.4\linewidth,
legend cell align={left},
legend style={
  fill opacity=0.8,
  draw opacity=1,
  text opacity=1,
  at={(0.03,0.97)},
  anchor=north west,
  draw=lightgrey204
},
tick align=outside,
tick pos=left,
x grid style={darkgrey176},
xmin=-0.635, xmax=5.635,
xtick style={color=black},
xtick={0,1,2,3,4,5},
xticklabel style={rotate=45.0,anchor=east},
xticklabels={RAD-DS,QoS-FK,VT-SH,EXP/PF,2L-FLS,WD-PS},
y grid style={darkgrey176},
ylabel={Avg. Bitrate (Mbps)},
ymin=0, ymax=25.3201447358636,
ytick style={color=black},
ymajorgrids
]
\draw[draw=black,fill=teal41120142,fill opacity=0.85] (axis cs:-0.35,0) rectangle (axis cs:0,20.1247520693965);
\addlegendimage{ybar,ybar legend,draw=black,fill=mediumseagreen34167132,fill opacity=0.85}
\addlegendentry{RT UEs (21, 24)}
\draw[draw=black,fill=teal41120142,fill opacity=0.85] (axis cs:0.65,0) rectangle (axis cs:1,14.6576497885463);
\draw[draw=black,fill=teal41120142,fill opacity=0.85] (axis cs:1.65,0) rectangle (axis cs:2,16.9096260202093);
\draw[draw=black,fill=teal41120142,fill opacity=0.85] (axis cs:2.65,0) rectangle (axis cs:3,7.07308175685358);
\draw[draw=black,fill=teal41120142,fill opacity=0.85] (axis cs:3.65,0) rectangle (axis cs:4,8.19527717106535);
\draw[draw=black,fill=teal41120142,fill opacity=0.85] (axis cs:4.65,0) rectangle (axis cs:5,8.49916847621424);
\draw[draw=black,fill=mediumseagreen34167132,fill opacity=0.85] (axis cs:0,0) rectangle (axis cs:0.35,10.6377048196738);
\addlegendimage{ybar,ybar legend,draw=black,fill=teal41120142,fill opacity=0.85}
\addlegendentry{NRT UEs (22, 23, 25)}
\draw[draw=black,fill=mediumseagreen34167132,fill opacity=0.85] (axis cs:1,0) rectangle (axis cs:1.35,23.7407263750672);
\draw[draw=black,fill=mediumseagreen34167132,fill opacity=0.85] (axis cs:2,0) rectangle (axis cs:2.35,14.9655632257283);
\draw[draw=black,fill=mediumseagreen34167132,fill opacity=0.85] (axis cs:3,0) rectangle (axis cs:3.35,24.1120713652172);
\draw[draw=black,fill=mediumseagreen34167132,fill opacity=0.85] (axis cs:4,0) rectangle (axis cs:4.35,24.1144235579653);
\draw[draw=black,fill=mediumseagreen34167132,fill opacity=0.85] (axis cs:5,0) rectangle (axis cs:5.35,23.8779246353313);
\end{axis}
\end{tikzpicture}
    }
    \vspace{-16pt}
    \caption{Average throughput for \gls{qos}-Aware schedulers (iPerf scenario)}
    \label{fig:throughput_aware}
\end{figure}
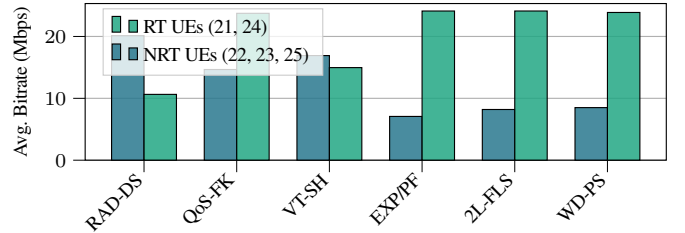
We plot the aggregated throughput of the iPerf scenario in Fig. \ref{fig:throughput_aware}. The first observation we make is that with RAD-DS and VT-SH, the throughput of RT \glspl{ue} is below the rate transmitted by iPerf of $24$ Mbps. This means despite supposed prioritization, those \glspl{ue} experience packet loss, which is typically unacceptable for RT traffic. The other algorithms successfully allocate resources to RT \glspl{ue} first before letting NRT \gls{ue} transmit.

%
\begin{figure}[t]
    \centering
    \includegraphics[width=\linewidth]{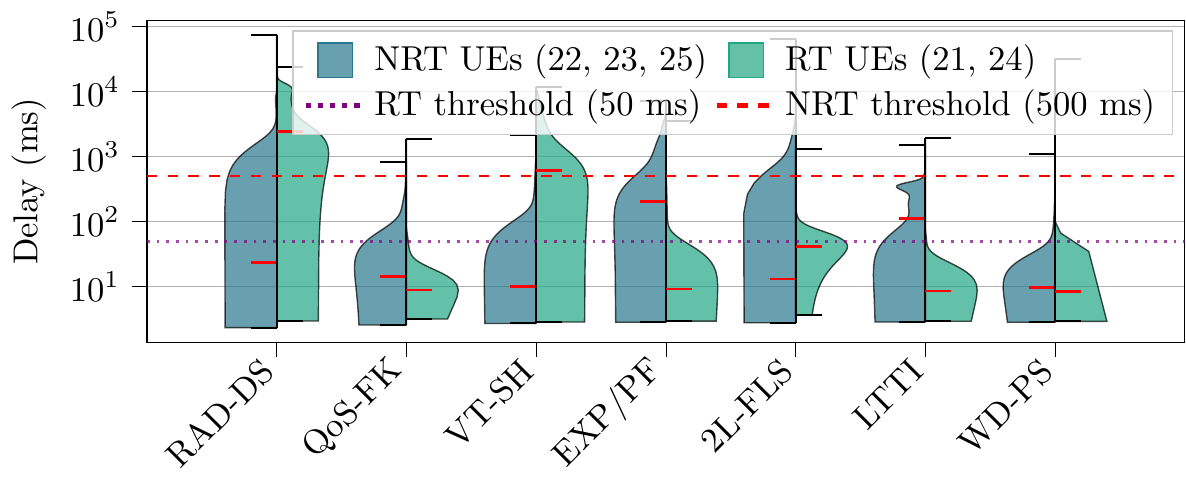}
    \vspace*{-22pt}
    \caption{Delay for \gls{qos}-aware schedulers (bursty scenario)}
    \label{fig:latency_aware}
\end{figure}

\begin{figure}[t]
    \centering
    \includegraphics[width=\linewidth]{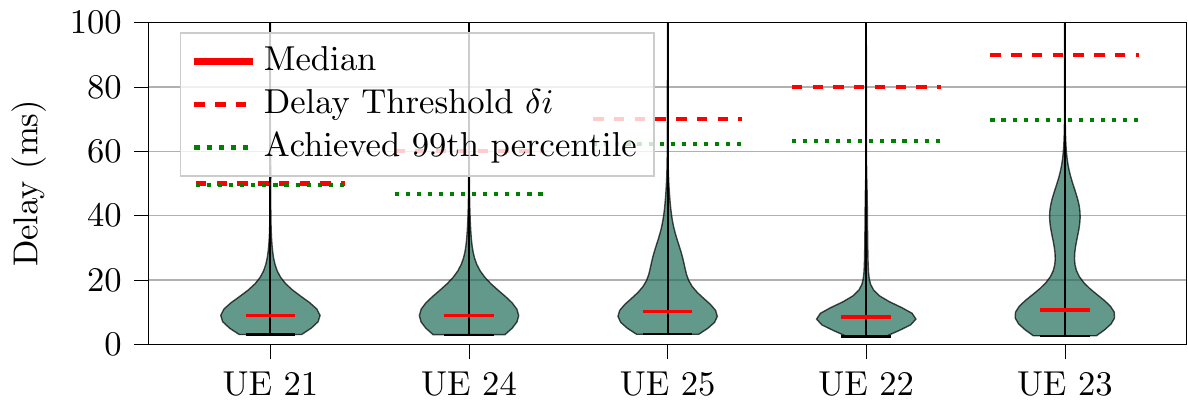}
    \vspace*{-15pt}
    \caption{Delay distribution for LTTI with per-\gls{ue} thresholds}
    \label{fig:res_LTTI}
    \vspace{-10pt}
\end{figure}

The main result in the bursty scenario is depicted in Fig. \ref{fig:latency_aware}, where we observe that most schedulers do not meet the required latency for the RT users. Out of all the schedulers, QoS-FK and LTTI stand out since they are the only schedulers which succeed at satisfying the thresholds. 
For this reason, we run extra tests for those two schedulers. In this new scenario, the arrival rate during bursts is set to 1500 packets/s and we set a different 5QI for each \gls{ue}, each mapping to a different delay target of 50, 60, 70, 80, and 90 ms. 
During this experiment, all runs were starved for QoS-FK, hence we only show the results of LTTI, in Fig. \ref{fig:res_LTTI}. The key takeaway of this experiment is that LTTI successfully accommodates various constraints. We also note that for UE 23, and also for the NRT \glspl{ue} of Fig. \ref{fig:latency_aware}, the distribution of delays is multimodal. This is due to (i) prioritization, which causes unpriviledged users to accumulate larger buffers more often and (ii) the exponential factor of the algorithm (see Table \ref{tab:scheduling_algorithms_qos}), which causes surges of priority when the $\HOLdelayuser{i}$ of a \gls{ue} becomes close to its threshold. Overall, this set of tests has enabled us to weed out most \gls{qos}-aware schedulers due to their various flaws, and to ensure that, by leveraging LTTI for delay-related queries, \framework will be able to accommodate complex intents involving various delay thresholds without issue.

%

\section{\gls{ibs} Use Cases}
\label{sec:ibs-results}

In this section, we showcase our \gls{ibs} framework, introduced in Section~\ref{sec:ibs}. 
Recall that, in order to leverage the schedulers selected, we will map their different elements into a Knapsack Problem. Hence, the first part of this section is dedicated to the mapping of some of the tested algorithms into the \gls{ibs} framework, along with additional grouping and limits functions. We then evaluate \gls{ibs} on use cases that require dynamic grouping of the \glspl{ue}, which is not effectively addressed by existing slicing-based solutions.

\subsection{\gls{ibs} Scheduling Elements Library}
\label{sec:ibs-scheds}

We discuss here the scheduling elements of the  library (see in Fig.~\ref{fig:IBS_agent}) based on the insights from Section~\ref{sec:num-results}.

\textbf{Value Functions $v$.} For \gls{qos}-unaware schedulers, we generalize GPF (for its flexibility) and PFB (for cost-free low-latency) into a new \gls{gpfb} reward function:
$${[\Rinstuser{i}]^{\beta}}/{[\Raveuser{i}]^{\gamma}} \times\Buffuser{i}.$$
We evaluate \gls{gpfb} with the same parameters as GPF, and present the throughput similarity metric (from the iPerf experiment) in Table \ref{tab:GPFB}, along with the 95th percentile for the latency (from the bursty traffic experiments). We observe that when trying to achieve reasonable levels of fairness (first two cases in the table), GPFB is extremely similar to GPF. 
However, for the more unfair case of $\beta = 0.6, \gamma = 0.2$, it tends to achieve a significantly higher level of fairness than the GPF. We argue this is due to the $\Buffuser{i}$ factor, which incentivizes serving \glspl{ue} with low $\Raveuser{i}$, as, they have a worst $\Rinstuser{i}$, which leads to more buffering.
We include GPFB with an option to deactivate the $\Buffuser{i}$ factor, for when high unfairness is desired, and implement it as a value function within the \gls{ibs} framework.

\begin{table}[b]
    \vspace{-10pt}
    \scriptsize
    \centering

    \begin{tabular}{ll ccc ccc}
    \toprule
        \multirow{2}{*}{Parameters} & \multirow{2}{*}{$\Rave$ RMSE} & \multicolumn{2}{c}{Gini Index} & \multicolumn{2}{c}{Latency 95th Quantile} \\
        \cmidrule(lr){3-4} \cmidrule(lr){5-6}
        & & GPF & GPFB & GPF & GPFB \\
    \midrule
        $\beta = 0.2, \gamma = 0.7$ & $0.51$ Mbps & $0.037$ & $0.047$ & $722.00$ & $118.31$ \\ 
        $\beta = 0.6, \gamma = 0.7$ & $1.15$ Mbps & $0.112$ & $0.115$ & $1596.31$ & $108.44$ \\
        $\beta = 0.6, \gamma = 0.2$ & $4.87$ Mbps & $0.331$ & $0.221$ & $5339.74$ & $85.79$ \\
    \bottomrule
    \end{tabular}
\caption{Statistics on GPF vs. GPFB.}
    \label{tab:GPFB}
\end{table}

Among the \gls{qos}-aware schedulers, we include LTTI as it is the only algorithm that reliably respects the delay thresholds in the bursty scenario. This algorithm is also implemented as a value function, in two versions that can either handle a single delay target or, similar to the original paper, arbitrary subgroups of a group can be defined to handle multiple delay targets.

\textbf{Grouping Functions.} We implement multiple grouping function. First, that of EUFS, which separates \glspl{ue} based on their $\Raveuser{i}$ rank, along with an extra function based on the \gls{cqi}. We also provide a bursty traffic grouping function, which demonstrates how \gls{ibs} could be used in conjunction with real-time traffic classification based on the buffer size. For the scope of the paper, we emulate it using 5QI, with further traffic classification left as future work. 
Finally, we propose a grouping function based on a \gls{lstm} model classifying  mobile and static \glspl{ue}. It is trained on \glspl{kpi} (\verb|pusch_snrx10|, \verb|pucch_snrx10|, \verb|dl_rsrp|, \verb|bler|, \verb|rssi|) collected \gls{ota} using a Samsung S23 phone held by a user alternatively moving and staying static, forming a labeled dataset. We obtain a 98\% validation accuracy for this classification task, with 20\% of the traces held as validation. We use PyTorch for training and uses ONNX for inference within \gls{ibs}.

\textbf{Limit Functions.} We provide a limit function, setting the \gls{prb} limit to zero if the \gls{ue} $\Raveuser{i}$ is above a pre-defined threshold and to the maximum number of \glspl{prb} otherwise, which enables throughput throttling. We also provide a default limit function, which sets the upper limit to the number of \glspl{prb} allocated to the group. 
\vspace{-10pt} 

\subsection{Flexible Delay-Aware Scheduling}
\label{sec:use-case-flexible-delay-aware-scheduling}

In this use case, we task the \gls{llm} with adapting the delay objective to the number of bursty \glspl{ue} and to the network load, making the requirements more stringent for \glspl{ue} with a better channel. While we will showcase the results with our static \glspl{ue}, which eases interpretability, such an intent can be useful for high mobility cases where delays are important but it is acceptable to temporarily degrade the SLA while the channel is poor. We also make sure the \gls{llm} allocates a fixed portion of resources to non-bursty \glspl{ue}. Our intent is:
\begin{mdframed}[backgroundcolor=gray!10, linewidth=1pt, linecolor=black]
\scriptsize
\textbf{Intent:} Generate a scheduler that: \\
-uses 80\% of PRBs to give to bursty UEs \\
-if there are 2 UEs or less with bursty traffic, guarantee them a delay of 50ms \\
-else, guarantee 50ms of delay to 2 bursty UEs with best RSRP and 300ms to others \\
-makes sure other UEs use spectrum efficiently with some degree of fairness
\end{mdframed}
The resulting scheduler code is shown in Listing \ref{listing:ibs1}, where we see that the \gls{llm} successfully formulates a new dynamic value function. To test this scheduler, we use the bursty scenario with 1500 packets/s during bursts for bursty \glspl{ue}. We first run the test with all \glspl{ue} receiving bursty traffic except for \gls{ue} 25. Then, only \glspl{ue} 21 and 24 receive bursty traffic. We depict the results in Table \ref{tab:results_ibs} where we can observe that the 
scheduler successfully meets the delay thresholds of the different \glspl{ue} in both scenarios, and that when 4 bursty \glspl{ue} are present, the two with a higher RSRP have a significantly lower delay. Furthermore, when there is only one non-bursty \gls{ue} (\gls{ue} 25), its delay is an order of magnitude lower ($528$ ms vs $14$ s) than in the scenario with 2 bursty \glspl{ue}, since, in that case, three \glspl{ue} have to share 20\% of the resources, leading to congestion.


\begin{table}[]
    \centering
    \caption{Delay results for Flexible Delay Aware IBS scenarios with either 2 or 4 bursty \glspl{ue}. RSRP is measured in dBm.}
    \begin{tabular}{ll | ll | ll}
    \toprule
        \multirow{2}{*}{UE} & Avg & Intent $\delta_i$ & Delay & Intent $\delta_i$ & Delay \\

        &   RSRP & (2 bursty UEs) &  95th \%ile & (4 bursty UEs) & 95th \%ile \\
    \midrule
        21 & 58 & $50$ ms & $25.11$ ms & $300$ ms & $130.68$ ms \\
        22 & 74 & N/A & $29$ s & $50$ ms & $37.66$ ms\\
        23 & 76 & N/A & $10.8$ s & $50$ ms & $32.35$ ms\\
        24 & 58 & $50$ ms & $20.89$ ms & $300$ ms & $153.34$ ms \\
        25 & 59 & N/A & $14$ s & N/A & $528$ ms \\
    \bottomrule
    \end{tabular}
    \label{tab:results_ibs}
    \vspace{-10pt}
\end{table}

\begin{lstlisting}[float=ht,floatplacement=h,language=json,style=LuaStyle,
caption={Resulting code from \gls{llm} for first \gls{ibs} usecase}, 
label={listing:ibs1}]
function get_schedulers()
    return {
        { {set_limits=set_default_limits, set_values=generate_conditional_value_function( adaptive_delay_handler)}
        },
        {   {set_limits=set_default_limits, set_values=generate_set_values_GPFB(0.8, 0.6, 1)}
        }
    }
end

function adaptive_delay_handler(group_metrics)
    local total_ues = #group_metrics
    if total_ues <= 2 then
        return generate_set_values_LTTI(50000, 0.95)
    else
        local high_rsrp_rntis = {}
        local low_rsrp_rntis = {}
        local ues_with_rsrp = {}
        for _, ue in ipairs(group_metrics) do
            if ue.dl_rsrp then
                table.insert(ues_with_rsrp, ue)
            else
                table.insert(low_rsrp_rntis, ue.rnti)
            end
        end
        table.sort(ues_with_rsrp, function(a, b) return a.dl_rsrp > b.dl_rsrp end)
        for i = 1, math.min(2, #ues_with_rsrp) do
            table.insert(high_rsrp_rntis, ues_with_rsrp[i].rnti)
        end
        for i = 3, #ues_with_rsrp do
            table.insert(low_rsrp_rntis, ues_with_rsrp[i].rnti)
        end
        return generate_set_values_LTTI_multi({50000, 300000}, {0.95, 0.95}, {high_rsrp_rntis, low_rsrp_rntis})
    end
end
compute_targets = generate_compute_targets( generate_groups_by_burstiness(0.8), get_schedulers())

\end{lstlisting}

\subsection{Mobility-aware scheduling}
\begin{figure}[htbp]
    \vspace{-3pt}
    \centering
    \scriptsize
    \resizebox{\linewidth}{!}{%
        \input{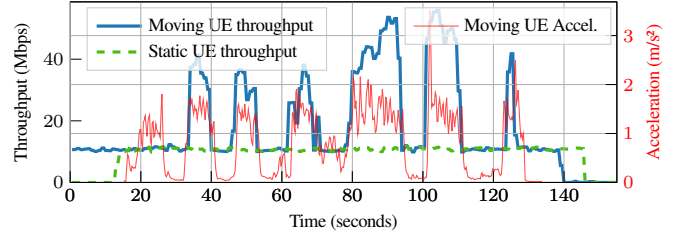}
    }
    \vspace{-10pt} 
    \caption{Throughput over time with mobility-aware scheduler.}
    \label{fig:mobility}
    \vspace{-5pt}
\end{figure}

In this use case, we consider scenarios such as industry 4.0 in which \glspl{ue} are embedded on mobile robots. Such robots might use congestion-control protocols (such as TCP for general sensor data and webRTC to stream camera feeds). In such a case, it may be desirable to throttle static robots and favor currently moving ones, as performing tasks while moving may typically require more precise control, i.e., more granular data, which means more bandwidth. Our intent is hence:
\begin{mdframed}[backgroundcolor=gray!10, linewidth=1pt, linecolor=black]
\scriptsize
\textbf{Intent:} Generate a scheduler that equally allocates resources to static and dynamic UEs, but make sure static UEs do not get more than 10 Mbps.
\end{mdframed}
We obtain a scheduler which leverages the grouping by mobility feature along with the limit-based throughput throttling to satisfy the intent. We test the scheduler using one of the Sierra Wireless \glspl{ue} (UE 22) as a static \gls{ue} along with a Samsung S23, held by a user who alternatively walks around the lab and stops. We also use the \textit{phyphox} App to collect accelerometer data (which are displayed in Fig.~\ref{fig:mobility} for reference but not consumed by the scheduler). As depicted in Fig. \ref{fig:mobility}, the grouping function successfully classifies the Sierra \gls{ue} as static, while the moving \gls{ue} is classified as static or dynamic depending on the mobility, which directly impacts the throughput.

\section{Conclusions}
\label{sec:conclusions}

In this paper, we presented \framework, a novel approach that bridges the gap between intent-based network management and real-time scheduling in \gls{oran} architectures. By leveraging \glspl{llm} to interpret operator intents, automatically generate schedulers from research literature, and match intents to appropriate scheduling algorithms, we enable dynamic customization of \gls{ran} behavior at unprecedented granularity. Our extensive experimental evaluation across 18 different scheduling algorithms in a production-ready 5G network demonstrated significant variations between theoretical performance claims and actual \gls{ota} deployment results, underscoring the importance of our approach for practical implementation of intent-based scheduling.
Our framework also demonstrated successfully the ability to adapt scheduling policies based on fine-grained conditions such as buffer status, mobility patterns, and traffic classification—all without requiring extensive retraining for each new intent. 

\vspace{-5pt}
\balance
\bibliographystyle{IEEEtran}
\bibliography{biblio}

\begin{thebibliography}{10}
\providecommand{\url}[1]{#1}
\csname url@samestyle\endcsname
\providecommand{\newblock}{\relax}
\providecommand{\bibinfo}[2]{#2}
\providecommand{\BIBentrySTDinterwordspacing}{\spaceskip=0pt\relax}
\providecommand{\BIBentryALTinterwordstretchfactor}{4}
\providecommand{\BIBentryALTinterwordspacing}{\spaceskip=\fontdimen2\font plus
\BIBentryALTinterwordstretchfactor\fontdimen3\font minus \fontdimen4\font\relax}
\providecommand{\BIBforeignlanguage}[2]{{%
\expandafter\ifx\csname l@#1\endcsname\relax
\typeout{** WARNING: IEEEtran.bst: No hyphenation pattern has been}%
\typeout{** loaded for the language `#1'. Using the pattern for}%
\typeout{** the default language instead.}%
\else
\language=\csname l@#1\endcsname
\fi
#2}}
\providecommand{\BIBdecl}{\relax}
\BIBdecl

\bibitem{tataria20216g}
H.~Tataria, M.~Shafi, A.~F. Molisch, M.~Dohler, H.~Sjöland, and F.~Tufvesson, ``{6G Wireless Systems: Vision, Requirements, Challenges, Insights, and Opportunities},'' \emph{Proceedings of the IEEE}, vol. 109, no.~7, pp. 1166--1199, July 2021.

\bibitem{polese2022understanding}
M.~Polese, L.~Bonati, S.~D’Oro, S.~Basagni, and T.~Melodia, ``{Understanding O-RAN: Architecture, Interfaces, Algorithms, Security, and Research Challenges},'' \emph{{IEEE Communications Surveys \& Tutorials}}, 2023.

\bibitem{lacava2025dapps}
\BIBentryALTinterwordspacing
A.~Lacava, L.~Bonati, N.~Mohamadi, R.~Gangula, F.~Kaltenberger, P.~Johari, S.~D'Oro, F.~Cuomo \emph{et~al.}, ``{dApps: Enabling Real-Time AI-Based Open RAN Control},'' pp. 1--31, 2025. [Online]. Available: \url{https://arxiv.org/pdf/2501.16502}
\BIBentrySTDinterwordspacing

\bibitem{kundu2025ai}
L.~Kundu, X.~Lin, R.~Gadiyar, J.-F. Lacasse, and S.~Chowdhury, ``{AI-RAN: Transforming RAN with AI-driven Computing Infrastructure},'' \emph{arXiv preprint arXiv:2501.09007}, 2025.

\bibitem{leivadeas2023survey}
A.~Leivadeas and M.~Falkner, ``{A Survey on Intent-Based Networking},'' \emph{IEEE Communications Surveys \& Tutorials}, 2023.

\bibitem{nahum2024intent}
C.~V. Nahum, V.~H.~L. Lopes, R.~M. Dreifuerst, P.~Batista, I.~Correa, K.~V. Cardoso, A.~Klautau, and R.~W. Heath, ``{Intent-Aware Radio Resource Scheduling in a RAN Slicing Scenario Using Reinforcement Learning},'' \emph{IEEE Transactions on Wireless Communications}, vol.~23, no.~3, pp. 2253--2267, March 2024.

\bibitem{zhong2017heterogeneous}
Y.~Zhong, T.~Q.~S. Quek, and X.~Ge, ``{Heterogeneous Cellular Networks With Spatio-Temporal Traffic: Delay Analysis and Scheduling},'' \emph{IEEE Journal on Selected Areas in Communications}, vol.~35, no.~6, pp. 1373--1386, June 2017.

\bibitem{capozzi2013downlink}
F.~Capozzi, G.~Piro, L.~Grieco, G.~Boggia, and P.~Camarda, ``{Downlink Packet Scheduling in LTE Cellular Networks: Key Design Issues and a Survey},'' \emph{IEEE Communications Surveys \& Tutorials}.

\bibitem{mamane2022scheduling}
A.~Mamane, M.~Fattah, M.~E. Ghazi, M.~E. Bekkali, Y.~Balboul, and S.~Mazer, ``{Scheduling Algorithms for 5G Networks and Beyond: Classification and Survey},'' \emph{IEEE Access}, 2022.

\bibitem{kaltenberger2019openairinterface}
F.~Kaltenberger, G.~De~Souza, R.~Knopp, and H.~Wang, ``{The OpenAirInterface 5G new radio implementation: Current status and roadmap},'' in \emph{ITG WS}, 2019.

\bibitem{kelly1997charging}
F.~Kelly, ``{Charging and rate control for elastic traffic},'' \emph{European transactions on Telecommunications}, vol.~8, no.~1, pp. 33--37, 1997.

\bibitem{saglam20195g}
M.~I. Saglam and M.~Kartal, ``{5G enhanced mobile broadband downlink scheduler},'' in \emph{ELECO}, 2019.

\bibitem{sadiq2010delay}
B.~Sadiq, S.~J. Baek, and G.~De~Veciana, ``{Delay-optimal opportunistic scheduling and approximations: The log rule},'' \emph{ToN}, 2010.

\bibitem{mamane2021proportional}
A.~Mamane, M.~Fattah, M.~El~Ghazi, Y.~Balboul, M.~El~Bekkali, and S.~Mazer, ``{Proportional fair buffer scheduling algorithm for 5G enhanced mobile broadband.}'' \emph{International Journal of Electrical \& Computer Eng.}, 2021.

\bibitem{ramjee2006generalized}
T.~Ramjee, T.~Bu, and L.~Li, ``{Generalized proportional fair scheduling in third generation wireless data networks},'' in \emph{IEEE INFOCOM}, 2006.

\bibitem{afifi2021novel}
W.~S. Afifi, A.~A. El-Moursy, M.~Saad, S.~M. Nassar, and H.~M. El-Hennawy, ``{A novel scheduling technique for improving cell-edge performance in 4G/5G systems},'' \emph{Ain Shams Engineering Journal}, 2021.

\bibitem{sadiq2009downlink}
B.~Sadiq, R.~Madan, and A.~Sampath, ``{Downlink scheduling for multiclass traffic in LTE},'' \emph{EURASIP Wireless Com. and Net.}, 2009.

\bibitem{monghal2010dynamic}
G.~Monghal, D.~Laselva, P.-H. Michaelsen, and J.~Wigard, ``{Dynamic packet scheduling for traffic mixes of best effort and VoIP users in E-UTRAN downlink},'' in \emph{Vehicular Technology Conference}, 2010.

\bibitem{brehm2013overload}
M.~Brehm and R.~Prakash, ``{Overload-state downlink resource allocation in LTE MAC layer},'' \emph{Wireless networks}, vol.~19, pp. 913--931, 2013.

\bibitem{iturralde2012resource}
M.~Iturralde, A.~Wei, T.~A. Yahiya, and A.-L. Beylot, ``{Resource allocation for real time services using cooperative game theory and a virtual token mechanism in LTE networks},'' in \emph{CCNC}, 2012.

\bibitem{basukala2009performance}
R.~Basukala, H.~M. Ramli, and K.~Sandrasegaran, ``{Performance analysis of EXP/PF and M-LWDF in downlink 3GPP LTE system},'' in \emph{2009 First Asian Himalayas International Conference on Internet}.\hskip 1em plus 0.5em minus 0.4em\relax IEEE, 2009.

\bibitem{rhee2003scheduling}
J.-H. Rhee, J.~M. Holtzman, and D.-K. Kim, ``{Scheduling of real/non-real time services: adaptive EXP/PF algorithm},'' in \emph{VTC}, 2003.

\bibitem{piro2011two}
G.~Piro, L.~A. Grieco, G.~Boggia, R.~Fortuna, and P.~Camarda, ``{Two-level downlink scheduling for real-time multimedia services in LTE networks},'' \emph{IEEE Transactions on Multimedia}, 2011.

\bibitem{mahfoudi2015new}
M.~Mahfoudi, M.~E. Bekkali, A.~Najd, M.~El~Ghazi, and S.~Mazer, ``{A new downlink scheduling algorithm proposed for real time traffic in LTE system},'' \emph{International Journal of Electronics and Telecom.}, 2015.

\bibitem{husain2020efficient}
M.~I. Husain, M.~E. Haque, and F.~Tariq, ``{An efficient packet scheduling algorithm for URLLC systems},'' in \emph{UCET}, 2020.

\bibitem{leivadeas2022survey}
A.~Leivadeas and M.~Falkner, ``{A survey on intent-based networking},'' \emph{IEEE Communications Surveys \& Tutorials}, 2022.

\bibitem{wei2020intent}
Y.~Wei, M.~Peng, and Y.~Liu, ``{Intent-based networks for 6G: Insights and challenges},'' \emph{Digital Communications and Networks}, 2020.

\bibitem{mijumbi2015network}
R.~Mijumbi, J.~Serrat, J.-L. Gorricho, N.~Bouten, F.~De~Turck, and R.~Boutaba, ``{Network function virtualization: State-of-the-art and research challenges},'' \emph{IEEE Communications surveys \& tutorials}, 2015.

\bibitem{xia2014survey}
W.~Xia, Y.~Wen, C.~H. Foh, D.~Niyato, and H.~Xie, ``{A survey on software-defined networking},'' \emph{IEEE Com. Surveys \& Tutorials}, 2014.

\bibitem{han2016intent}
Y.~Han, J.~Li, D.~Hoang, J.-H. Yoo, and J.~W.-K. Hong, ``{An intent-based network virtualization platform for SDN},'' in \emph{CNSM}, 2016.

\bibitem{pham2016sdn}
M.~Pham and D.~B. Hoang, ``{SDN applications-The intent-based Northbound Interface realisation for extended applications},'' in \emph{NetSoft}, 2016.

\bibitem{szyrkowiec2018automatic}
T.~Szyrkowiec, M.~Santuari, M.~Chamania, D.~Siracusa, A.~Autenrieth, V.~Lopez, J.~Cho, and W.~Kellerer, ``{Automatic intent-based secure service creation through a multilayer SDN network orchestration},'' \emph{JOCN}, 2018.

\bibitem{foukas2017network}
X.~Foukas, G.~Patounas, A.~Elmokashfi, and M.~K. Marina, ``{Network slicing in 5G: Survey and challenges},'' \emph{Communications Magazine}, 2017.

\bibitem{cheng2024oranslice}
H.~Cheng, S.~D'Oro, R.~Gangula, S.~Velumani, D.~Villa, L.~Bonati, M.~Polese, T.~Melodia \emph{et~al.}, ``{ORANSlice: An Open Source 5G Slicing Platform for O-RAN},'' in \emph{MobiCom}, 2024.

\bibitem{chen2023flexslice}
C.-C. Chen, C.-Y. Chang, and N.~Nikaein, ``{FlexSlice: Flexible and real-time programmable RAN slicing framework},'' in \emph{GLOBECOM 2023}.

\bibitem{tsampazi2024pandora}
M.~Tsampazi, S.~D'Oro, M.~Polese, L.~Bonati, G.~Poitau, M.~Healy, M.~Alavirad, and T.~Melodia, ``{PandORA: Automated design and comprehensive evaluation of deep reinforcement learning agents for Open RAN},'' \emph{TMC}, 2024.

\bibitem{polese2022colo}
M.~Polese, L.~Bonati, S.~D'Oro, S.~Basagni, and T.~Melodia, ``Colo-ran: Developing machine learning-based xapps for open ran closed-loop control on programmable experimental platforms,'' \emph{TMC}, 2022.

\bibitem{yeh2023deep}
S.-P. Yeh, S.~Bhattacharya, R.~Sharma, and H.~Moustafa, ``{Deep learning for intelligent and automated network slicing in 5G open RAN (ORAN) deployment},'' \emph{Open Journal of the Communications Society}, 2023.

\bibitem{kouchaki2022actor}
M.~Kouchaki and V.~Marojevic, ``{Actor-critic network for O-RAN resource allocation: xApp design, deployment, and analysis},'' in \emph{Globecom WS}, 2022.

\bibitem{filali2023communication}
A.~Filali, B.~Nour, S.~Cherkaoui, and A.~Kobbane, ``{Communication and computation O-RAN resource slicing for URLLC services using deep reinforcement learning},'' \emph{Communications Standards Magazine}, 2023.

\bibitem{fattah2002overview}
H.~Fattah and C.~Leung, ``{An overview of scheduling algorithms in wireless multimedia networks},'' \emph{Wireless Communications}, 2002.

\bibitem{hu2021distributed}
S.~Hu, X.~Chen, W.~Ni, E.~Hossain, and X.~Wang, ``{Distributed Machine Learning for Wireless Communication Networks: Techniques, Architectures, and Applications},'' \emph{IEEE Com. Surveys \& Tutorials}, 2021.

\bibitem{apostolakis2023athena}
N.~Apostolakis, M.~Gramaglia, L.~E. Chatzieleftheriou, T.~Subramanya, A.~Banchs, and H.~Sanneck, ``Athena: Machine learning and reasoning for radio resources scheduling in vran systems,'' \emph{JSAC}, 2023.

\bibitem{chinchali2018cellular}
S.~Chinchali, P.~Hu, T.~Chu, M.~Sharma, M.~Bansal, R.~Misra, M.~Pavone, and S.~Katti, ``Cellular network traffic scheduling with deep reinforcement learning,'' in \emph{Proceedings of the AAAI Conference on Artificial Intelligence}, vol.~32, no.~1, 2018.

\bibitem{eschmann2021reward}
J.~Eschmann, ``Reward function design in reinforcement learning,'' \emph{Reinforcement learning algorithms: Analysis and Applications}, pp. 25--33, 2021.

\bibitem{nikbakht2024tspec}
\BIBentryALTinterwordspacing
R.~Nikbakht, M.~Benzaghta, and G.~Geraci, ``{TSpec-LLM: An Open-source Dataset for LLM Understanding of 3GPP Specifications},'' 2024. [Online]. Available: \url{https://arxiv.org/abs/2406.01768}
\BIBentrySTDinterwordspacing

\bibitem{saraiva2024telco}
T.~Saraiva, M.~Sousa, P.~Vieira, and A.~Rodrigues, ``{Telco-DPR: A Hybrid Dataset for Evaluating Retrieval Models of 3GPP Technical Specifications},'' \emph{arXiv preprint arXiv:2410.19790}, 2024.

\bibitem{zou2024telecomgpt}
H.~Zou, Q.~Zhao, Y.~Tian, L.~Bariah, F.~Bader, T.~Lestable, and M.~Debbah, ``{TelecomGPT: A framework to build telecom-specfic large language models},'' \emph{arXiv preprint arXiv:2407.09424}, 2024.

\bibitem{wu2025llm}
X.~Wu, J.~Farooq, Y.~Wang, and J.~Chen, ``{LLM-xApp: A Large Language Model Empowered Radio Resource Management xApp for 5G O-RAN},'' in \emph{NDSS Workshop (FutureG)}, 2025.

\bibitem{bao2025llmguidedopenranempowering}
\BIBentryALTinterwordspacing
L.~Bao, S.~Yun, J.~Lee, and T.~Q.~S. Quek, ``{LLM-Guided Open RAN: Empowering Hierarchical RAN Intelligent Control},'' 2025. [Online]. Available: \url{https://arxiv.org/abs/2504.18062}
\BIBentrySTDinterwordspacing

\bibitem{maxenti2025autoranautomatedzerotouchopen}
\BIBentryALTinterwordspacing
S.~Maxenti, R.~Shirkhani, M.~Elkael, L.~Bonati, S.~D'Oro, T.~Melodia, and M.~Polese, ``{AutoRAN: Automated and Zero-Touch Open RAN Systems},'' 2025. [Online]. Available: \url{https://arxiv.org/abs/2504.11233}
\BIBentrySTDinterwordspacing

\bibitem{mekrache2024llm}
A.~Mekrache and A.~Ksentini, ``{LLM-enabled intent-driven service configuration for next generation networks},'' in \emph{NetSoft}, 2024.

\bibitem{mekrache2024intent}
A.~Mekrache, A.~Ksentini, and C.~Verikoukis, ``{Intent-based management of next-generation net.: An LLM-centric approach},'' \emph{IEEE Network}, 2024.

\bibitem{changjie2024netconfeval}
W.~Changjie, ``{NetConfEval: Can LLMs Facilitate Network Configuration?}'' in \emph{CoNext}, 2024.

\bibitem{ifland2024genet}
\BIBentryALTinterwordspacing
B.~Ifland, E.~Duani, R.~Krief, M.~Ohana, A.~Zilberman, A.~Murillo, O.~Manor, O.~Lavi \emph{et~al.}, ``{GeNet: A Multimodal LLM-Based Co-Pilot for Net. Topology and Conf.}'' 2024. [Online]. Available: \url{https://arxiv.org/abs/2407.08249}
\BIBentrySTDinterwordspacing

\bibitem{seo2025paper2code}
M.~Seo, J.~Baek, S.~Lee, and S.~J. Hwang, ``{Paper2Code: Automating Code Generation from Scientific Papers in ML},'' \emph{arXiv preprint}, 2025.

\bibitem{dapp-ngrg-report}
{O-RAN next Generation Research Group (nGRG)}, ``{dApps for Real-Time RAN Control: Use Cases and Requirements},'' 2024.

\bibitem{3GPP-38214}
\BIBentryALTinterwordspacing
{3GPP}, ``{NR; Physical layer procedures for data},'' {3rd Generation Partnership Project (3GPP)}, {Technical Specification (TS)} 38.214, Jun. 2025, version 19.0.0. [Online]. Available: \url{https://www.3gpp.org/DynaReport/38214.htm}
\BIBentrySTDinterwordspacing

\bibitem{SCF2021FAPI}
{Small Cell Forum}, ``{5G} {FAPI}: {PHY} {API} {S}pecification,'' Tech. Rep., 2021.

\bibitem{luajit}
\BIBentryALTinterwordspacing
M.~Pall, ``{LuaJIT}.'' [Online]. Available: \url{https://luajit.org/}
\BIBentrySTDinterwordspacing

\bibitem{blecher2023nougat}
L.~Blecher, G.~Cucurull, T.~Scialom, and R.~Stojnic, ``{Nougat: Neural Optical Understanding for Academic Documents},'' 2023.

\bibitem{villa2024x5g}
D.~Villa, I.~Khan, F.~Kaltenberger, N.~Hedberg, R.~S. da~Silva, S.~Maxenti, L.~Bonati, A.~Kelkar \emph{et~al.}, ``{X5G: An Open, Programmable, Multi-vendor, End-to-end, Private 5G O-RAN Testbed with NVIDIA ARC and OpenAirInterface},'' \emph{arXiv}, 2024.

\end{thebibliography}
\bstctlcite{IEEEexample:BSTcontrol}



\end{document}